\documentclass[12pt,a4paper]{article}
\usepackage[T2A]{fontenc}
\usepackage[cp1251]{inputenc}
\usepackage{cite}
\usepackage{mathtext}
\usepackage{amssymb,amsthm,amsmath}
\usepackage{array}
\usepackage[dvips]{epsfig}
\usepackage[russian, english]{babel}
\usepackage[title,titletoc]{appendix}

\begin{document}
\author{\bf Yu.A. Markov$\!\,$\thanks{e-mail:markov@icc.ru}$\,$,
M.A. Markova$\!\,$\thanks{e-mail:markova@icc.ru}$\,$,
and A.I. Bondarenko$\!\,$\thanks{e-mail:370omega@mail.ru}}
\title{Third order wave equation\\
in Duffin-Kemmer-Petiau theory. Massive case}
%
%
\date{\it\normalsize $^{\ast}\!$Institute for System Dynamics and Control Theory SB RAS\\
P.O. Box 1233, 664033 Irkutsk, Russia\\
\vspace{0.2cm}
$^{\dagger}\!$Irkutsk State
University, Department of Theoretical Physics,\\ 664003, Gagarin blrd,
20, Irkutsk, Russia}
\thispagestyle{empty}
\maketitle{}
\def\theequation{\arabic{section}.\arabic{equation}}
\[
{\bf Abstract}
\]

{
\noindent
Within the framework of the Duffin-Kemmer-Petiau (DKP) formalism a more consistent approach to the derivation of the third order wave equation obtained earlier by M. Nowakowski [\hspace{0.02cm}Phys.Lett.A {\bf 244} (1998) 329\hspace{0.02cm}] on the basis of heuristic considerations is suggested. For this purpose an additional algebraic object, the so-called $q$\hspace{0.02cm}-\hspace{0.02cm}commutator ($q$ is a primitive cubic root of unity) and a new set of matrices $\eta_{\mu}$ instead of the original matrices $\beta_{\mu}$ of the DKP algebra are introduced. It is shown that in terms of these $\eta_{\mu}$ matrices we have succeeded in reducing a procedure of the construction of cubic root of the third order wave operator to a few simple algebraic transformations and to a certain operation of the passage to the limit $z \rightarrow q$, where $z$ is some complex deformation parameter entering into the definition of the $\eta$\hspace{0.02cm}-\hspace{0.02cm}matrices. A corresponding generalization of the result obtained to the case of the interaction with an external electromagnetic field introduced through the minimal coupling scheme is carried out and a comparison with M. Nowakowski's result is performed.
A detailed analysis of the general structure for a solution of the first order differential equation for the wave function $\psi(x; z)$ is performed and it is shown that the solution is singular in the $z \rightarrow q$ limit. The application to the problem of construction within the DKP approach of the path integral representation in parasuperspace for the propagator of a massive vector particle in a background gauge field is discussed.
}
{}


\newpage

\section{Introduction}
\setcounter{equation}{0}

In the paper by Nowakowski \cite{nowakowski_1998} devoted to the problem of electromagnetic coupling in the Duffin-Kemmer-Petiau (DKP) theory several rather unusual circumstances relating to a second order DKP equation have been pointed out. The first of them is connected with the fact that the second order Kemmer equation \cite{kemmer_1939} in the presence of an external electromagnetic field is only one member of a class of second order equations which, in principle, can be derived from the first order DKP equation. Their physical meaning is therefore not entirely clear. Another circumstance is connected with the fact that the second order Kemmer equation lacks a back-transformation which would allow one to obtain solutions of the first order DKP equation from solutions of the second order equation. The reason for the latter is that the Klein-Gordon-Fock divisor \cite{umezawa_1956, takahashi_book} in the spin-1 case (throughout this work we put $\hbar = c = 1$)
\begin{equation}
d(\partial) = \frac{1}{m}\,(\hspace{0.02cm}\Box + m^{2}\hspace{0.02cm})I  + i\hspace{0.02cm}\beta_{\mu}\hspace{0.02cm}\partial^{\mu}
-\frac{1}{m}\,\beta_{\mu}\beta_{\nu}\hspace{0.02cm}\partial^{\mu}\partial^{\nu}
\label{eq:1q}
\end{equation}
ceases to be commuted with the original DKP operator
\begin{equation}
L(\partial) \equiv i\hspace{0.02cm}\beta_{\mu}\partial^{\mu} - m\hspace{0.02cm} I,
\label{eq:1w}
\end{equation}
when we introduce the interaction with an external electromagnetic field within the framework of the minimal coupling scheme $\partial^{\mu}\rightarrow D^{\mu}\equiv\partial^{\mu} + i\hspace{0.015cm}eA^{\mu}$, i.e.
\[
[\hspace{0.04cm}d(D), L(D)\hspace{0.03cm}] \neq 0.
\]
Here $I$ is the unity matrix; $\Box \equiv \partial_{\mu} \partial^{\mu}, \, \partial_{\mu} = \partial/\partial x^{\mu}$, and the matrices $\beta_{\mu}$ obey the famous trilinear relation
\begin{equation}
\beta_{\mu}\beta_{\nu}\beta_{\lambda} + \beta_{\lambda}\beta_{\nu}\beta_{\mu} =
g_{\mu\nu}\beta_{\lambda} + g_{\lambda\nu}\beta_{\mu}
\label{eq:1e}
\end{equation}
with the metric $g_{\mu \nu} = {\rm diag}\hspace{0.03cm}(1, -1, -1, -1)$. As a result, the analogy of the second order Kemmer equation to a similar looking Dirac equation is very limited. Whereas in the Dirac case one can transform solutions of the second order equation to solutions of the Dirac equation and vice versa, such a one-to-one correspondence is not possible in the Kemmer case.\\
\indent
Nowakowski has suggested a way this problem may be circumvented. To achieve the commutativity of the reciprocal operator $d(D)$ and the DKP operator $L(D)$ in the presence of an external gauge field we have to give up the requirement that the product of these two operators is an operator of the Klein-Gordon-Fock type, i.e.
\[
d(D) L(D) \neq - (D^2 + m^2)I + {\cal G}\hspace{0.02cm}[A_{\mu}],
\]
where ${\cal G}\hspace{0.02cm}[A_{\mu}]$ is a functional of the potential $A_{\mu}$, which vanishes in the  interaction free case. In other words it is necessary to introduce into consideration not the second order, but a higher order wave equation which would have the same virtue as the second order Dirac equation, i.e. a back-transformation to the solutions of the first order equation. In the paper \cite{nowakowski_1998} from heuristic considerations such a higher (third) order wave equation possessing a necessary property of the reversibility was proposed. However, by virtue of the fact that the higher order equation is not reduced to the Klein-Gordon-Fock equation in the interaction free case, this leads to the delicate question of physical interpretation of the terms in such a higher order equation (it is known that even for the second order Kemmer equation there exists this kind of problem). In particular, this is concerned with the interpretation of parameter $m$ as the mass of a particle since this is the only possible interpretation in the free case when the following equality,
\begin{equation}
d(\partial) L(\partial) = -\hspace{0.03cm} (\hspace{0.02cm}\Box + m^2)I
\label{eq:1ee}
\end{equation}
holds.\\
\indent
It should be noted that the divisors with a minimal electromagnetic coupling and, in particular, for the spin-1 case
\begin{equation}
d(D) = \frac{1}{m}\,(\hspace{0.02cm}D^{2} + m^{2}\hspace{0.02cm})I  + i\hspace{0.02cm}\beta_{\mu}\hspace{0.02cm}D^{\mu}
-\frac{1}{m}\,\beta_{\mu}\beta_{\nu}\hspace{0.02cm}D^{\mu}D^{\nu}
\label{eq:1r}
\end{equation}
were first introduced into consideration in the earlier papers by Nagpal \cite{nagpal_1974}, Cox \cite{cox_1976}, and Krajcik and Nieto \cite{krajcik_1976}. The divisors have been intensively used in analysis of causality violation in higher spin theories in the presence of an electromagnetic field. The suggested divisors represent merely a straightforward  generalization of the well-known operators of Takahashi and Umezawa \cite{umezawa_1956, takahashi_book} by the replacement $\partial_{\mu}\rightarrow D_{\mu}(A)$. However, a question of commutativity of the generalized divisors with the initial first order operators $L(D)$ in these papers was not discussed at all, although this can be of certain importance. Further, in the papers mentioned above the questions of causality were discussed on the basis of analysis of a product of two operators $d(D)$ and $L(D)$. In particular, in the spin-1 case when we take the divisor in the form (\ref{eq:1r}), in the product $d(D)L(D)$ the principle part of interacting and free wave operators remains the same as it was defined by Eq.\,(\ref{eq:1ee}). This is connected with the fact that the terms of the third order in derivatives reduce to the terms of the first order by using the trilinear relation (\ref{eq:1e}) and thus the effect of electromagnetic interactions (or nonderivative coupling) occurs only in lower derivatives. The resulting field equation remains equivalent to a hyperbolic system with light cone as ray cone, the same holds in the interacting and free cases. Therefore, it is concluded that the spin-1 field even in the presence of electromagnetic field in the system possesses only causal modes of propagation.\\
\indent
The situation can qualitatively change if as $d(D)$ one takes a divisor such that
\[
[\hspace{0.02cm}d(D), L(D)\hspace{0.02cm}]=0,
\]
for example, the divisor suggested by Nowakovski \cite{nowakowski_1998}. In this case in the product $d(D)L(D)$ in accordance with formula (\ref{eq:7p})  in Section 7, the principle part of the interacting and free wave operators will be already the third order in derivatives, instead of (\ref{eq:1ee}), and the terms with the nonderivative coupling remain the terms of the first order. The question of whether a change of the order of the principle part of wave operator leads to a change of the propagation properties of the equation
\[
d(D) L(D) \psi(x)=0,
\]
generally speaking, has to be the subject of separate research.\\
\indent
Further, in addition to the absence of required one-to-one correspondence between solutions of the second order Kemmer equation and the DKP equation, one can point out one more negative consequence  of the noncommutativity of the divisor (\ref{eq:1r}) with the operator $L(D)= i\hspace{0.02cm} \beta_{\mu} D^{\mu} - m I$. The lack of commutativity does not give a possibility within the framework of the DKP theory to construct the path integral representation for the Green's function of a spin-1 particle in a background gauge field in a spirit of the approaches developed for a spin-1/2 particle (see for example \cite{borisov_1982, fradkin_1991}).
Having obtained all the necessary expressions, this very interesting question will be discussed in more detail  in section 8.\\
\indent
The purpose of this paper is to give a systematic way of deriving the third order wave equation within the framework of the massive Duffin-Kemmer-Petiau theory in the free and interacting cases. However,  first of all it should be noted that the wave equations of the third order in derivatives, as applied to the problems of classical and quantum field theories, for any length of time have drawn attention of researchers for various reasons. Below, we give a number of examples related somehow to our problem.\\
\indent
In the papers by Finkelstein {\it et al.} \cite{finkelstein_1986} in constructing the theory of the nine-dimensional ternary hyperspin manifold the so-called {\it trine-Gordon} equation, the unique scalar wave equation of least differential order\footnote{All formulas cited below, up to Eq.\,(\ref{eq:1u}), are given in the notations of the authors of the corresponding works.}
\[
\bigl[\hspace{0.02cm}\det(\partial) - i\hspace{0.03cm} m^{3}\hspace{0.02cm}\bigr]\varphi = 0
\]
was suggested. Here, $\det(\partial)$ is the determinant of a $3\times3$ matrix composed of partial derivatives in coordinates. The authors have also performed an analysis of the corresponding dispersion relation for plane waves and have suggested the generalization to the case of a minimal interaction with a gauge field. The questions close to this research were considered independently in the papers by Solov'yov {\it et al.} \cite{solovyov_2001}, where the general algebraic theory of the Finslerian spinors was constructed. The generalized Duffin-Kemmer-Petiau equation for a Finslerian 3-spinor wave function of a free particle in the momentum representation was also suggested there and it was shown that each of these 3-spinor components of the wave function ($i^{r},\, \beta_{\dot{s}}$) satisfies the Finslerian analog of the Klein-Gordon-Fock equation
\[
\bigl(G_{ABC}\hspace{0.02cm}P^{A\!}\hspace{0.02cm}P^{B\!}\hspace{0.02cm}P^{C} - M^{\hspace{0.02cm}3}\hspace{0.02cm}\bigr)i^{r} = 0, \quad r = 1,2,3
\]
and a similar equation holds for the $\beta_{\dot{s}}$ components. Here, $G_{ABC}$ is a symmetric covariant tensor of the third order rank that plays a role of the metric tensor in the nine-dimensional linear Finslerian space.\\
\indent
Further, in the works by Yamaleev \cite{yamaleev_1987, yamaleev_1988, yamaleev_1989} an attempt has been made to construct in a systematic way the foundations of quantum mechanics on cubic forms (or even more generally, polylinear forms). The mathematical basis of the construction would become the cyclic algebras of $N>2$ degree. The cyclic algebra with respect to cubic forms here plays a role like the Clifford algebra with respect to the quadratic forms. In particular, the cubic generalization of the standard relativistic relation between energy ${\cal E}$, momentum $\vec{p}=(p_1, p_2, p_3)$ and mass $m$ was suggested in the following form:
\begin{equation}
({\cal E} - q\hspace{0.03cm}m)({\cal E} - q^{2}m)({\cal E} - m) = \sum_{i\hspace{0.02cm} = \hspace{0.01cm}1}^{3}p_{i}^{\hspace{0.03cm}3} - 3\hspace{0.02cm} p_{1}\hspace{0.01cm}p_{2}\hspace{0.01cm}p_{3},
\label{eq:1t}
\end{equation}
where $q$ is a primitive cubic root of unity
\begin{equation}
\begin{split}
&q = {\rm e}^{2\hspace{0.02cm}\pi i/3} = -\frac{1}{2} + i\,\frac{\sqrt{3}}{2}\hspace{0.02cm},\\
&q^2 = {\rm e}^{4\hspace{0.005cm}\pi i/3} = -\frac{1}{2} - i\,\frac{\sqrt{3}}{2}\hspace{0.02cm}.
\end{split}
\label{eq:1y}
\end{equation}
As an analog of the Klein-Gordon-Fock equation (for any of three possible correspondences:
${\cal E}\rightarrow \theta_{k}\hspace{0.02cm} \partial/\partial t, p_i \rightarrow \theta_k \partial/\partial x_i,$ where $\theta_k=(q, q^2, 1),\,k=1, 2, 3$ in Eq.\,(\ref{eq:1t})) the following equation of the third order
\[
\left(\frac{\partial}{\partial\hspace{0.02cm}  t^{3}} - \sum_{i\hspace{0.02cm} =\hspace{0.02cm} 1}^{3}\frac{\partial^{\hspace{0.03cm} 3}}{\partial x_{i}^{3}}\right)\varphi
\hspace{0.02cm} +\hspace{0.02cm}
\frac{\partial^{\hspace{0.03cm}3}}{\partial x_{1}\hspace{0.02cm} \partial x_{2}\hspace{0.02cm}
\partial x_{3}}\,\varphi
-
m^{3}\varphi = 0
\]
was used. The author also suggested cubic analogs of the Dirac equation and the nonrelativistic Schr\"odinger equation.\\
\indent
The third order wave equations arise, however, not only in the generalization of quantum mechanics and quantum field theory to more abstract spaces in the foundation of which not the quadratic forms of various type (an interval, relativistic relationship between mass, energy and momentum and so on), but the forms of one degree higher are laid. These equations arise also within the framework of generally accepted physical theories for solving quite concrete problems. Thus, one of the first generalizations of this kind in the context of quantum electrodynamics can be found in the paper by Pais and Uhlenbeck \cite{pais_1950}. The latter have considered the generalization of the Dirac equation to the multimass Dirac equation like
\[
\prod_{j = 1}^{N}\hspace{0.03cm}(\gamma_{\mu}\partial_{\mu} + m_{j}) \hspace{0.02cm} \psi(x) = 0,
\]
where $N$ is an integer, which in particular can take the value 3. The purpose of this generalization of the spinor field equation to equation of higher order was to eliminate by this means the divergent features in quantum field theory.\\
\indent
Further, in the paper by Barut {\it et al.} \cite{barut_1970} another version of the generalized Dirac equation of the third order in derivatives describing particles with spin 1/2 and three mass states was suggested. The aim of this paper was to treat in a unified fashion all known at that time leptonic matter. The case when one of the states is massless (neutrino) and the corresponding generalized equation takes the form
\[
\bigl[i\hspace{0.03cm}\alpha_{1}(\gamma\cdot\partial) + \alpha_{2}\hspace{0.03cm}\Box - i\hspace{0.02cm}\alpha_{3}\hspace{0.02cm}(\gamma\cdot\partial)\hspace{0.03cm}\Box
\hspace{0.03cm}\bigr] \psi(x) = 0
\]
was studied in more detail. The parameters $\alpha_{1}, \alpha_2$, and $\alpha_3$ are related to the electron mass $m_{e}$ and muon one $m_{\mu}$ through the relations $\alpha_1/\alpha_3 = m_{e}\hspace{0.02cm}m_{\mu}, \, \alpha_2{}/\alpha_{3} = m_{e} + m_{\mu}$. It is interesting to observe that the term of  the third order in derivatives in the equation above has the structure similar to the corresponding term on the right-hand side of Eq.\,(\ref{eq:1p}) for the spin-1 case, which will be discussed further. In the paper by Kruglov \cite{kruglov_2007} this model was analyzed for the case when all three states are massive. As was noted in the last paper, such higher order differential equations may be treated as effective equations and represent a nonperturbative approach to quantum field theory.\\
\indent
In the spin-$\frac{3}{2}$ theories by Joos \cite{joos_1962}, Weinberg \cite{weinberg_1964} and Shay {\it et al.} \cite{shay_1965} it was shown that the corresponding wave function $\psi(x)$ in the interaction free case must satisfy component by component not only the second order Klein-Gordon-Fock equation, but also the third order wave equation of the type
\[
\gamma_{\mu\nu\lambda}\,\frac{\partial^{3}\psi(x)}{\partial x_{\mu}\partial x_{\nu}\partial x_{\lambda}} - m^{3}\psi(x) = 0.
\]
Here the $8 \times 8$ matrices $\gamma_{\mu \nu \lambda}$ are defined in terms of the spin-$\frac{3}{2}$  matrices $s_{i}$, $i=1, 2, 3$ and obey algebraic relation representing the spin-$\frac{3}{2}$ generalization of relation for the Dirac matrices.\\
\indent
Finally, we can also mention  that in the familiar formulation of Bhabha \cite{bhabha_1945} (see also \cite{krajcik_1974}) of the multimass high-spin theory, for the spin-1 case we have instead of the Klein-Gordon-Fock equation the third order one
\[
(\alpha\cdot\partial)(\hspace{0.02cm}\Box - m^{2}\hspace{0.03cm})\hspace{0.02cm}\psi = 0.
\]
The extra differential factor $(\alpha \cdot \partial)$ comes from the subsidiary components.\\
\indent
The examples given above show that the higher order systems, in particular the third order ones, might themselves have some applicability in field theories.\\
\indent
Before proceeding with the formal development of the construction of the third order wave equation within the framework of the DKP approach in the presence of an electromagnetic field, it is necessary to ask what form the equation should have in the interaction free case. It is necessary to have at hand a certain simple rule of deriving this equation (and perhaps, the equations of higher order for high-spin cases). Here, we attempt to follow as close as possible the free Dirac theory added by some considerations of algebraic character. For this purpose let us introduce a set of the square roots of unity: $(\lambda, 1)$, where $\lambda\!\equiv\! -1$ is the primitive square root. Then it is obvious that \begin{equation}
(\hspace{0.02cm}i\hspace{0.02cm}\gamma_{\mu}\hspace{0.02cm}\partial^{\mu} - \lambda\hspace{0.015cm} m\hspace{0.02cm}I)
\hspace{0.02cm}
(\hspace{0.02cm}i\hspace{0.02cm}\gamma_{\nu}\hspace{0.02cm}\partial^{\nu} - m\hspace{0.02cm}I) =
-\hspace{0.03cm} (\hspace{0.02cm}\Box + m^2)I.
\label{eq:1u}
\end{equation}
Let us state a question of defining such a matrix ${\cal O}$ that the following equality holds
\begin{equation}
\bigl[\hspace{0.01cm}
{\cal O}\hspace{0.03cm}(\hspace{0.02cm}
i\hspace{0.02cm}\gamma_{\mu}\hspace{0.02cm}\partial^{\mu} - m\hspace{0.02cm}I)\bigr]
\bigl[
{\cal O}\hspace{0.04cm}(\hspace{0.02cm}
i\hspace{0.02cm}\gamma_{\nu}\hspace{0.02cm}\partial^{\nu} - m\hspace{0.02cm}I)\bigr] =
-\hspace{0.03cm} (\hspace{0.02cm}\Box + m^2)I.
\label{eq:1i}
\end{equation}
In fact it represents a solution of the problem of constructing the square root of the Klein-Gordon-Fock operator. Here, the answer is known. As such a matrix one has to take
\[
{\cal O} = \pm\hspace{0.03cm} i\hspace{0.03cm}\gamma_{5},
\]
where
\begin{equation}
\gamma_{5} = \frac{i}{4!}\;\epsilon^{\mu\nu\lambda\sigma}\gamma_{\mu}\gamma_{\nu}
\gamma_{\lambda}\gamma_{\sigma}, \quad
\gamma_{5}^{2} = 1.
\label{eq:1o}
\end{equation}
Thus we can consider that the expression on the left-hand side of (\ref{eq:1u}) gives us the rule for the determination of the right form of the second order wave operator (the right-hand side of (\ref{eq:1u})) and in turn the expression on the left-hand side of Eq.\,(\ref{eq:1i}) gives its square root and thereby the problem is reduced to the construction of an algorithm of calculating the matrix ${\cal O}$.\\
\indent
If one consider as a guiding principle the considerations above, then the next step will be the following extension: as a basis we take the cubic roots of unity $(q, q^2, 1)$, where the primitive roots $q$ and $q^2$ are given by the formulas (\ref{eq:1y}), and as the spin matrices we take the $\beta$\hspace{0.02cm}-\hspace{0.02cm}matrices of the DKP algebra. It is an easy matter to verify that an analog of equation (\ref{eq:1u}) will be that in the following form (cp. with (\ref{eq:1t})):
\begin{equation}
(\hspace{0.02cm}i\hspace{0.02cm}\beta_{\mu}\hspace{0.02cm}\partial^{\mu} - q\hspace{0.02cm}m\hspace{0.02cm}I)
\hspace{0.02cm}
(\hspace{0.02cm}i\hspace{0.01cm}\beta_{\nu}\hspace{0.02cm}\partial^{\nu} -  q^{2}\hspace{0.01cm}m\hspace{0.02cm}I)
\hspace{0.02cm}
(\hspace{0.02cm}i\hspace{0.01cm}\beta_{\lambda}\hspace{0.02cm}\partial^{\lambda} - m\hspace{0.02cm}I)
=
-i\,\Box\hspace{0.03cm}\beta_{\mu}\hspace{0.02cm}\partial^{\mu} - m^{3\!}\hspace{0.02cm}I.
\label{eq:1p}
\end{equation}
On the right-hand side of (\ref{eq:1p}) we now have the differential operator of the third order, which we take as a ``genuine'' expression for the third order wave operator. It is precisely this expression that arises from the Nowakowski's third order wave equation \cite{nowakowski_1998} in the limit when we switch off an external electromagnetic field. In deriving (\ref{eq:1p}) one of the properties of the roots of unity, namely
\begin{equation}
1 + q + q^{2} = 0
\label{eq:1a}
\end{equation}
and the identity
\begin{equation}
\beta_{\mu}\beta_{\nu}\beta_{\lambda}\partial^{\mu}\partial^{\nu}\partial^{\lambda} =
\Box\hspace{0.02cm}\beta_{\mu}\partial^{\mu}
\label{eq:1s}
\end{equation}
valid in view of the algebra of the $\beta$-matrices, Eq.\,(\ref{eq:1e}), were used. Besides, we have taken into account the fact that the mass term $m I$ is diagonal and commutes with everything. We note that such an approach was used in the papers by Kerner \cite{kerner_1992} devoted to a generalization of supersymmetry based on $Z_3$-\hspace{0.02cm}qraded algebras, more exactly in the construction of the operators whose trilinear combinations yield the supersymmetric generators (cubic root of the supersymmetry (SUSY) translations).\\
\indent
Further we can state a question of defining a matrix $A$ such that the following relation holds:
\begin{equation}
\bigl[\hspace{0.03cm}A(\hspace{0.02cm}i\hspace{0.02cm}\beta_{\mu}\hspace{0.02cm}\partial^{\mu} - m\hspace{0.02cm}I)\bigr]
\bigl[\hspace{0.03cm}A(\hspace{0.02cm}i\hspace{0.02cm}\beta_{\nu}\hspace{0.02cm}\partial^{\nu} - m\hspace{0.02cm}I)\bigr]
\bigl[\hspace{0.03cm}A(\hspace{0.02cm}i\hspace{0.02cm}\beta_{\lambda}\hspace{0.02cm}
\partial^{\lambda} - m\hspace{0.02cm}I)\bigr]
=
-i\,\frac{1}{m}\,\Box\hspace{0.03cm}\beta_{\mu}\hspace{0.02cm}\partial^{\mu} - m^{2\!}\hspace{0.02cm}I.
\label{eq:1d}
\end{equation}
The latter solves the problem of calculating the cubic root of the third order wave operator. In this paper we have attempted to answer this question by using a very rich apparatus of the matrix algebra in the DKP theory added by new structures generated by algebra of the cubic roots of unity. We
have also performed a generalization of the resulting equations to the case of the presence in the system of an external electromagnetic field.\\
\indent
The paper is organized as follows. In Section 2 the construction of cubic root of the second order Klein-Gordon-Fock operator within the framework of DKP formalism is considered. This problem has a purely auxiliary character. However, a number of expressions derived here are of decisive importance for subsequent research. Section~3 is devoted to the construction of the cubic root of the third order wave operator. For this purpose an additional algebraic object, the $q$-commutator representing a generalization of the usual commutator by entering a primitive cubic root of unity $q$ into the initial definition, is introduced. This new algebraic object has allowed us to remove not only the terms linear in derivatives, but also the quadratic terms as it is required by virtue of the definition of the third order wave operator, Eq.\,(\ref{eq:1d}). However, at the same time it is found that the necessary term cubic in derivatives, on the symmetrization, vanishes identically.\\
\indent
In Section 4 a new set of matrices $\eta_{\mu}$ instead of the original matrices $\beta_{\mu}$ is introduced. It is shown that these matrices possess rather nontrivial commutation relations which enable us to reduce the problem of the construction of the desired cubic root to a number of simple algebraic operations. On the basis of these matrices the reason for vanishing the term of the third order in derivatives is analyzed and a way to overcome this problem is suggested. Section 5 is concerned with the discussion of various properties of the $\eta$\hspace{0.02cm}-\hspace{0.02cm}matrices: commutation relations, the trilinear relation (the analog of the trilinear relation for the $\beta$\hspace{0.02cm}-\hspace{0.02cm}matrices), the behavior on Hermitian conjugation, etc. In Section 6 an extension of the results of the previous sections to the case of the presence in the system of an external electromagnetic field is performed. The detailed comparison of the expression for the third order wave operator with a similar expression earlier obtained by Nowakowski \cite{nowakowski_1998} is given. In Section 7 an analysis of the general structure for a solution of the first order differential equation for the wave function $\psi(x; z)$, where $z$ is the deformation parameter is performed. It is shown that this solution is the singular one in the limit $z \rightarrow q$. In Section 8 a question of a possible application of the results obtained to the problem of the construction within the framework of the Duffin-Kemmer-Petiau formalism of the path integral representation for the propagator of a vector particle in a background gauge field is discussed. In the concluding Section 9 the key points of our work are specified and the massless limit of the third order wave operator is briefly discussed.\\
\indent
In Appendix \ref{appendix_A} all of the necessary formulas of the DKP algebra of the matrices $\beta_{\mu}$ are listed. In Appendix \ref{appendix_B} a procedure of the construction of a certain matrix ${\cal A}$ is presented. This matrix formally can be considered as a matrix analog of the primitive cubic roots of unity, i.e. a set of the matrices $({\cal A}, {\cal A}^{\hspace{0.02cm}2}, {\cal A}^{\hspace{0.02cm}3}\!\equiv\! \frac{\!1}{m}\hspace{0.02cm} I)$ satisfies the properties identical to those for a set of the cubic roots of unity: $(q, q^{\hspace{0.03cm}2}, q^{\hspace{0.03cm}3}\!\equiv\!1)$. In Appendix \ref{appendix_B1} a complete proof of vanishing cube of matrix differential operator, where matrix is defined through the deformed commutator is produced. In Appendix \ref{appendix_C} the details of the proof of trilinear relation of the type (\ref{eq:1e}) for a new set of the matrices $\eta_{\mu}$ are given. Finally, in Appendix \ref{appendix_D} the proof of the identity (\ref{eq:7e}) for a product of three covariant derivatives is presented.


\section{\bf Cubic root of the Klein-Gordon-Fock equation}
\setcounter{equation}{0}

Before proceeding to the problem stated in Introduction, we first to consider a question
of the construction of cubic root of the second order massive Klein-Gordon-Fock operator. This
problem in the general statement has been investigated by Plyushchay and Rausch de Traubenberg in the paper \cite{plyushchay_2000}. Here, we examine it again and look how far we can proceed in solving this problem while remaining within the framework of DKP formalism only.\\
\indent
Let us now turn to Eq.\,(\ref{eq:1d}), but instead of the third order operator we put the
Klein-Gordon-Fock operator on the right-hand side
\begin{equation}
\bigl[\hspace{0.03cm}A(\hspace{0.02cm}i\hspace{0.02cm}\beta_{\mu}\hspace{0.02cm}\partial^{\mu}
- m\hspace{0.02cm}I)\bigr]
\bigl[\hspace{0.03cm}A(\hspace{0.02cm}i\hspace{0.02cm}\beta_{\nu}\hspace{0.02cm}\partial^{\nu}
- m\hspace{0.02cm}I)\bigr]
\bigl[\hspace{0.03cm}A(\hspace{0.02cm}i\hspace{0.02cm}\beta_{\lambda}\hspace{0.02cm}
\partial^{\lambda} - m\hspace{0.02cm}I)\bigr]
=
-\hspace{0.03cm} (\hspace{0.02cm}\Box + m^2)I.
\label{eq:2q}
\end{equation}
One can somewhat simplify the problem if one takes the operator on the right-hand side in the
factorized form
\[
-\hspace{0.03cm} (\hspace{0.02cm}\Box + m^2)I = d(\partial)\hspace{0.02cm}
\bigl(i\hspace{0.02cm}\beta_{\mu}\hspace{0.02cm}\partial^{\mu} - m\hspace{0.02cm} I\bigr).
\]
We recall that
\[
d(\partial) = m\hspace{0.02cm} I + i\hspace{0.02cm}\beta_{\mu}\hspace{0.02cm}\partial^{\mu} +
\bigl(\hspace{0.01cm}2\hspace{0.02cm}g_{\mu\nu} -\hspace{0.02cm} \{\beta_{\mu},\beta_{\nu}\}\bigr)\frac{\partial^{\mu}\partial^{\nu}}{2\hspace{0.02cm}m}
\]
is the Klein-Gordon-Fock divisor in the spin-1 case; $\{\,, \}$ designates anticommutator. By virtue of this factorization we can examine instead of (\ref{eq:2q}) the following equation:
\begin{equation} A\hspace{0.02cm}\bigl(i\hspace{0.02cm}\beta_{\mu\,}\partial^{\mu} - m\hspace{0.02cm} I\bigr) A\hspace{0.02cm}\bigl(i\hspace{0.02cm}\beta_{\nu\,}\partial^{\nu} - m\hspace{0.02cm} I\bigr)A
= d(\partial).
\label{eq:2w}
\end{equation}
By equating the coefficients of partial derivatives we obtain a system of algebraic equations for the
unknown matrix $A$:
\begin{align}
&A^3 = \frac{1}{m}\,I,  &\label{eq:2e}\\
&A\hspace{0.02cm}\beta_{\mu}A^2 + A^{2}\beta_{\mu}A =
-\,\frac{1}{m}\,\beta_{\mu}, \label{eq:2r}\\
&A\hspace{0.02cm}\beta_{\mu}A\hspace{0.02cm}\beta_{\nu}A
+ A\hspace{0.02cm}\beta_{\nu}A\hspace{0.02cm}\beta_{\mu}A
=
-\,\frac{1}{m}\,\bigl[\hspace{0.02cm}2\hspace{0.02cm}g_{\mu\nu}\hspace{0.02cm}I
- \{\beta_{\mu},\beta_{\nu}\}\bigr]. \label{eq:2t}
\end{align}
These equations can be paired with the corresponding equations in the paper \cite{plyushchay_2000},
if we correlate the generators $g_{\mu}$ and $\tilde{g}$ introduced in \cite{plyushchay_2000} with the matrices $A$ and $\beta_{\mu}$ by the rules
\[
g_{\mu} \sim m A\hspace{0.02cm} \beta_{\mu}, \quad  \tilde{g} \sim m A.
\]
In this case, Eq.\,(\ref{eq:2e}) (up to a sign) will correspond to the first equation
of the system (\ref{eq:2e}) in the paper \cite{plyushchay_2000},Eq.\,(\ref{eq:2r}) will correspond to the second equation of the same system (or Eq.\,(\ref{eq:2y})), and (\ref{eq:2t}) corresponds to the third equation.\\
\indent
Before turning to solving the matrix equations (\ref{eq:2e})\,--\,(\ref{eq:2t}), we make a few comments of a general character. Equations (\ref{eq:2e}) and (\ref{eq:2r}) are universal in the determinate sense. The former defines the mass term on the right-hand side of the equality (\ref{eq:2q}) (or (\ref{eq:1d})), and the latter enables us to get rid of the term of the first order in the derivatives in (\ref{eq:2q}) (or (\ref{eq:1d})). The universality of these matrix equations lies in the fact that they must be satisfied in any case irrespective of that we take as the right part: either the right-hand side of (\ref{eq:2q}) or the right-hand side of (\ref{eq:1d}). As will be shown below, Eqs.\,(\ref{eq:2e}) and (\ref{eq:2r}) uniquely define the required matrix $A$ (more exactly, to within the choice of one of three roots of the cubic equation for some parameter $\alpha$, see Eq.\,(\ref{eq:2i}) below). An explicit form of the matrix $A$ and also the equalities (\ref{eq:2e}) and (\ref{eq:2r}) to which it satisfies, are of fundamental importance for further presentation. The third equation (\ref{eq:2t}) is not already universal and completely depends on the specific choice of the right-hand side in the equalities of  the (\ref{eq:2q}) type. This equation must be identically satisfied. If not, we come to the contradiction.\\
\indent
Let us now introduce the matrix $\omega$ setting by definition
\begin{equation} \omega = \frac{i}{4}\; \epsilon^{\mu\nu\lambda\sigma\!}\beta_{\mu}\beta_{\nu}\beta_{\lambda}\beta_{\sigma}.
\label{eq:2y}
\end{equation}
This matrix plays an important part in further consideration. It was introduced into DKP theory for the first time by E. Schr\"odinger \cite{schrodinger_1943}. Here, we follow the notation used in the works
by Harish-Chandra \cite{harish-chandra_1946}, where the properties of the $\omega$ matrix were studied in detail. In Appendix A we give all necessary relations for the
$\omega$\hspace{0.005cm}-\hspace{0.005cm}$\beta_{\mu}$ algebra. Let us note only that the matrix
$\omega$ is identically zero for the spin 0 (five-dimensional irreducible representation of the DKP algebra). Therefore, only the ten-row representation needs to be considered.\\
\indent
In spite of a formal similarity between definitions $\gamma_5$ and $\omega$ matrices (it is worthy of special emphasis that in the latter case the factor $1/4$  stands rather than $1/4!$), Eqs.\,(\ref{eq:1o}) and (\ref{eq:2y}), as the matrix $A$ we cannot simply take $\frac{1}{m^{1/3}}\,\omega$. Really, for example, on the left-hand side of Eq.\,(\ref{eq:2e}) in view of (\ref{ap:A1}) we will have
$\frac{1}{m}\,\omega^3 = \frac{1}{m}\,\omega \neq \frac{1}{m}\,I$.\\
\indent
We seek the matrix $A$ in the form of the most general expansion in powers of $\omega$:
\[
A = \alpha\hspace{0.02cm}I  + \beta\hspace{0.02cm}\omega
+ \gamma\hspace{0.03cm}\omega^{2},
\]
where $\alpha,\,\beta$, and $\gamma$ are unknown, generally speaking complex, scalar constants.
By virtue of the property (\ref{ap:A1}) it is easy to find that
\begin{equation}
A^2 = \alpha^{2}I + (2\alpha\beta + 2\beta\gamma)\hspace{0.02cm}\omega + (\beta^{2} + \gamma^{2} + 2\alpha\gamma)\hspace{0.03cm}\omega^{2}
\label{eq:2u}
\end{equation}
and further
\[
\begin{split}
A^3 &= \alpha^{3}I \\
&+\bigl[\hspace{0.02cm}\alpha\hspace{0.02cm}(2\alpha\beta + 2\beta\gamma) +
\alpha^{2}\beta +
\gamma\hspace{0.02cm}(2\alpha\beta + 2\beta\gamma) +
\beta\hspace{0.02cm}(\beta^2 + \gamma^2 + 2\alpha\gamma)\hspace{0.02cm}\bigr]
\hspace{0.02cm}\omega\\
&+\bigl[\hspace{0.02cm}\alpha\hspace{0.02cm}(\beta^2 + \gamma^2 + 2\alpha\gamma) +
\beta(2\alpha\beta + 2\beta\gamma) + \alpha^2\gamma +
\gamma\hspace{0.02cm}(\beta^2 + \gamma^2 + 2\alpha\gamma)\hspace{0.02cm}\bigr]
\hspace{0.02cm}\omega^{2}\\
&=\frac{1}{m}\,I.
\end{split}
\]
The foregoing expression enables us to reduce Eq.\,(\ref{eq:2e}) to a system of three algebraic
equations for unknown scalar constants, the first of which defines the parameter $\alpha$:
\begin{equation}
\alpha^{3} = \frac{1}{m}\,.
\label{eq:2i}
\end{equation}
Two other equations follow from vanishing the expressions in square brackets. However, instead of these equations it is convenient to consider their sum and difference, which after simple algebraic transformations can be recast in a more convenient form
\[
\begin{split}
&(\beta + \gamma)\bigl[\hspace{0.02cm}(\beta + \gamma)^2
+ 3\hspace{0.02cm}\alpha\hspace{0.02cm}(\beta + \gamma) +
3\hspace{0.02cm}\alpha^2\hspace{0.02cm}\bigr] = 0, \vspace{0.04cm}\\
&(\beta - \gamma)\bigl[\hspace{0.02cm}(\beta - \gamma)^2
- 3\hspace{0.02cm}\alpha\hspace{0.02cm}(\beta - \gamma) +
3\hspace{0.02cm}\alpha^2\hspace{0.02cm}\bigr] = 0.
\end{split}
\]
As solutions of these equations we take solutions of the quadratic equations in square brackets
for the variables $(\beta + \gamma)$ and $(\beta - \gamma)$, namely
\begin{equation}
\begin{split}
&(\beta + \gamma)_{\pm} = \biggl(\!-\frac{3}{2} \,\pm\, i\,\frac{\sqrt{3}}{2}\biggr)\alpha,\\
&(\beta - \gamma)_{\pm} = \biggl(\,\frac{3}{2} \,\pm\, i\,\frac{\sqrt{3}}{2}\biggr)\alpha.
\end{split}
\label{eq:2o}
\end{equation}
We return to the obtained solutions just below, and now we pass on the second matrix
equation (\ref{eq:2r}). By using the properties (\ref{ap:A1})\,--\,(\ref{ap:A2}), we get
\[
A\hspace{0.02cm}\beta_{\mu}A^2 + A^{2}\beta_{\mu}\hspace{0.02cm}A =
\]
\[
= \bigl[\hspace{0.02cm}2\alpha^3 + \alpha^2\gamma + \alpha\hspace{0.02cm}(\beta^2 +
\gamma^2 + 2\hspace{0.02cm}\alpha\hspace{0.02cm}\gamma)\bigr]\beta_{\mu} +
\bigl[\hspace{0.03cm}\alpha^2\beta + \alpha\hspace{0.02cm}
(2\hspace{0.02cm}\alpha\hspace{0.01cm}\beta + 2\hspace{0.02cm}\beta\hspace{0.02cm}\gamma)\bigr]
(\omega\beta_{\mu} + \beta_{\mu}\omega) \equiv -\,\frac{1}{m}\,\beta_{\mu},
\]
which due to (\ref{eq:2i}) gives us the second system of algebraic equations:
\[
\begin{split}
&\alpha\hspace{0.03cm}(\hspace{0.02cm}\beta^2 + \gamma^2
+ 3\hspace{0.02cm}\alpha\hspace{0.02cm}\gamma) =
-\,\displaystyle\frac{3}{m}\,, \\
&\alpha\hspace{0.03cm}(\hspace{0.02cm}3\hspace{0.02cm}\alpha\hspace{0.02cm}\beta
+ 2\hspace{0.02cm}\beta\hspace{0.02cm}\gamma\hspace{0.02cm}) = 0.
\end{split}
\]
Since $\alpha \neq 0$ and considering $\beta \neq 0$, from the very last equation we obtain
\[
\gamma = -\,\frac{3}{2}\,\alpha.
\]
Taking into account this fact, from the first equation we derive
$\beta^2= -\hspace{0.02cm}(3/4)\,\alpha^2$ or
$\beta= \pm\,i\hspace{0.03cm}(\sqrt{3}/2)\hspace{0.02cm}\alpha$. The solution obtained for the
parameter $\gamma$ is not in contradiction with the solutions (\ref{eq:2o}). As the $\beta$ parameter one can take either of the two solutions
$\pm\,i\hspace{0.02cm}(\sqrt{3}/2)\hspace{0.02cm} \alpha$. It is also consistent with the solutions (\ref{eq:2o}). For definiteness let us fix the sign $+$\,, i.e., we set
\[
\beta = i\,\frac{\sqrt{3}}{2}\,\alpha.
\]
The matrix $A$ in this case takes the following form:
\begin{equation}
A = \alpha\biggl(I + i\,\frac{\sqrt{3}}{2}\,\omega - \frac{3}{2}\;\omega^2\biggr).
\label{eq:2p}
\end{equation}
An explicit form of the matrix $A^{2}$ can be obtained by squaring (\ref{eq:2p}) or by making use of (\ref{eq:2u}). Here we have
\begin{equation}
A^{2} = \alpha^{2}\biggl(I - i\,\frac{\sqrt{3}}{2}\,\omega - \frac{3}{2}\;\omega^2\biggr).
\label{eq:2a}
\end{equation}
Note that $A^{2}$ is different from $A$ not only by the extra dimension factor $\alpha$, but also by the opposite sign before the second term (in fact, here we have the second possible value of the
parameter $\beta$). An explicit form of the matrices $A$ and $A^{2}$ hints that they are
mutually conjugated. Indeed, if one takes into account the Hermitian character of the $\omega$-matrix:
\[
\omega^{\dagger} = \omega,
\]
then the following two relations are true
\begin{equation}
A^{\dagger} = m^{1/3\!}A^{2}, \quad AA^{\dagger} = A^{\dagger}A = \frac{1}{m^{2/3}}\,I.
\label{eq:2s}
\end{equation}
In principle, one can avoid the mass multipliers if one overdetermines the matrix $A$, but we do not do it.\\
\indent
Furthermore, we can draw an interesting parallel between a set of matrices $(A,\hspace{0.02cm}A^2, A^3\!\equiv\!\frac{1}{m}\,I)$ and a set of cubic roots of unity:
$(q,\hspace{0.02cm}q^2, q^{3}\!\equiv\! 1)$, Eq.\,(\ref{eq:1y}). In the latter case the following relations, which are similar to (\ref{eq:2s}), hold,
\[
\quad q^{\ast} = q^{2}, \quad q\hspace{0.03cm}q^{\ast} = 1,
\]
where $q^{\ast}$ is the complex conjugate of $q$. However, the cubic roots
$(q,\hspace{0.02cm}q^2, 1)$ possess one more important property (\ref{eq:1a}), whereas for the matrix set $(A,\hspace{0.02cm}A^2, \frac{1}{m}\,I)$ we have
\[
I + \frac{1}{\alpha}\,A + \frac{1}{\alpha^2}\,A^{2} = 3\hspace{0.025cm}
(I - \omega^2).
\]
Nevertheless, it is possible to redefine the matrix $A \rightarrow {\cal A}$ such that the following
equality will be held
\[
I + \frac{1}{\alpha}\,{\cal A} + \frac{1}{\alpha^2}\,{\cal A}^{2} = 0
\]
and at the same time the properties
\[
{\cal A}^3 = \frac{1}{m}\,I,\quad {\cal A}^{\dagger} = m^{1/3\!}{\cal A}^{2}
\]
will be survived. Since the matrix ${\cal A}$ will not play any role later, we give its explicit form in Appendix B.\\
\indent
Now we turn to analysis of the remaining equation (\ref{eq:2t}). By making use of an explicit form of the matrix $A$, Eq.\,(\ref{eq:2p}), and the properties (\ref{ap:A4})\,--\,(\ref{ap:A7}), we obtain for the
left-hand side of (\ref{eq:2t})
\begin{equation}
A\beta_{\mu}A\beta_{\nu}A + A\beta_{\nu}A\beta_{\mu}A =
-\,\frac{1}{m}\,\biggl[\,\frac{1}{2}\,\{\beta_{\mu},\beta_{\nu}\} +
i\,\frac{\sqrt{3}}{2}\;g_{\mu\nu}\hspace{0.03cm}\omega + \frac{3}{2}\;g_{\mu\nu}\hspace{0.03cm}
\omega^{2} - \frac{3}{2}\,\{\beta_{\mu},\beta_{\nu}\}\hspace{0.04cm}\omega^{2}\biggr].
\label{eq:2d}
\end{equation}
Comparing the expression with the right-hand side of (\ref{eq:2t}), we see that their matrix structure is sufficiently close to each other. The main difference with (\ref{eq:2t}) is the presence of the term in (\ref{eq:2d}) linear in the matrix $\omega$. We can remove this term if we slightly complicate the left-hand side of the initial expression (\ref{eq:2w}), namely, we present it in the following form:
\begin{equation}
\frac{1}{2}\,\bigl(A\hspace{0.02cm}L(\partial)A\hspace{0.02cm}L(\partial)A \,+\, m\hspace{0.02cm}
A^2L(\partial)A^2L(\partial)A^2\bigr),
\label{eq:2f}
\end{equation}
where the operator $L(\partial)$ was specified by Eq.\,(\ref{eq:1w}). It is not difficult to see that the first two equations (\ref{eq:2e}) and (\ref{eq:2r}) remain unchanged and instead of (\ref{eq:2t}) now we have
\begin{equation} \frac{1}{2}\,\bigl[\hspace{0.02cm}(A\hspace{0.02cm}\beta_{\mu}A\hspace{0.02cm}\beta_{\nu}A
+ A\hspace{0.02cm}\beta_{\nu}A\hspace{0.02cm}\beta_{\mu}A) + m\hspace{0.02cm}
(A^2\beta_{\mu}A^2\beta_{\nu}A^2 + A^2\beta_{\nu}A^{2\!}\beta_{\mu}A^2)\hspace{0.02cm}\bigr] =
-\,\frac{1}{m}\,\bigl(\hspace{0.02cm}2\hspace{0.02cm}g_{\mu\nu} -
\{\hspace{0.02cm}\beta_{\mu},\beta_{\nu}\}\bigr).
\label{eq:2g}
\end{equation}
By using an explicit form of the matrix $A^2$ one can see that the expression on the left-hand side of (\ref{eq:2g}) completely coincides with (\ref{eq:2d}) except for cancellation of the term linear in $\omega$. Thus, the matrix equation (\ref{eq:2g}) leads to fulfilment of the following equality:
\[
\frac{1}{2}\,\{\hspace{0.02cm}\beta_{\mu},\beta_{\nu}\} + \frac{3}{2}\;g_{\mu\nu}
\hspace{0.02cm}\omega^{2} -\,
\frac{3}{2}\,\{\hspace{0.02cm}\beta_{\mu},\beta_{\nu}\}\,\omega^{2} =
2\hspace{0.02cm}g_{\mu\nu} - \{\beta_{\mu},\beta_{\nu}\}.
\]
The relation is inconsistent. To verify this, it is sufficient to contract it with $g^{\mu \nu}$,
for example. With the relations (\ref{ap:A9}) we result in a contradiction
\[
B - \frac{8}{3} = 0.
\]
\indent
One can look at the problem in a different way. As we know in the interaction free case the divisor
$d(\partial)$ commutes with the operator $L(\partial)$, i.e.
\[
[\hspace{0.04cm}d(\partial), L(\partial)\hspace{0.03cm}] = 0.
\]
Let us substitute now the operator (\ref{eq:2f}) instead of the divisor $d(\partial)$. The result of
calculations is very simple, namely
\[
\frac{1}{2}\,\bigl[\hspace{0.02cm}(A\hspace{0.02cm}L(\partial)
A\hspace{0.02cm}L(\partial)A \,+\, m\hspace{0.02cm}
A^{2\!}\hspace{0.02cm}L(\partial)A^{2\!}\hspace{0.02cm}L(\partial)A^2),
L(\partial)\hspace{0.03cm}\bigr]
=
-\,\frac{3\hspace{0.02cm}i}{m}\;\Box\,
[\hspace{0.03cm}\omega^{2},\beta_{\mu}\hspace{0.02cm}\hspace{0.02cm}]
\hspace{0.02cm}\partial^{\mu}.
\]
Most of the terms in expression (\ref{eq:2f}) in calculating the commutator vanish. The only term of the third order in partial derivatives survives by virtue of
$[\hspace{0.02cm}\omega^2, \beta_{\mu}\hspace{0.02cm}] \neq 0$.\\
\indent
Finally, we note that one can get rid of the matrix $\omega^2$ before the higher order derivative
if instead of the initial equation (\ref{eq:2q}) (more exactly, its left-hand side) one considers the most general and more symmetric expression
\begin{equation}
\frac{1}{4}\,\bigl\{\hspace{0.02cm}\bigl(A\hspace{0.02cm}L(\partial)A\hspace{0.02cm}L(\partial)A \,+
\, m\hspace{0.02cm}
A^2L(\partial)A^2L(\partial)A^2\bigr), L(\partial)\bigr\}
\label{eq:2h}
\end{equation}
\[
= \frac{i}{8\hspace{0.02cm}m}\;\Box\hspace{0.03cm} \beta_{\mu}\partial^{\mu}
- \frac{3}{4}\,\Bigl(\hspace{0.02cm}
\{\hspace{0.02cm}\beta_{\mu},\beta_{\nu}\} + g_{\mu\nu}\hspace{0.04cm} \omega^{2} -\,
\{\hspace{0.02cm}\beta_{\mu},\beta_{\nu}\}\hspace{0.04cm}\omega^{2}\Bigl)\hspace{0.02cm}
\partial^{\mu}\partial^{\nu}
\,-\, m^{2}I.
\]
The matrix $\omega^2$ in the first term disappears by virtue of the property (\ref{ap:A2}). We cannot eliminate this term within the framework of the Duffin-Kemmer-Petiau formalism in principle. On the other hand its structure up to a numerical factor coincides with the corresponding term on the right-hand side of the equation (\ref{eq:1d}). Here the other question arises whether one could remove the term of the second order in $\partial^{\mu}$ in Eq.\,(\ref{eq:2h}). The remainder  of the paper will be devoted to answering this question.

\section{\bf Cubic root of the third order wave equation}
\setcounter{equation}{0}

Let us consider now the construction of the cubic root of the third order wave equation. It is clear that the ``na\"{\i}ve'' representation of the cubic root as was defined on the left-hand side of expression (\ref{eq:1d}) is unsuitable. Even with the use of the most general representation (the left-hand side of (\ref{eq:2h})) the undesirable term of the second order in the derivatives survives. Besides, the coefficient of the operator
\[
\frac{i}{m}\;\Box\hspace{0.03cm}\beta_{\mu}\hspace{0.02cm}\partial^{\mu}
\]
differs from the corresponding coefficient on the right-hand side of Eq.\,(\ref{eq:1d}) by the factor $(-1/8)$ and to correct it is also by no means easy.  This imply that we cannot get rid of the unwanted term and correct the coefficient mentioned above by making use of the properties of the matrices $A$ and $\beta_{\mu}$ only. Here, it is necessary to involve some additional considerations of algebraic character. In this section we attempt to outline a general approach to the stated problem.\\
\indent
Let us introduce the following {\it deformed} commutator,
\begin{equation}
\Xi^{(z)}_{\mu} \equiv A\hspace{0.02cm}\beta_{\mu} - z\hspace{0.02cm}\beta_{\mu}\hspace{0.02cm}
A \equiv [\hspace{0.03cm}A\hspace{0.02cm},\beta_{\mu}\hspace{0.03cm}]_{\hspace{0.01cm}z},
\label{eq:4q}
\end{equation}
where $z$ is an arbitrary complex number and perform an analysis of the following expression
\begin{equation}
\bigl(\hspace{0.03cm}i\hspace{0.04cm}\Xi^{(z)}_{\mu\,}\partial^{\mu} -
A\hspace{0.02cm} m\hspace{0.02cm}\bigr)^{3}
=
-i\hspace{0.03cm}\bigl(\hspace{0.02cm}\Xi^{(z)}_{\mu\,}\Xi^{(z)}_{\nu\,}\Xi^{(z)}_{\lambda}\bigr)
\partial^{\mu}\partial^{\nu}\partial^{\lambda} - A^{3}\hspace{0.02cm}m^{3\!}
\label{eq:3r}
\end{equation}
\[
+\,m\bigl(\hspace{0.02cm}\Xi^{(z)}_{\mu}\Xi^{(z)}_{\nu}A + A\hspace{0.02cm}\Xi^{(z)}_{\mu}\Xi^{(z)}_{\nu} +
\Xi^{(z)}_{\mu}\hspace{0.02cm}A\hspace{0.03cm}\Xi^{(z)}_{\nu}\bigr)\hspace{0.02cm}
\partial^{\mu}\partial^{\nu}
\]
\[
+\,\hspace{0.02cm}i\hspace{0.03cm}m^{2\!}\left(\hspace{0.02cm}\Xi^{(z)}_{\mu\,}A^{2} +
A\hspace{0.04cm}\Xi^{(z)}_{\mu}\hspace{0.02cm}A + A^{2}\hspace{0.03cm}\Xi^{(z)}_{\mu}\hspace{0.02cm}\right)\!\hspace{0.02cm}\partial^{\mu}.
\]
First, we consider the contribution linear in $\partial^{\mu}$ in (\ref{eq:3r}), namely
\[
i\hspace{0.03cm}m^{2\!}\left(\hspace{0.03cm}\Xi^{(z)\!}_{\mu\!\!}A^{2} +
A\,\Xi^{(z)\!}_{\mu\!\!}A + A^{2\,}\Xi^{(z)}_{\mu}\hspace{0.01cm}\right)\!\hspace{0.03cm}\partial^{\mu},
\]
where by virtue of the definition (\ref{eq:4q}) we have
\[
\begin{split}
&\Xi^{(z)}_{\mu\!}\!A^{2} = A\hspace{0.02cm}\beta_{\mu}\hspace{0.02cm}A^{2} - z\,\frac{1}{m}\,\beta_{\mu}, \\
&A^{2\,}\Xi^{(z)}_{\mu} = \frac{1}{m}\,\beta_{\mu} - z\hspace{0.02cm}A^{2}\beta_{\mu}\hspace{0.02cm}A, \\
&A\,\Xi^{(z)\!}_{\mu\!}A = A^{2}\beta_{\mu}\hspace{0.02cm}A - z\hspace{0.02cm}A\hspace{0.02cm}\beta_{\mu}\hspace{0.02cm}A^{2}.
\end{split}
\]
A sum of these three expressions gives
\[
\bigl(A\hspace{0.02cm}\beta_{\mu}\hspace{0.02cm}A^{2} + A^{2}\beta_{\mu}\hspace{0.02cm}A
+ \frac{1}{m}\,\beta_{\mu}\bigr)
-z\hspace{0.02cm}\bigl(A\hspace{0.02cm}\beta_{\mu}\hspace{0.02cm}A^{2} + A^{2}\beta_{\mu}\hspace{0.02cm}A
+ \frac{1}{m}\,\beta_{\mu}\bigr).
\]
We see that for {\it any} value of the parameter $z$ this expression vanishes by virtue of (\ref{eq:2r}).\\
\indent
Let us consider the contribution quadratic in $\partial^{\mu}$:
\begin{equation}
m\hspace{0.02cm}\bigl(\hspace{0.03cm}\Xi^{(z)}_{\mu\,}\hspace{0.03cm}\Xi^{(z)}_{\nu\!\!}A + A\,\Xi^{(z)}_{\mu\,}\hspace{0.03cm}\Xi^{(z)}_{\nu} + \Xi^{(z)}_{\mu}\!A\,\Xi^{(z)}_{\nu}\hspace{0.02cm}\bigr)\hspace{0.02cm}
\partial^{\mu}\partial^{\nu} .
\label{eq:4w}
\end{equation}
Note that we have written out the expression (\ref{eq:4w}) with no explicit symmetrization with respect to the vector indices $\mu$ and ${\nu}$. Analysis of the expression in parentheses in (\ref{eq:4w}) is now more cumbersome. Our first step is to write out explicitly each term in the expression (\ref{eq:4w}),
\[
\begin{split}
&\Xi^{(z)}_{\mu\,}\hspace{0.03cm}\Xi^{(z)\!}_{\nu\!}A =
A\hspace{0.02cm}\beta_{\mu}\hspace{0.02cm}A\hspace{0.02cm}\beta_{\nu}\hspace{0.02cm}A
-z\hspace{0.02cm}\beta_{\mu}A^{2}\beta_{\nu}\hspace{0.02cm}A
-z\hspace{0.02cm}A\hspace{0.02cm}\beta_{\mu}\hspace{0.02cm}\beta_{\nu}\hspace{0.02cm}A^{2}
+z^{2}\beta_{\mu}A\hspace{0.02cm}\beta_{\nu}\hspace{0.02cm}A^{2},\\
&A\,\Xi^{(z)}_{\mu\,}\hspace{0.03cm}\Xi^{(z)}_{\nu} =
A^{2}\beta_{\mu}\hspace{0.02cm}A\hspace{0.02cm}\beta_{\nu}
- z\hspace{0.02cm}A\hspace{0.02cm}\beta_{\mu}\hspace{0.02cm}A^{2}\beta_{\nu}
- z\hspace{0.02cm}A^{2}\beta_{\mu}\hspace{0.02cm}\beta_{\nu}\hspace{0.02cm}A
+ z^{2}A\hspace{0.02cm}\beta_{\mu}\hspace{0.02cm}A\hspace{0.02cm}\beta_{\nu}
\hspace{0.02cm}A,\\
&\Xi^{(z)\!}_{\mu\!}A\,\Xi^{(z)}_{\nu} =
A\hspace{0.02cm}\beta_{\mu}\hspace{0.02cm}A^{2}\beta_{\nu}
- z\hspace{0.02cm}A\hspace{0.02cm}\beta_{\mu}\hspace{0.02cm}A\hspace{0.02cm}\beta_{\nu}
\hspace{0.02cm}A
- \frac{1}{m}\,z\hspace{0.02cm}\beta_{\mu}\hspace{0.02cm}\beta_{\nu}
+ z^{2}\beta_{\mu}A^{2}\beta_{\nu}\hspace{0.02cm}A.
\end{split}
\]
A sum of these three expressions after collecting similar terms is
\[
(1 - z + z^{2}) \hspace{0.02cm}A\hspace{0.02cm}\beta_{\mu}\hspace{0.02cm}A\hspace{0.02cm}\beta_{\nu}
\hspace{0.02cm}A
- \frac{1}{m}\,z\hspace{0.02cm}\beta_{\mu}\hspace{0.02cm}\beta_{\nu}
- z\hspace{0.02cm}\bigl(
\hspace{0.02cm}A\hspace{0.02cm}\beta_{\mu}\hspace{0.02cm}\beta_{\nu}\hspace{0.02cm}A^{2}
+
A^{2}\beta_{\mu}\hspace{0.02cm}\beta_{\nu}\hspace{0.02cm}A\hspace{0.02cm}\bigr)
\]
\[
+\,\bigl[\hspace{0.03cm}(-z + z^{2})\beta_{\mu}A^{2}\beta_{\nu}\hspace{0.02cm}A
+ z^{2}\beta_{\mu}A\hspace{0.02cm}\beta_{\nu}\hspace{0.02cm}A^{2}\hspace{0.02cm}\bigr]
+
\bigl[\hspace{0.03cm}(1 - z)\hspace{0.02cm}A\hspace{0.02cm}
\beta_{\mu}\hspace{0.02cm}A^{2}\beta_{\nu}
+
A^{2}\beta_{\mu}\hspace{0.02cm}A\hspace{0.02cm}\beta_{\nu}\hspace{0.02cm}\bigr].
\]
Further the use of the identity
\[
A\hspace{0.02cm}\beta_{\mu}\hspace{0.02cm}A\hspace{0.02cm}\beta_{\nu}\hspace{0.02cm}A
=
-\,\frac{1}{2}\,\Bigl[\hspace{0.01cm}\bigl(
\hspace{0.01cm}A\hspace{0.02cm}\beta_{\mu}\hspace{0.02cm}\beta_{\nu}\hspace{0.02cm}A^{2}
+
A^{2}\beta_{\mu}\hspace{0.02cm}\beta_{\nu}\hspace{0.02cm}A\hspace{0.02cm}\bigr)
+
\bigl(\hspace{0.02cm}A\hspace{0.02cm}\beta_{\mu}\hspace{0.02cm}A^{2}\beta_{\nu}
+
\beta_{\mu}A^{2}\beta_{\nu}\hspace{0.02cm}A\hspace{0.02cm}\bigr)\hspace{0.01cm}\Bigr]
\]
enables us to rewrite the sum in a more compact form
\[
\varepsilon\hspace{0.01cm}(z)\,\frac{1}{2}\,\biggl[\hspace{0.02cm}
\frac{1}{m}\,\beta_{\mu}\hspace{0.02cm}\beta_{\nu}
-
\bigl(
\hspace{0.01cm}A\hspace{0.02cm}\beta_{\mu}\hspace{0.02cm}\beta_{\nu}\hspace{0.02cm}A^{2}
+
A^{2}\beta_{\mu}\hspace{0.02cm}\beta_{\nu}\hspace{0.02cm}A\hspace{0.02cm}\bigr)
+
\bigl(\hspace{0.01cm}A^{2}\beta_{\mu}\hspace{0.02cm}A\hspace{0.02cm}\beta_{\nu}
+
\beta_{\mu}A\hspace{0.02cm}\beta_{\nu}\hspace{0.02cm}A^{2}\bigr)\biggr].
\]
Here, we have introduced the function
\begin{equation}
\varepsilon\hspace{0.01cm}(z) = 1 + z + z^{2} \equiv (z - q)(z - q^{2}),
\label{eq:4e}
\end{equation}
which is of great importance for the subsequent discussion. From the expression obtained we see that the quadratic contribution (\ref{eq:4w}) may vanish if as the parameter $z$ one takes a primitive root of equation $z^3-1=0$, i.e., $q$ or $q^2$, Eq.\,(\ref{eq:1y}). In addition, it should be noted especially that the expression (\ref{eq:4w}) vanishes without any symmetrization over the vector indices.\\
\indent
Now we need to analyze the term cubic in $\partial^{\mu}$ in (\ref{eq:3r}). The initial expression is
\begin{equation}
-\hspace{0.02cm}i\,\hspace{0.02cm}\Xi^{(z)\,}_{\mu\,}\Xi^{(z)\,}_{\nu\,}\Xi^{(z)}_{\lambda}
\partial^{\mu}\partial^{\nu}\partial^{\lambda},
\label{eq:4r}
\end{equation}
where by virtue of the definition of the matrix $\Xi_{\mu}^{(z)}$ we have
\begin{equation}
\Xi^{(z)\,}_{\mu}\hspace{0.02cm}\Xi^{(z)\,}_{\nu}\hspace{0.02cm}\Xi^{(z)}_{\lambda}
=
A\hspace{0.02cm}\beta_{\mu}\hspace{0.02cm}A\hspace{0.02cm}\beta_{\nu}
\hspace{0.02cm}A\hspace{0.02cm}\beta_{\lambda}
-
\beta_{\mu}A\hspace{0.02cm}\beta_{\nu}\hspace{0.02cm}A\hspace{0.02cm}\beta_{\lambda}
\hspace{0.02cm}A
-
z\hspace{0.01cm}A\hspace{0.02cm}\beta_{\mu}\hspace{0.02cm}\beta_{\nu}\hspace{0.02cm}A^{2}
\beta_{\lambda}
-
z\hspace{0.02cm}\beta_{\mu}A^{2}\beta_{\nu}\hspace{0.02cm}A\hspace{0.02cm}\beta_{\lambda}
\label{eq:4t}
\end{equation}
\[
+\,z^{2}\beta_{\mu}A\hspace{0.02cm}\beta_{\nu}\hspace{0.02cm}A^{2}\beta_{\lambda}
-
z\hspace{0.02cm}A\hspace{0.02cm}\beta_{\mu}\hspace{0.02cm}A\hspace{0.02cm}\beta_{\nu}
\hspace{0.02cm}\beta_{\lambda}\hspace{0.02cm}A
+
z^{2\!}A\hspace{0.02cm}\beta_{\mu}\beta_{\nu}A\hspace{0.02cm}\beta_{\lambda}\hspace{0.02cm}A
+
z^{2}\beta_{\mu}A^{2}\beta_{\nu}\hspace{0.02cm}\beta_{\lambda}\hspace{0.02cm}A.
\]
A somewhat lengthy computation has shown (see Appendix \ref{appendix_B1}) that the contribution cubic in $\partial^{\mu}$, Eq.\,(\ref{eq:4r}), in the choice $z=q$ and {\it symmetrization} over the vector indices turns to zero. Here we only note that the following two equalities,
\begin{equation}
\begin{split}
&A\hspace{0.02cm}\beta_{\mu}\hspace{0.02cm}A^{2} = -\,\frac{1}{2\hspace{0.02cm}m}\,
\bigl(\hspace{0.02cm}\beta_{\mu} - i\hspace{0.02cm}\sqrt{3}\,\xi_{\mu}
\bigr), \\
&A^{2\!}\hspace{0.02cm}\beta_{\mu}A = -\,\frac{1}{2\hspace{0.02cm}m}\,
\bigl(\hspace{0.02cm}\beta_{\mu} + i\hspace{0.02cm}\sqrt{3}\,\xi_{\mu}
\bigr)
\end{split}
\label{eq:4u}
\end{equation}
are rather useful in the analysis given in Appendix \ref{appendix_B1}. In the equalities (\ref{eq:4u}) we have introduced the matrices $\xi_{\mu}$ setting by definition
\[
\xi_{\mu} \equiv [\hspace{0.03cm}\omega,\beta_{\mu}\hspace{0.03cm}]
= -\,\frac{i}{2}\;\epsilon_{\mu\nu\lambda\sigma}\beta^{\nu}\beta^{\lambda}\beta^{\sigma}.
\]
These matrices have already been considered in the paper by Azimov and Ryndin \cite{azimov_1995}, where they played a role of a spin pseudovector for a spin-1 particle. Besides, the matrices were intensively used by Fushchich {\it et al.} \cite{nikitin_1976} for establishment of the complementary (non-Lie) symmetry of the Duffin-Kemmer-Petiau equation.

\section{The $\eta_{\mu}$ matrices}
\setcounter{equation}{0}

Let us analyze the results of the previous section from a slightly different point of view. For this purpose we introduce a new set of matrices $\eta_{\mu}$ that would satisfy the following condition,
\begin{equation}
A\hspace{0.02cm}\eta_{\mu} = {\rm w}\hspace{0.02cm}\eta_{\mu}\hspace{0.02cm}A,
\label{eq:5q}
\end{equation}
and as an immediate consequence
\begin{equation}
A^{2}\eta_{\mu} = {\rm w}^{2}\eta_{\mu}A^{2},
\label{eq:5w}
\end{equation}
where ${\rm w}$ is some complex number. We return to the expression (\ref{eq:1d}). Here, on the left-hand side, instead of the original matrices $\beta_{\mu}$, we set $\eta_{\mu}$:
\begin{equation}
\begin{split}
\bigl[\hspace{0.03cm}A\hspace{0.02cm}(\hspace{0.02cm}i\hspace{0.03cm}\eta_{\mu}
\hspace{0.02cm}\partial^{\mu} &- m\hspace{0.02cm}I)\bigr]^{3}
=
-\hspace{0.02cm}i\hspace{0.02cm}\bigl(A\hspace{0.02cm}\eta_{\mu}
\hspace{0.02cm}A\hspace{0.02cm}\eta_{\nu}\hspace{0.02cm}
A\hspace{0.02cm}\eta_{\lambda}\bigr)\hspace{0.02cm}\partial^{\mu}\partial^{\nu}\partial^{\lambda}
- m^{3\!}A^{3} \\
&+ m\hspace{0.02cm}\bigl(A\hspace{0.02cm}\eta_{\mu}A\hspace{0.02cm}\eta_{\nu}A
+
A\hspace{0.02cm}\eta_{\mu}A^{2}\eta_{\nu} + A^{2}\eta_{\mu}A\hspace{0.02cm}\eta_{\nu}
\bigr)\partial^{\mu}\partial^{\nu} \\
&+
i\hspace{0.02cm}m^{2}
\bigl(A^{3}\eta_{\mu} + A\hspace{0.02cm}\eta_{\mu}\hspace{0.02cm}A^{2}
+
A^{2}\eta_{\mu}\hspace{0.02cm}A\bigr)\partial^{\mu}.
\end{split}
\label{eq:5e}
\end{equation}
We use the rules of the rearrangements (\ref{eq:5q}) and (\ref{eq:5w}) to bring the matrix coefficients preceding the partial derivatives into a simple form:
\begin{equation}
\begin{split}
&A\hspace{0.02cm}\eta_{\mu}\hspace{0.02cm}A\hspace{0.02cm}
\eta_{\nu}\hspace{0.02cm}A\hspace{0.02cm}\eta_{\lambda}
=
{\rm w}^{3}\hspace{0.02cm}\frac{1}{m}\,
\eta_{\mu}\hspace{0.02cm}\eta_{\nu}\hspace{0.02cm}\eta_{\lambda},\\
&A\hspace{0.02cm}\eta_{\mu}A\hspace{0.02cm}\eta_{\nu}A
+
A\hspace{0.02cm}\eta_{\mu}A^{2}\eta_{\nu} + A^{2}\eta_{\mu}A\hspace{0.02cm}\eta_{\nu}
=
{\rm w}\hspace{0.02cm}\varepsilon({\rm w})\hspace{0.02cm}
\frac{1}{m}\,\eta_{\mu}\hspace{0.02cm}\eta_{\nu},\\
&A^{3}\eta_{\mu} + A\hspace{0.02cm}\eta_{\mu}\hspace{0.02cm}A^{2}
+
A^{2}\eta_{\mu}\hspace{0.02cm}A
=
\varepsilon({\rm w})\hspace{0.02cm}\frac{1}{m}\,\eta_{\mu},
\end{split}
\label{eq:5r}
\end{equation}
where the function $\varepsilon({\rm w})$ is defined by the expression (\ref{eq:4e}). It is evident that if we set the complex number ${\rm w}$ equal to $q$ (or $q^2$), then (\ref{eq:5e}) reduces to
\begin{equation}
-\hspace{0.01cm}i\,\frac{1}{m}\;\eta_{\mu}\hspace{0.02cm}\eta_{\nu}\hspace{0.02cm}
\eta_{\lambda\,}\partial^{\mu}\partial^{\nu}\partial^{\lambda} - m^{2}I.
\label{eq:5t}
\end{equation}
Further, if the matrices $\eta_{\mu}$ satisfied the identity of the form (\ref{eq:1s}) we could reproduce the right-hand side of the relation (\ref{eq:1d}) (with the replacement $\beta_{\mu}$ by $\eta_{\mu}$).\\
\indent
Let us now turn to the construction of an explicit form of the matrices $\eta_{\mu}$. To this end, we return to the generalized commutator (\ref{eq:4q}) in which for definiteness we set $z=q$. We rearrange the matrix $A$ to the left
\[
[\hspace{0.03cm}A,\beta_{\mu}\hspace{0.03cm}]_{\hspace{0.01cm}q}
 \equiv A\hspace{0.02cm}\beta_{\mu} - q\hspace{0.02cm}\beta_{\mu}\hspace{0.02cm}A
 =
A\bigl(\hspace{0.02cm}\beta_{\mu} - m\hspace{0.02cm}q\hspace{0.01cm}A^{2\!}
\beta_{\mu}\hspace{0.02cm}A\bigr).
\]
Here, we have taken into account that $A^{-1}=m A^2$. On the other hand we can rearrange the same matrix to the right
\[
[\hspace{0.03cm}A,\beta_{\mu}\hspace{0.03cm}]_{\hspace{0.01cm}q}
 \equiv A\hspace{0.02cm}\beta_{\mu} - q\hspace{0.02cm}\beta_{\mu}\hspace{0.02cm}A
 =
\bigl(m\hspace{0.02cm}A\hspace{0.02cm}\beta_{\mu}\hspace{0.02cm}A^{2\!} -
q\hspace{0.02cm}\beta_{\mu}\bigr)A.
\]
Finally, with the use of an explicit form of the matrices $A \beta_{\mu} A^2$ and $A^2 \beta_{\mu} A$, Eq.\,(\ref{eq:4u}), we derive the final form of two equivalent representations of the $q$-commutator,
\[
[\hspace{0.03cm}A,\beta_{\mu}\hspace{0.03cm}]_{\hspace{0.02cm}q}
=
A\biggl[\biggl(1 + \frac{1}{2}\,q\biggr)\beta_{\mu} + \biggl(\frac{i\sqrt{3}}{2}\biggr)\hspace{0.02cm}q\hspace{0.03cm}\xi_{\mu}\biggr]
=
\biggl[\hspace{0.02cm}-\biggl(\,\frac{1}{2} + q\biggr)\beta_{\mu} + \biggl(\frac{i\sqrt{3}}{2}\biggr)\hspace{0.02cm}\xi_{\mu}\biggr] A.
\]
The expressions in square brackets are related with each other by a simple relation
\[
\biggl[\hspace{0.02cm}-\biggl(\,\frac{1}{2} + q\biggr)\beta_{\mu} + \biggl(\frac{i\sqrt{3}}{2}\biggr)\hspace{0.02cm}\xi_{\mu}\biggr]
=
q^{2}
\biggl[\biggl(1 + \frac{1}{2}\,q\biggr)\beta_{\mu} + \biggl(\frac{i\sqrt{3}}{2}\biggr)\hspace{0.02cm}q\hspace{0.03cm}\xi_{\mu}\biggr].
\]
It is clear that as the matrix $\eta_{\mu}$ in (\ref{eq:5q}) it is necessary to take the following expression\footnote{\label{foot:2}\,The notation $\eta_{\mu}$ we have introduced for the matrices (\ref{eq:5y}), is not quite appropriate. In the general theory of the DKP algebra \cite{kemmer_1939, fujiwara_1953, fischbach_1973} usually by this symbol the specific expression, namely $\eta_{\mu}\equiv2\hspace{0.02cm} \beta_{\mu}^{\hspace{0.01cm}2} - g_{\mu \mu}$, is meant. However, by virtue of the fact that we do not use these matrices in the text, this should not mislead. The only exception is Section 5, where we will need a particular value of the expression written out just above for $\mu=0$. To avoid confusion, we set off the  symbol $\boldsymbol{\eta}_0\,(\hspace{0.02cm}\equiv2\hspace{0.02cm} \beta_0^2-1)$ in bold.}
\begin{equation}
\eta_{\mu} = \biggl(1 + \frac{1}{2}\,q\biggr)\beta_{\mu} + \biggl(\frac{i\sqrt{3}}{2}\biggr)\hspace{0.02cm}q\hspace{0.03cm}\xi_{\mu},
\label{eq:5y}
\end{equation}
and the complex parameter ${\rm w}$ should be set equal to $q^2$. Thus, the rules of the rearrangements of the matrices $A$ and $\eta_{\mu}$ can be written in the final form,
\begin{equation}
A\hspace{0.02cm}\eta_{\mu} = q^{2}\hspace{0.02cm}\eta_{\mu}\hspace{0.02cm}A,
\quad
A^{2}\eta_{\mu} = q\hspace{0.03cm}\eta_{\mu}A^{2}.
\label{eq:5u}
\end{equation}
\indent
In the choice ${\rm w} = q^{2}$ according to (\ref{eq:5r}) the linear and quadratic in ${\partial}^{\mu}$ contributions in (\ref{eq:5e}) vanish. However, as we already know from the results of the previous section, the contribution in (\ref{eq:5t}) cubic in the derivatives after symmetrization with respect to the vector indices also vanishes. Here we can trace in more detail the reason of this strange fact. By using an explicit form of the $\eta$-matrices, it is not difficult to see that now instead of the identity (\ref{eq:1s}) we have
\begin{equation}
\eta_{\mu}\hspace{0.02cm}\eta_{\nu}\hspace{0.02cm}
\eta_{\lambda\,}\partial^{\mu}\partial^{\nu}\partial^{\lambda}
=
\biggl[\biggl(1 + \frac{1}{2}\,q\biggr)^{\!2}
-\,
\biggl(\frac{i\sqrt{3}}{2}\biggr)^{\!2\!}q^{2}\biggr]\hspace{0.03cm}
\Box\hspace{0.03cm}\eta_{\mu}\hspace{0.02cm}\partial^{\mu}.
\label{eq:5i}
\end{equation}
The expression in square brackets is formally equal to
\begin{equation}
\biggl[\biggl(1 + \frac{1}{2}\;q\biggr)^{\!2} -\, \biggl(\frac{i\sqrt{3}}{2}\biggr)^{\!2\!}q^{2}
\biggr]
=
1 + q +q^{2} \equiv \lim_{z\hspace{0.02cm}\rightarrow\hspace{0.02cm} q}\hspace{0.02cm}\varepsilon(z) = 0.
\label{eq:5o}
\end{equation}
Let the function $\varepsilon(z)$ be a small but finite quantity. It is clear that the quantity is defined correctly to an arbitrary numeric factor. For example, the left-hand side of (\ref{eq:5o}) can be formally represented as a product of two multipliers
\[
\biggl[\biggl(1 + \frac{1}{2}\;q\biggr)^{\!2} -\, \biggl(\frac{i\sqrt{3}}{2}\biggr)^{\!2\!}q^{2}
\biggr]
=
\biggl[\biggl(1 + \frac{1}{2}\,q\biggr) - \biggl(\frac{i\sqrt{3}}{2}\biggr)\hspace{0.02cm}q\,\biggr]
\biggl[\biggl(1 + \frac{1}{2}\,q\biggr) + \biggl(\frac{i\sqrt{3}}{2}\biggr)\hspace{0.02cm}q\,\biggr]
\]
\[
=
\biggl[1 - q\biggl(-\frac{1}{2} + i\,\frac{\sqrt{3}}{2}\biggr)\biggr]
\biggl[1 - q\biggl(-\frac{1}{2} - i\,\frac{\sqrt{3}}{2}\biggr)\biggr]
\equiv
(1 - q^{2})(1 - q^{3})
\]
\[
= (1 - q^{2})(1 - q)(1 + q + q^{2}) \equiv 3\!\lim_{z\hspace{0.02cm}\rightarrow\hspace{0.03cm} q}\hspace{0.02cm}\varepsilon(z) = 0.
\]
Here, we have used the definition of cubic roots of unity (\ref{eq:1y}). The expression obtained differs from (\ref{eq:5o}) by the factor 3.\\
\indent
Instead of the operator $(i\hspace{0.015cm}\eta_{\mu}\hspace{0.03cm}\partial^{\mu} -
m\hspace{0.01cm}I\hspace{0.02cm})$, we introduce the following operator:
\[
\biggl(\frac{\!i}{\,\varepsilon^{1/3}(z)}\,\eta_{\mu}(z)
\hspace{0.03cm}\partial^{\mu} - m\hspace{0.02cm}I\biggr)
\]
and correspondingly instead of the expression on the left-hand side of (\ref{eq:5e}) we put
\[
\biggl[\hspace{0.03cm}A\hspace{0.02cm}\biggl(\frac{\!i}{\,\varepsilon^{1/3}(z)}\,\eta_{\mu}(z)
\hspace{0.03cm}\partial^{\mu} - m\hspace{0.02cm}I\biggr)\biggr]^{3},
\]
where we have introduced the notation
\[
\eta_{\mu}(z) \equiv \biggl(1 + \frac{1}{2}\,z\biggr)\beta_{\mu} + z\hspace{0.03cm}\biggl(\frac{i\sqrt{3}}{2}\biggr)\hspace{0.02cm}\hspace{0.02cm}\xi_{\mu}.
\]
In the limit $z \rightarrow q$ according to the formulas (\ref{eq:5r}), the contribution that is linear in $\partial^{\mu}$ behaves as $\varepsilon^{2/3}(z) \rightarrow 0$ and the one that is quadratic in $\partial^{\mu}$ behaves as $\varepsilon^{1/3}(z) \rightarrow 0$. On the strength of Eqs.\,(\ref{eq:5i}) and (\ref{eq:5o}), nonvanishing contribution gives us only the term cubic in ${\partial}^{\mu}$ and thus we finally obtain the desired expression,
\begin{equation}
\lim_{z\hspace{0.02cm}\rightarrow\hspace{0.03cm} q}\hspace{0.02cm}
\biggl[\hspace{0.03cm}A\hspace{0.02cm}\biggl(\frac{\!i}{\,\varepsilon^{1/3}(z)}\,\eta_{\mu}(z)
\hspace{0.03cm}\partial^{\mu} - m\hspace{0.02cm}I\biggr)\biggr]^{3}
=
\biggl(
-i\,\frac{1}{m}\;\Box\hspace{0.04cm}\eta_{\mu}\hspace{0.02cm}\partial^{\mu} - m^{2\!}\hspace{0.02cm}I
\biggr),
\label{eq:5p}
\end{equation}
where
\[
\lim_{z\hspace{0.02cm}\rightarrow\hspace{0.03cm} q}\eta_{\mu}(z) = \eta_{\mu}(q)\equiv \eta_{\mu}
\]
and $\eta_{\mu}$ is defined by the expression (\ref{eq:5y}).

\section{Properties of the $\eta\hspace{0.02cm}$-matrices}
\setcounter{equation}{0}

Let us derive a number of relations to which the matrices $\eta_{\mu}$ satisfy. Our first step is to consider the commutator of two $\eta$\hspace{0.02cm}-\hspace{0.02cm}matrices. In view of the original definition (\ref{eq:5y}) we have
\[
\begin{split}
[\hspace{0.02cm}\eta_{\mu},\eta_{\nu}]
&=
\biggl(1 + \frac{1}{2}\,q\biggr)^{\!\!2\,}[\hspace{0.02cm}\beta_{\mu},\beta_{\nu}\hspace{0.02cm}]
+
q^{2\,}\biggl(\frac{i\sqrt{3}}{2}\biggr)^{\!\!2\,}[\hspace{0.03cm}\xi_{\mu},\xi_{\nu}\hspace{0.02cm}] \\
&+
q\hspace{0.02cm}\biggl(1 + \frac{1}{2}\,q\biggr)\biggl(\frac{i\sqrt{3}}{2}\biggr)
\bigl(\hspace{0.02cm}[\hspace{0.03cm}\xi_{\mu},\beta_{\nu}\hspace{0.02cm}] + [\hspace{0.02cm}\beta_{\mu},\xi_{\nu}\hspace{0.02cm}]\hspace{0.02cm}\bigr).
\end{split}
\]
We recall that $\xi_{\mu} = [\hspace{0.02cm}\omega, \beta_{\mu}\hspace{0.02cm}]$. By making use of the formulas of the $\omega$\hspace{0.02cm}-\hspace{0.01cm}$\beta_{\mu}$ algebra in Appendix \ref{appendix_A}, it is not difficult to obtain the following relations:
\[
[\hspace{0.03cm}\xi_{\mu},\xi_{\nu}\hspace{0.03cm}] = - \hspace{0.02cm} [\hspace{0.03cm}\beta_{\mu},\beta_{\nu}\hspace{0.03cm}]
\]
and
\[
[\hspace{0.02cm}\xi_{\mu},\beta_{\nu}] + [\hspace{0.02cm}\beta_{\mu},\xi_{\nu}] =
[\hspace{0.04cm}\omega,[\hspace{0.02cm}\beta_{\mu},\beta_{\nu}]\hspace{0.02cm}],
\]
whereupon
\[
 i\hspace{0.04cm}[\hspace{0.02cm}\eta_{\mu},\eta_{\nu}]
=
\biggl[\biggl(1 + \frac{1}{2}\,q\biggr)^{\!2} -\, \biggl(\frac{i\sqrt{3}}{2}\biggr)^{\!2\!}q^{2}
\biggr]S^{(\beta)}_{\mu\nu}
+
q\hspace{0.02cm}\biggl(1 + \frac{1}{2}\,q\biggr)\biggl(\frac{i\sqrt{3}}{2}\biggr)
[\hspace{0.03cm}\omega,S^{(\beta)}_{\mu\nu}\hspace{0.02cm}].
\]
Here, we have denoted
\begin{equation}
S^{(\beta)}_{\mu\nu} \equiv i\hspace{0.035cm}[\hspace{0.02cm}\beta_{\mu},\beta_{\nu}].
\label{eq:6q}
\end{equation}
In view of the property (\ref{ap:A8}) the last term on the right-hand side of the above expression is equal to zero. In the first term by the coefficient preceding the matrix $S_{\mu \nu}^{(\beta)}$, the expression (\ref{eq:5o}) is meant.  Thus, as the commutation relation for the $\eta$\hspace{0.02cm}-\hspace{0.02cm}matrices we take the following expression:
\begin{equation}
\lim_{z\rightarrow\hspace{0.03cm} q}\hspace{0.01cm}
\frac{1}{\,\varepsilon(z)}\,i\hspace{0.04cm}[\hspace{0.03cm}\eta_{\mu}(z),\eta_{\nu}(z)
\hspace{0.02cm}]  = S^{(\beta)}_{\mu\nu}.
\label{eq:6w}
\end{equation}
This relation will be deeply used in the next section in analysis of the interacting case.\\
\indent
In addition, the relation (\ref{eq:6w}) enables us to clear up a question about the relativistic invariance of equation
\begin{equation}
A\biggl(\frac{\!i}{\,\varepsilon^{1/3}(z)}\,\eta_{\mu}(z)
\hspace{0.03cm}\partial^{\mu} - m\hspace{0.02cm}I\biggr)\psi(x;z) = 0
\label{eq:6ww}
\end{equation}
(in the notation of  the wave function $\psi$ we have explicitly separated out the dependence on the deformation parameter $z$). In fact, let us consider the double commutation relation with the $\eta$-\hspace{0.01cm}matrices. By using (\ref{eq:6w}) we have
\begin{equation}
\lim_{z\rightarrow\hspace{0.03cm} q}\hspace{0.01cm}\frac{1}{\,\varepsilon(z)}\,i\hspace{0.05cm}
[\hspace{0.03cm}[\hspace{0.03cm}\eta_{\mu}(z),\eta_{\nu}(z)],\eta_{\lambda}(z)\hspace{0.02cm}]
=
[\hspace{0.02cm}S^{(\beta)}_{\mu\nu},\eta_{\lambda}\hspace{0.02cm}].
\label{eq:6e}
\end{equation}
On the strength of the definition of the $\eta$\hspace{0.03cm}-\hspace{0.02cm}matrices we write out the right-hand side
\[
[\hspace{0.02cm}S^{(\beta)}_{\mu\nu},\eta_{\lambda}\hspace{0.02cm}]
=
\biggl(1 + \frac{1}{2}\,q\biggr)
[\hspace{0.02cm}S^{(\beta)}_{\mu\nu},\beta_{\lambda}\hspace{0.02cm}]
+
q\hspace{0.02cm}\biggl(\frac{i\sqrt{3}}{2}\biggr)
[\hspace{0.02cm}S^{(\beta)}_{\mu\nu},\xi_{\lambda}\hspace{0.02cm}].
\]
The first commutator, by virtue of the trilinear algebra of $\beta$\hspace{0.02cm}-\hspace{0.02cm}matrices, equals
\[
[\hspace{0.02cm}S^{(\beta)}_{\mu\nu},\beta_{\lambda}\hspace{0.02cm}]
=
i\hspace{0.03cm}(\hspace{0.02cm}\beta_{\mu}\hspace{0.03cm}g_{\lambda\nu} - \beta_{\nu}\hspace{0.03cm}g_{\lambda\mu}\hspace{0.02cm}),
\]
and the second one by using the same algebra and the property (\ref{ap:A8}) does
\[
[\hspace{0.02cm}S^{(\beta)}_{\mu\nu},\xi_{\lambda}\hspace{0.02cm}]
=
i\hspace{0.03cm}(\xi_{\mu}\hspace{0.02cm}g_{\lambda\nu}
-
\xi_{\nu}\hspace{0.02cm}g_{\lambda\mu}\hspace{0.02cm}).
\]
Gathering the expressions obtained, we finally find
\begin{equation}
[\hspace{0.02cm}S^{(\beta)}_{\mu\nu},\eta_{\lambda}\hspace{0.02cm}]
=
i\hspace{0.04cm}(\hspace{0.02cm}\eta_{\mu}\hspace{0.02cm}g_{\lambda\nu} - \eta_{\nu}\hspace{0.02cm}g_{\lambda\mu}\hspace{0.01cm}).
\label{eq:6r}
\end{equation}
If by analogy with (\ref{eq:6q}) we introduce a new matrix $S_{\mu \nu}^{(\eta)}(z)$ setting by definition
\[
S^{(\eta)}_{\mu\nu}(z) \equiv \frac{\!1}{\varepsilon^{2/3}(z)}\,i\,
[\hspace{0.02cm}\eta_{\mu}(z),\eta_{\nu}(z)\hspace{0.02cm}],
\]
then the double commutator can be presented in a more customary form
\[
\lim_{z\rightarrow\hspace{0.03cm} q}\,\hspace{0.02cm}
[\hspace{0.02cm}S^{(\eta)}_{\mu\nu}(z),\hspace{0.02cm}
\frac{\!1}{\varepsilon^{1/3}(z)}\,\eta_{\lambda}(z)\hspace{0.02cm}]
=
i\hspace{0.04cm}(\hspace{0.02cm}\eta_{\mu}\hspace{0.02cm}g_{\lambda\nu} - \eta_{\nu}\hspace{0.02cm}g_{\lambda\mu}\hspace{0.01cm}).
\]
This relationship ensures us the invariance of equation for the wave function $\psi(x;z)$ under a Lorentz transformation when we pass to the limit as $z$ tends to $q$.\\
\indent
Further, let us consider the question of a trilinear relation to which the matrices $\eta_{\mu}$ have to satisfy. In other words, what is analog of the relation (\ref{eq:1e}) for these matrices? Here, we skip the calculational details and give only the final result
\begin{equation}
\lim_{z\rightarrow\hspace{0.03cm} q}\hspace{0.02cm}
\frac{1}{\varepsilon(z)}\,\bigl(\hspace{0.03cm}\eta_{\mu}(z)\eta_{\lambda}(z)\eta_{\nu}(z) + \eta_{\nu}(z)\eta_{\lambda}(z)\eta_{\mu}(z)\bigr)
=
g_{\mu\lambda}\hspace{0.02cm}\eta_{\nu} + g_{\nu\lambda}\hspace{0.02cm}\eta_{\mu}.
\label{eq:6t}
\end{equation}
The proof of the trilinear relation is presented in Appendix \ref{appendix_C}.\\
\indent
One more interesting question is connected with the behavior of the
$\eta$\hspace{0.02cm}-\hspace{0.02cm}matrices under the operation of Hermitian conjugation (denoted by the symbol $\dagger$). First of all we note that instead of the expression (\ref{eq:5p}) for  the cube of the first order differential operator, an equivalent expression could be used
\[
\lim_{z\hspace{0.02cm}\rightarrow\hspace{0.03cm} q}\hspace{0.02cm}
\biggl[\hspace{0.03cm}m^{1/3\!}A^{2}\biggl(\frac{\!1}{\,\varepsilon^{1/3}(z)}\,i\hspace{0.03cm}
\bar{\eta}_{\mu}(z)\hspace{0.03cm}\partial^{\mu} - m\hspace{0.02cm}I\biggr)\biggr]^{3}
=
\biggl(
-i\,\frac{1}{m}\;\Box\hspace{0.05cm}\bar{\eta}_{\mu}\hspace{0.02cm}\partial^{\mu} - m^{2}I
\biggr),
\]
where the matrices $\bar{\eta}_{\mu}$\,$\bigl(\hspace{0.02cm}\equiv\lim_{z\hspace{0.02cm}\rightarrow\hspace{0.03cm} q} \bar{\eta}_{\mu}(z)\bigr)$ satisfy the $q$-commutation relations
\[
A^{2}\hspace{0.02cm}\bar{\eta}_{\mu} = q\hspace{0.02cm}\bar{\eta}_{\mu}\hspace{0.02cm}A^{2},
\quad
A\hspace{0.02cm}\bar{\eta}_{\mu} = q^{2}\hspace{0.02cm}\bar{\eta}_{\mu}\hspace{0.02cm}A,
\]
and their explicit form is defined by the following expression
\[
\bar{\eta}_{\mu} = \biggl(1 + \frac{1}{2}\,q^{2}\biggr)\beta_{\mu} - \biggl(\frac{i\sqrt{3}}{2}\biggr)\hspace{0.02cm}q^{2}\hspace{0.02cm}\xi_{\mu}.
\]
The matrices $\bar{\eta}_{\mu}$ and $\eta_{\mu}$ are related by the simple relationship
\begin{equation}
\bar{\eta}_{\mu} = \frac{\!\!\!1 - q}{1 - q^{2}}\,\hspace{0.02cm}\eta_{\mu} \equiv -\hspace{0.02cm}q\hspace{0.03cm}\eta_{\mu}.
\label{eq:6y}
\end{equation}
\indent
By virtue of the initial definition (\ref{eq:5y}) we have
\begin{equation}
\eta^{\dagger}_{\mu} = \biggl(1 + \frac{1}{2}\,q^{\ast}\biggr)\beta^{\dagger}_{\mu} - \biggl(\frac{i\sqrt{3}}{2}\biggr)\hspace{0.02cm}q^{\ast}\hspace{0.02cm}\xi^{\dagger}_{\mu}.
\label{eq:6u}
\end{equation}
On the strength of the properties of cubic roots of unity the equality $q^{\ast}=q^2$ holds. Further, for the $\beta$\hspace{0.02cm}-\hspace{0.02cm}matrices the following relation,
\[
\beta^{\dagger}_{\mu} = \boldsymbol{\eta}_{0}\beta_{\mu}\boldsymbol{\eta}_{0}
\]
is true. Here $\boldsymbol{\eta}_0 = 2\hspace{0.02cm}(\beta_{0})^{2} - 1$ (see footnote \ref{foot:2} in the preceding section). Besides, by making use of the Hermitian character of the $\omega$\hspace{0.02cm}-\hspace{0.02cm}matrix, we find
\[
\xi^{\dagger}_{\mu} = (\boldsymbol{\eta}_{0}\hspace{0.02cm}\beta_{\mu}\hspace{0.01cm}\boldsymbol{\eta}_{0})
\hspace{0.03cm}\omega
-
\omega\hspace{0.02cm}(\boldsymbol{\eta}_{0}\hspace{0.02cm}\beta_{\mu}\hspace{0.01cm}
\boldsymbol{\eta}_{0}).
\]
Under these circumstances, multiplying the expression (\ref{eq:6u}) on both sides by $\boldsymbol{\eta}_0$ and taking into account the properties
\[
\boldsymbol{\eta}_{0}\hspace{0.02cm}\beta_{i} =
- \beta_{i}\hspace{0.02cm}\boldsymbol{\eta}_{0},
\quad
\boldsymbol{\eta}_{0}\hspace{0.02cm}\beta_{0} =
\beta_{0}\hspace{0.02cm}\boldsymbol{\eta}_{0},
\quad
\boldsymbol{\eta}_{0}\hspace{0.02cm}\omega\hspace{0.03cm}\boldsymbol{\eta}_{0} = -\hspace{0.02cm} \omega,
\quad
\boldsymbol{\eta}_{0}^{2} = 1,
\]
we finally obtain
\[
\boldsymbol{\eta}_{0}\hspace{0.02cm}\eta^{\dagger}_{\mu}\hspace{0.02cm}
\boldsymbol{\eta}_{0}
=
\biggl(1 + \frac{1}{2}\,q^{2}\biggr)\beta_{\mu} -
\biggl(\frac{i\sqrt{3}}{2}\biggr)\hspace{0.02cm}q^{2}\hspace{0.02cm}\xi_{\mu}
\equiv \bar{\eta}_{\mu}.
\]
Comparing the expression above with (\ref{eq:6y}), we derive the desired rule of Hermitian conjugation
\[
\boldsymbol{\eta}_{0}\hspace{0.02cm}\eta^{\dagger}_{\mu}\hspace{0.02cm}
\boldsymbol{\eta}_{0}
=
-\hspace{0.02cm}q\hspace{0.02cm}\eta_{\mu}.
\]
\indent
In closing this section we discuss the question of the existence of such a nonsingular transformation $\mathbb{T}$ that would connect the matrices $\beta_{\mu}$ with the matrices $\eta_{\mu}$, i.e.,
\begin{equation}
\mathbb{T}\hspace{0.02cm}\beta_{\mu}\hspace{0.01cm}\mathbb{T}^{-1} = \eta_{\mu}.
\label{eq:6i}
\end{equation}
Let us seek the matrices $\mathbb{T}$ and $\mathbb{T}^{-1}$ in the form of an expansion in powers of $\omega$
\[
\begin{split}
&\mathbb{T} = a + b\,\omega + c\,\omega^{2},\\
&\mathbb{T}^{-1} = \bar{a} + \bar{b}\,\omega + \bar{c}\,\omega^{2},
\end{split}
\]
where $a,\hspace{0.02cm}b,\hspace{0.02cm}\ldots$ are some unknown constants. Substituting these expansions into the left-hand side of Eq.\,(\ref{eq:6i}) and using the formulas (\ref{ap:A1})\,--\,(\ref{ap:A3}), we derive the first system of algebraic equations for unknown quantities
\[
\left\{
\begin{array}{ll}
a\hspace{0.02cm}\bar{a} + a\hspace{0.02cm}\bar{c} =
\biggl(1 + \displaystyle\frac{1}{2}\;q\biggr),\\[2ex]
c\hspace{0.02cm}\bar{a} - a\hspace{0.02cm}\bar{c} = 0,
\end{array}
\right.
\qquad
\left\{
\begin{array}{ll}
b\hspace{0.03cm}\bar{a} = q\hspace{0.03cm}\biggl(\displaystyle\frac{i\sqrt{3}}{2}\biggr),\\[2.3ex]
a\hspace{0.03cm}\bar{b} = -\hspace{0.02cm}q\hspace{0.03cm}\biggl(\displaystyle\frac{i\sqrt{3}}{2}\biggr).
\end{array}
\right.
\]
We define a solution of this system as a function of two arbitrary quantities $\bar{a}$ and $\bar{c}$
\begin{equation}
\begin{array}{ll}
a = \displaystyle\frac{1}{(\bar{a} + \bar{c})}\,\biggl(1 + \displaystyle\frac{1}{2}\,q\biggr),\\[4ex]
\vspace{0.03cm}
c = \displaystyle\frac{\bar c}{\bar a}\,\displaystyle\frac{1}{(\bar{a} + \bar{c})}\,
\biggl(1 + \displaystyle\frac{1}{2}\,q\biggr),
\end{array}
\qquad
\begin{array}{ll}
\vspace{0.03cm}
b = q\,\displaystyle\frac{1}{\bar{a}}\,\biggl(\displaystyle\frac{i\sqrt{3}}{2}\biggr),\\[2ex]
\vspace{0.03cm}
\bar{b} = -\hspace{0.02cm}\displaystyle\frac{q}{\biggl(1 + \displaystyle\frac{1}{2}\,q\biggr)}\hspace{0.03cm}
(\bar{a} + \bar{c})\hspace{0.03cm}\biggl(\displaystyle\frac{i\sqrt{3}}{2}\biggr).
\end{array}
\label{eq:6o}
\end{equation}
Further, we require the fulfilment of the relation
\begin{equation}
\mathbb{T}\hspace{0.04cm}\mathbb{T}^{-1} = I
\label{eq:6p}
\end{equation}
that gives us the second algebraic system:
\begin{equation}
\left\{
\begin{array}{ll}
a\bar{a} = 1,\\
\vspace{0.03cm}
(a\hspace{0.01cm}\bar{b} + b\hspace{0.03cm}\bar{a}\hspace{0.02cm}) + (b\hspace{0.03cm}\bar{c}
+ c\hspace{0.03cm}\bar{b}\hspace{0.02cm}) = 0,\\
(a\hspace{0.01cm}\bar{c} + c\hspace{0.03cm}\bar{a}\hspace{0.02cm}) + (b\hspace{0.03cm}\bar{b}
+ c\hspace{0.03cm}\bar{c}\hspace{0.02cm}) = 0.
\end{array}
\right.
\label{eq:6a}
\end{equation}
The use of the first equation in (\ref{eq:6a}) enables us to express all coefficients through an arbitrary constant $\bar{a}$,
\begin{equation}
\begin{array}{ll}
a = \displaystyle\frac{1}{\bar{a}}\,,\\[1.7ex]
\phantom{\bar{b}
= \biggl(\displaystyle\frac{i\sqrt{3}}{2}\biggr)}
\end{array}
\begin{array}{ll}
\vspace{0.03cm}
b = \displaystyle\frac{1}{\bar{a}}\,\biggl(\displaystyle\frac{i\sqrt{3}}{2}\biggr)\hspace{0.02cm}q,\\[2ex]
\bar{b} = -\hspace{0.02cm}\bar{a}\hspace{0.02cm}
\biggl(\displaystyle\frac{i\sqrt{3}}{2}\biggr)\hspace{0.02cm}q,
\end{array}
\quad
\begin{array}{ll}
\vspace{0.05cm}
c = \displaystyle\frac{1}{2}\,\displaystyle\frac{1}{\bar a}\,q\\[2.4ex]
\vspace{0.05cm}
\bar{c} = \displaystyle\frac{1}{2}\,{\bar a}\,q
\end{array}
\label{eq:6s}
\end{equation}
and {\it ipso facto} the required transformation has the following structure:
\[
\begin{split}
&\mathbb{T}\, = \frac{1}{\bar{a}}\, \biggl[\hspace{0.02cm}I + \biggl(\displaystyle\frac{i\sqrt{3}}{2}\biggr)\hspace{0.02cm}q\hspace{0.035cm}\omega
+
\frac{1}{2}\,q\hspace{0.035cm}\omega\hspace{0.02cm}\biggr],\\
&\mathbb{T}^{-1\!} =  \bar{a}\,\biggl[\hspace{0.02cm}I - \biggl(\displaystyle\frac{i\sqrt{3}}{2}\biggr)\hspace{0.02cm}q\hspace{0.035cm}\omega
+
\frac{1}{2}\,q\hspace{0.035cm}\omega\hspace{0.02cm}\biggr].
\end{split}
\]
However, a straightforward multiplication of these two expressions leads to
\[
\mathbb{T}\hspace{0.04cm}\mathbb{T}^{-1} = I - \omega^{2},
\]
instead of the desired one (\ref{eq:6p}). This points to the fact that  there is a contradiction in two remaining equations of the system (\ref{eq:6a}). Substitution of the solution (\ref{eq:6s}) into the second equation of (\ref{eq:6a}) results in the identity, and the third equation gives
\[
\frac{1}{2}\;q + \hspace{0.01cm} \frac{1}{2}\;q
-  \biggl(\displaystyle\frac{i\sqrt{3}}{2}\biggr)\hspace{0.02cm}q^{2}
+ \hspace{0.03cm} \frac{1}{4}\;q^{2}
= q \hspace{0.03cm}+\hspace{0.03cm} q^{2} = -1.
\]
The equation does not vanish.  This tells us that there is no a nonsingular similarity transformation connecting the matrices $\beta_{\mu}$ with $\eta_{\mu}$ and in this sense, they are {\it nonequivalent}. However, it can be supposed that these matrices are related in a somewhat weak (limiting) sense.

\section{Interacting case}
\setcounter{equation}{0}

In the interaction free case we have derived the expression for the third order wave operator as a limit of cube of a certain first order operator, namely,
\begin{equation}
\lim_{z\hspace{0.02cm}\rightarrow\hspace{0.03cm} q}\hspace{0.02cm}
\biggl[\hspace{0.03cm}A\hspace{0.02cm}\biggl(\frac{\!i}{\,\varepsilon^{1/3}(z)}\,\eta_{\mu}(z)
\hspace{0.03cm}\partial^{\mu} - m\hspace{0.02cm}I\biggr)\biggr]^{3}
=
\biggl(
-i\,\frac{1}{m}\;\Box\hspace{0.04cm}\eta_{\mu}\hspace{0.02cm}\partial^{\mu} - m^{2\!}\hspace{0.02cm}I
\biggr).
\label{eq:7q}
\end{equation}
It remains to take up the question of a modification of this expression in the presence of an external electromagnetic field. We introduce the interaction via the minimal substitution:
\[
\partial^{\mu} \rightarrow D^{\mu}\equiv\partial^{\mu} + i\hspace{0.02cm}eA^{\mu}(x).
\]
With an external gauge field in the system the left-hand side of Eq.\,(\ref{eq:7q}) takes the form
\[
\lim_{z\hspace{0.02cm}\rightarrow\hspace{0.03cm} q}\hspace{0.02cm}
\biggl[\hspace{0.03cm}A\hspace{0.02cm}\biggl(\frac{\!i}{\,\varepsilon^{1/3}(z)}\,\eta_{\mu}(z)
\hspace{0.03cm}D^{\mu} - m\hspace{0.02cm}I\biggr)\biggr]^{3}.
\]
From the last two terms in (\ref{eq:5e}) and from the corresponding relations in (\ref{eq:5r}) it is not difficult to see that the contributions linear and quadratic in the derivatives vanish in the limit $z \rightarrow q$ in the interacting case also. This is independent of the eventual noncommutativity of $D$'s and in doing so, as in the interaction free case, we have the following expression\footnote{\,In the subsequent discussion for simplicity the $z$-dependence of the various quantities under the limit sign is understood although not written out explicitly.}:
\begin{equation}
i^{3\!}\lim_{z\hspace{0.02cm}\rightarrow\hspace{0.03cm} q}\hspace{0.02cm}\frac{1}{\varepsilon}\hspace{0.02cm}
\Bigl(A\hspace{0.02cm}
\eta_{\mu}\hspace{0.02cm}A\hspace{0.02cm}\eta_{\nu}\hspace{0.02cm}A\hspace{0.02cm}
\eta_{\lambda\,}D^{\mu}D^{\nu}D^{\lambda}\Bigr) - A^{3}m^{3}
=
-\hspace{0.01cm}i\,\frac{1}{m}\;\lim_{z\hspace{0.02cm}\rightarrow\hspace{0.03cm} q}\hspace{0.02cm}
\frac{1}{\varepsilon}\hspace{0.02cm}
\Bigl(
\eta_{\mu}\hspace{0.02cm}\eta_{\nu}\hspace{0.02cm}
\eta_{\lambda\,}D^{\mu}D^{\nu}D^{\lambda\!}\Bigr) - m^{2}.
\label{eq:7w}
\end{equation}
However, here one can already expect that by virtue of noncommutativity of the covariant derivative this limit will have overwhelmingly more complicated structure in comparison with the right-hand side of  (\ref{eq:7q}).\\
\indent
For analysis of the expression (\ref{eq:7w}) we make use the following identity for a product of three covariant derivatives
\begin{equation}
6\hspace{0.01cm}D^{\mu}D^{\nu}D^{\lambda} = \{D^{\mu\!},D^{\nu\!},D^{\lambda}\}
\label{eq:7e}
\end{equation}
\[
+\,
\bigl[\hspace{0.025cm}3\hspace{0.01cm}D^{\mu}(i\hspace{0.01cm}e\hspace{0.005cm}F^{\nu\lambda})
+
D^{\lambda}(i\hspace{0.01cm}e\hspace{0.005cm}F^{\mu\nu})
+
D^{\nu}(i\hspace{0.01cm}e\hspace{0.005cm}F^{\mu\lambda})\bigr]
+
\bigl[\hspace{0.02cm}2\hspace{0.02cm}(i\hspace{0.01cm}e\hspace{0.005cm}F^{\mu\lambda})D^{\nu}
+
2\hspace{0.02cm}(i\hspace{0.01cm}e\hspace{0.005cm}F^{\mu\nu})D^{\lambda}\hspace{0.02cm}\bigr],
\]
where for brevity by the symbol $ \{D^{\mu\!},D^{\nu\!},D^{\lambda}\}$ we mean a product of three $D$\hspace{0.02cm}-\hspace{0.02cm}operators completely symmetrized over the vector indices
$\mu,\hspace{0.02cm}\nu$ and $\lambda$:
\begin{equation}
\{D^{\mu\!},D^{\nu\!},D^{\lambda}\} \equiv
\bigl(D^{\mu}D^{\nu}D^{\lambda} + D^{\lambda}D^{\nu}D^{\mu}\bigr)
+
\bigl(D^{\nu}D^{\mu}D^{\lambda} + D^{\lambda}D^{\mu}D^{\nu}\bigr)
+
\bigl(D^{\mu}D^{\lambda}D^{\nu} + D^{\nu}D^{\lambda}D^{\mu}\bigr).
\label{eq:7r}
\end{equation}
The Abelian strength tensor $F_{\mu \nu}(x)$ is defined by
\[
[\hspace{0.02cm}D^{\mu\!},D^{\nu}\hspace{0.02cm}] = i\hspace{0.02cm}e\hspace{0.005cm}F^{\mu\nu}(x).
\]
The proof of the identity (\ref{eq:7e}) is given in Appendix \ref{appendix_D}.\\
\indent
Our first step is to consider the contribution in (\ref{eq:7w}) due to the symmetrized part (\ref{eq:7r}). In view of a total symmetry over permutation of the indices, we get
\[
-\hspace{0.01cm}i\,\frac{1}{m}\;\lim_{z\hspace{0.02cm}\rightarrow\hspace{0.03cm} q}\hspace{0.02cm}
\frac{1}{\varepsilon}\;
\eta_{\mu}\hspace{0.02cm}\eta_{\nu}\hspace{0.02cm}
\eta_{\lambda}\hspace{0.02cm}\{D^{\mu\!},D^{\nu\!},D^{\lambda}\}
\]
\[
= -\hspace{0.01cm}i\,\frac{1}{6\hspace{0.02cm}m}\;\lim_{z\hspace{0.02cm}
\rightarrow\hspace{0.03cm} q}\hspace{0.02cm}
\frac{1}{\varepsilon}\hspace{0.02cm}
\bigl[\hspace{0.02cm}
(\eta_{\mu}\hspace{0.02cm}\eta_{\nu}\hspace{0.02cm}\eta_{\lambda}
+
\eta_{\lambda}\hspace{0.02cm}\eta_{\nu}\hspace{0.02cm}\eta_{\mu})
+
(\eta_{\nu}\hspace{0.02cm}\eta_{\mu}\hspace{0.02cm}\eta_{\lambda}
+
\eta_{\lambda}\hspace{0.02cm}\eta_{\mu}\hspace{0.02cm}\eta_{\nu})
+
(\eta_{\nu}\hspace{0.02cm}\eta_{\lambda}\hspace{0.02cm}\eta_{\mu}
+
\eta_{\mu}\hspace{0.02cm}\eta_{\lambda}\hspace{0.02cm}\eta_{\nu})\bigr]
\{D^{\mu\!},D^{\nu\!},D^{\lambda}\}
\]
\[
= -\hspace{0.01cm}i\,\frac{1}{3\hspace{0.02cm}m}\,
\bigl(\hspace{0.02cm}g_{\mu\nu}\hspace{0.02cm}\eta_{\lambda} + g_{\nu\lambda}\hspace{0.02cm}\eta_{\mu} + g_{\mu\lambda}\hspace{0.02cm}\eta_{\nu}\hspace{0.02cm}\bigr)
\{D^{\mu\!},D^{\nu\!},D^{\lambda}\}.
\]
In the last line we have taken into account the trilinear relation (\ref{eq:6t}). By making use of an explicit form of the symmetrized expression (\ref{eq:7r}) and collecting similar terms we obtain finally the desired limit
\[
-\hspace{0.01cm}i\,\frac{1}{m}\;\lim_{z\hspace{0.02cm}\rightarrow\hspace{0.03cm} q}\hspace{0.02cm}
\frac{1}{\varepsilon}\;
\eta_{\mu}\hspace{0.02cm}\eta_{\nu}\hspace{0.02cm}
\eta_{\lambda}\hspace{0.02cm}\{D^{\mu\!},D^{\nu\!},D^{\lambda}\}
=
-\hspace{0.01cm}i\,\frac{2}{m}\,
\bigl[\hspace{0.02cm}D^{2}(\eta_{\lambda}D^{\lambda}) +
D^{\mu}(\eta_{\lambda}D^{\lambda})D_{\mu} + (\eta_{\lambda}D^{\lambda}) D^{2}\hspace{0.03cm}\bigr].
\]
\indent
Let us next consider the contribution from the expression in the first square brackets in (\ref{eq:7e}). The initial expression for our analysis is
\begin{align}
&-i\,\frac{1}{m}\;\lim_{z\hspace{0.02cm}\rightarrow\hspace{0.03cm} q}\hspace{0.03cm}
\frac{1}{\varepsilon}\,\hspace{0.02cm}
\eta_{\mu}\hspace{0.02cm}\eta_{\nu}\hspace{0.02cm}
\eta_{\lambda}\hspace{0.02cm}
\bigl[\hspace{0.025cm}3D^{\mu}(i\hspace{0.01cm}e\hspace{0.005cm}F^{\nu\lambda})
+
D^{\lambda}(i\hspace{0.01cm}e\hspace{0.005cm}F^{\mu\nu})
+
D^{\nu}(i\hspace{0.01cm}e\hspace{0.005cm}F^{\mu\lambda})\hspace{0.02cm}\bigr] \label{eq:7t}\\
&= e\hspace{0.03cm}\frac{1}{2\hspace{0.01cm}m}\,
\lim_{z\hspace{0.02cm}\rightarrow\hspace{0.03cm} q}
\frac{1}{\varepsilon}\hspace{0.04cm}
D^{\mu\!}\hspace{0.01cm}F^{\nu\lambda}\hspace{0.02cm}
\bigl\{\hspace{0.01cm}3\hspace{0.02cm}
\eta_{\mu}\hspace{0.02cm}[\hspace{0.02cm}\eta_{\nu},\eta_{\lambda}]
+
[\hspace{0.02cm}\eta_{\nu},\eta_{\lambda}]\hspace{0.02cm}\eta_{\mu}
+
(\eta_{\nu}\hspace{0.02cm}\eta_{\mu}\hspace{0.02cm}\eta_{\lambda}
-
\eta_{\lambda}\hspace{0.02cm}\eta_{\mu}\hspace{0.02cm}\eta_{\nu})\!\hspace{0.02cm}\bigr\}.\notag
\end{align}
For the first two terms in braces by virtue of the property (\ref{eq:6w}) we can set at once
\begin{equation}
\lim_{z\hspace{0.02cm}\rightarrow\hspace{0.03cm} q}\hspace{0.01cm}
\frac{1}{\,\varepsilon}\,i\,[\hspace{0.02cm}\eta_{\mu},\eta_{\nu}]  = S^{(\beta)}_{\mu\nu} \;
(\hspace{0.02cm}\equiv i\hspace{0.03cm}[\hspace{0.03cm}\beta_{\mu},\beta_{\nu}\hspace{0.02cm}]\hspace{0.02cm}).
\label{eq:7y}
\end{equation}
For the contribution in parentheses we use the identity
\begin{equation}
\eta_{\nu}\hspace{0.02cm}\eta_{\mu}\hspace{0.02cm}\eta_{\lambda}
-
\eta_{\lambda}\hspace{0.02cm}\eta_{\mu}\hspace{0.02cm}\eta_{\nu}
=
\eta_{\mu}\hspace{0.03cm}[\hspace{0.02cm}\eta_{\nu},\eta_{\lambda}]
-
\bigl(\hspace{0.02cm}[\hspace{0.02cm}\eta_{\mu},\eta_{\nu}]\hspace{0.03cm}\eta_{\lambda}
-
[\hspace{0.02cm}\eta_{\mu},\eta_{\lambda}]\hspace{0.03cm}\eta_{\nu}\bigr)
\label{eq:7u}
\end{equation}
and then the relation (\ref{eq:7y}). Eventually the limit of expression (\ref{eq:7t}) takes the form
\[
-\hspace{0.02cm}\frac{1}{2\hspace{0.02cm}m}\,D^{\mu}
(i\hspace{0.01cm}e\hspace{0.006cm}F^{\nu\lambda})\hspace{0.02cm}
\bigl(\hspace{0.02cm}4\hspace{0.03cm}\eta_{\mu}\hspace{0.03cm}S^{(\beta)}_{\nu\lambda}
+
S^{(\beta)}_{\nu\lambda}\eta_{\mu}\hspace{0.02cm}\bigr)
+
\frac{1}{2\hspace{0.02cm}m}\,D^{\mu}
(i\hspace{0.01cm}e\hspace{0.006cm}F^{\nu\lambda})\hspace{0.01cm}
\bigl(\hspace{0.02cm}S^{(\beta)}_{\mu\nu}\eta_{\lambda}
-
S^{(\beta)}_{\mu\lambda}\eta_{\nu}\hspace{0.02cm}\bigr).
\]
\indent
The expression in the second square brackets in (\ref{eq:7e}) is analyzed in just the same way with the use of Eqs.\,(\ref{eq:7y}) and (\ref{eq:7u}). Collecting all of the above calculated in (\ref{eq:7w}) and recalling the factor 6 on the left-hand side of the identity (\ref{eq:7e}), we derive the desired expression for the cube of the linear operator in the presence of an electromagnetic field,
\begin{equation}
\lim_{z\hspace{0.02cm}\rightarrow\hspace{0.03cm} q}\hspace{0.02cm}
\biggl[\hspace{0.03cm}A\hspace{0.02cm}\biggl(\frac{\!i}{\,\varepsilon^{1/3}(z)}\,\eta_{\mu}(z)
\hspace{0.03cm}D^{\mu} - m\hspace{0.02cm}I\biggr)\biggr]^{3}
=
-\hspace{0.01cm}i\,\frac{1}{6\hspace{0.02cm}m}\,
\Bigl\{2\hspace{0.02cm}\bigl[\hspace{0.02cm}
D^{2}(\eta_{\lambda}D^{\lambda})
+
D^{\mu}(\eta_{\lambda}D^{\lambda})D_{\mu}
+
(\eta_{\lambda}D^{\lambda}) D^{2}\hspace{0.03cm}\bigr]
\label{eq:7i}
\end{equation}
\[
\begin{split}
+\,
\frac{1}{2}\,&e\hspace{0.02cm}\bigl[\hspace{0.02cm}D^{\mu\!}\hspace{0.02cm}
F^{\nu\lambda}
\hspace{0.02cm}
\bigl(\hspace{0.02cm}4\hspace{0.02cm}\eta_{\mu}\hspace{0.03cm}S^{(\beta)}_{\nu\lambda}
+
S^{(\beta)}_{\nu\lambda}\eta_{\mu}\hspace{0.02cm}\bigr)
-
D^{\mu\!}\hspace{0.02cm}F^{\nu\lambda}
\hspace{0.02cm}
\bigl(\hspace{0.02cm}S^{(\beta)}_{\mu\nu}\eta_{\lambda}
-
S^{(\beta)}_{\mu\lambda}\eta_{\nu}\hspace{0.02cm}\bigr)\hspace{0.01cm}\bigr]\\
+\,
&e\hspace{0.02cm}\bigl[\hspace{0.02cm}F^{\nu\lambda\!}\hspace{0.02cm}D^{\mu}
\hspace{0.02cm}
\bigl(\hspace{0.02cm}\eta_{\mu}\hspace{0.03cm}S^{(\beta)}_{\nu\lambda}
+
S^{(\beta)}_{\nu\lambda}\eta_{\mu}\hspace{0.02cm}\bigr)
-
F^{\nu\lambda\!}\hspace{0.01cm}D^{\mu}\hspace{0.02cm}
\bigl(\hspace{0.02cm}S^{(\beta)}_{\mu\nu}\eta_{\lambda}
-
S^{(\beta)}_{\mu\lambda}\eta_{\nu}\hspace{0.02cm}\bigr)\hspace{0.01cm}\bigr]\!\Bigr\}
- m^{2}I.
\end{split}
\]
The above expression has been presented in the most symmetric form. However, it can be rewritten in a slightly different form. This will enable us, in particular, to compare it with a similar expression suggested earlier by Nowakowski \cite{nowakowski_1998}. At the beginning we consider the expression in the first square brackets on the right-hand side of (\ref{eq:7i}). In the second term there we rearrange the operator $D^{\mu}$ to the right,
\[
\begin{split}
&D^{\mu}(\eta_{\lambda}D^{\lambda})D_{\mu}
\equiv
(\eta_{\lambda}D^{\lambda}) D^{2} + \eta_{\lambda}\hspace{0.02cm}[\hspace{0.02cm}D^{\mu\!},D^{\lambda}\hspace{0.02cm}]
\hspace{0.02cm}D_{\mu} \\
&=
(\eta_{\lambda}D^{\lambda}) D^{2} +
\frac{1}{2}\,i\hspace{0.01cm}e\hspace{0.005cm}F^{\nu\lambda}D^{\mu}\hspace{0.02cm}
\bigl(\hspace{0.02cm}g_{\mu\nu}\hspace{0.02cm}\eta_{\lambda}
- g_{\mu\lambda}\hspace{0.02cm}\eta_{\nu}\hspace{0.02cm}\bigr).
\end{split}
\]
Similarly, in the first term at the same place we rearrange the operator $D^2$ also to the right
\[
D^{2}(\eta_{\lambda}D^{\lambda})
=
(\eta_{\lambda}D^{\lambda}) D^{2} +
\frac{1}{2}\,i\hspace{0.007cm}e\hspace{0.005cm}F^{\nu\lambda\!}\hspace{0.01cm}
D^{\mu}\hspace{0.02cm}
\bigl(\hspace{0.02cm}g_{\mu\nu}\hspace{0.02cm}\eta_{\lambda}
- g_{\mu\lambda}\hspace{0.02cm}\eta_{\nu}\hspace{0.02cm}\bigr)
+
\frac{1}{2}\,i\hspace{0.01cm}e\hspace{0.01cm}D^{\mu}F^{\nu\lambda}\hspace{0.02cm}
\bigl(\hspace{0.02cm}g_{\mu\nu}\hspace{0.02cm}\eta_{\lambda}
- g_{\mu\lambda}\hspace{0.02cm}\eta_{\nu}\hspace{0.02cm}\bigr).
\]
Thus, instead of the expression in the first square brackets now we have
\[
3\hspace{0.02cm}(\eta_{\lambda}D^{\lambda}) D^{2}
+
i\hspace{0.01cm}e\hspace{0.005cm}F^{\nu\lambda}D^{\mu}\hspace{0.02cm}
\bigl(\hspace{0.02cm}g_{\mu\nu}\hspace{0.02cm}\eta_{\lambda}
- g_{\mu\lambda}\hspace{0.02cm}\eta_{\nu}\hspace{0.02cm}\bigr)
+
\frac{1}{2}\,i\hspace{0.01cm}e\hspace{0.005cm}D^{\mu\!}\hspace{0.02cm}
F^{\nu\lambda}\hspace{0.02cm}
\bigl(\hspace{0.02cm}g_{\mu\nu}\hspace{0.02cm}\eta_{\lambda}
- g_{\mu\lambda}\hspace{0.02cm}\eta_{\nu}\hspace{0.02cm}\bigr).
\]
\indent
Further we rearrange the matrix $\eta_{\mu}$ in the terms $S_{\nu \lambda}^{(\beta)} \eta_{\mu}$  (the second and third square brackets in (\ref{eq:7i})) to the left
\[
S^{(\beta)}_{\nu\lambda}\eta_{\mu} \equiv \eta_{\mu}\hspace{0.02cm}S^{(\beta)}_{\nu\lambda}
+
[\hspace{0.02cm}S^{(\beta)}_{\nu\lambda\!},\eta_{\mu}\hspace{0.02cm}]
=
\eta_{\mu}\hspace{0.02cm}S^{(\beta)}_{\nu\lambda} -
i\hspace{0.03cm}(\hspace{0.02cm}g_{\mu\nu}\hspace{0.02cm}\eta_{\lambda}
- g_{\mu\lambda}\hspace{0.02cm}\eta_{\nu}\hspace{0.02cm}).
\]
Here we have used the property (\ref{eq:6r}). Gathering all of the above calculated and collecting similar terms, finally, we derive instead of (\ref{eq:7i}) the following expression:
\begin{equation}
\begin{split}
&\lim_{z\hspace{0.02cm}\rightarrow\hspace{0.03cm} q}\hspace{0.02cm}
\biggl[\hspace{0.03cm}A\hspace{0.02cm}\biggl(\frac{\!i}{\,\varepsilon^{1/3}(z)}\,\eta_{\mu}(z)
\hspace{0.03cm}D^{\mu} - m\hspace{0.02cm}I\biggr)\biggr]^{3}\\
&=
-\hspace{0.01cm}i\,\frac{1}{m}\,(\eta_{\mu}\hspace{0.02cm}D^{\mu})\hspace{0.02cm}D^{2}
- m^{2}I\\
&-\frac{5}{6}\,\biggl(\frac{i\hspace{0.015cm}e}{2\hspace{0.02cm}m}\biggr)
\bigl(\eta_{\mu}\hspace{0.02cm}S^{(\beta)}_{\nu\lambda}\bigr)
\hspace{0.01cm}D^{\mu\!}\hspace{0.02cm}F^{\nu\lambda\,}
+
\frac{1}{3}\,\biggl(\frac{i\hspace{0.015cm}e}{2\hspace{0.02cm}m}\biggr)
\bigl(\hspace{0.02cm}S^{(\beta)}_{\mu\nu}\eta_{\lambda} - g_{\mu\nu}\eta_{\lambda}\bigr)
\hspace{0.01cm}D^{\mu\!}\hspace{0.02cm}F^{\nu\lambda}\\
&-\frac{4}{6}\,\biggl(\frac{i\hspace{0.015cm}e}{2\hspace{0.02cm}m}\biggr)
\bigl(\eta_{\mu}\hspace{0.02cm}S^{(\beta)}_{\nu\lambda}\bigr)
\hspace{0.01cm}F^{\nu\lambda\!}\hspace{0.02cm}D^{\mu}
+
\frac{2}{3}\,\biggl(\frac{i\hspace{0.015cm}e}{2\hspace{0.02cm}m}\biggr)
\bigl(\hspace{0.02cm}S^{(\beta)}_{\mu\nu}\eta_{\lambda} - g_{\mu\nu}\eta_{\lambda}\bigr)
\hspace{0.01cm}F^{\nu\lambda\!}\hspace{0.02cm}D^{\mu}.
\end{split}
\label{eq:7o}
\end{equation}
Let us compare the right-hand side of this expression with that of Nowakowski (Eq.\,(\ref{eq:6y}) in \cite{nowakowski_1998}). In our notation, we have here
\begin{equation}
\begin{split}
&-\hspace{0.01cm}i\,\frac{1}{m}\,(\beta_{\mu}\hspace{0.025cm}D^{\mu}) D^{2} - m^{2\!}\hspace{0.02cm}I\\
&-\biggl(\frac{i\hspace{0.015cm}e}{2\hspace{0.02cm}m}\biggr)
\bigl(\beta_{\mu}\hspace{0.02cm}S^{(\beta)}_{\nu\lambda}\bigr)
\hspace{0.01cm}D^{\mu\!}\hspace{0.02cm}F^{\nu\lambda\,}\,
\\
&-\!\frac{1}{2}\,\biggl(\frac{i\hspace{0.015cm}e}{2\hspace{0.02cm}m}\biggr)
\bigl(\beta_{\mu}\hspace{0.02cm}S^{(\beta)}_{\nu\lambda}\bigr)
\hspace{0.01cm}F^{\nu\lambda\!}\hspace{0.02cm}D^{\mu}
+
\biggl(\frac{i\hspace{0.015cm}e}{2\hspace{0.02cm}m}\biggr)
\bigl(\hspace{0.02cm}S^{(\beta)}_{\mu\nu}\beta_{\lambda} - g_{\mu\nu}\beta_{\lambda}\bigr)
\hspace{0.01cm}F^{\nu\lambda\!}\hspace{0.02cm}D^{\mu}.
\end{split}
\label{eq:7p}
\end{equation}
In rewriting this expression we have used the identity for the $\beta$-\hspace{0.02cm}matrices
\[
i\hspace{0.03cm}(\hspace{0.02cm}\beta_{\nu}\hspace{0.02cm}\beta_{\mu}\hspace{0.02cm}
\beta_{\lambda}
-
\beta_{\lambda}\hspace{0.02cm}\beta_{\mu}\hspace{0.02cm}\beta_{\nu})
=
\beta_{\mu}S^{(\beta)}_{\nu\lambda}
-
\bigl(S^{(\beta)}_{\mu\nu}\beta_{\lambda}
-
S^{(\beta)}_{\mu\lambda}\beta_{\nu}\bigr).
\]
The first difference between (\ref{eq:7o}) and (\ref{eq:7p}) is that in (\ref{eq:7o}) in all terms instead of the matrices $\beta_{\mu}$ we have the matrices $\eta_{\mu}$ (except for the spin structure
$S_{\nu \lambda}^{(\beta)}=i\hspace{0.02cm}[\hspace{0.03cm}\beta_{\nu}, \beta_{\lambda}\hspace{0.02cm}]$, which is the same in both cases). Besides, in (\ref{eq:7o}) the third term has no analog at all. It is also interesting to note that the numerical coefficients in the last four terms in (\ref{eq:7o}) were generated in such a manner that if the operator $D^{\mu}$ commuted with the function $F^{\nu \lambda}(x)$ the structure of two expressions (\ref{eq:7o}) and (\ref{eq:7p}) (up to the corresponding matrices) might have perfectly coincided.\\
\indent
Thus, in this section we have shown that our approach correctly reproduces the general structure of the third order wave operator in the interacting DKP theory as it was suggested by Nowakowski in \cite{nowakowski_1998} on the basis of purely heuristic considerations. The expression obtained has a more symmetric form in comparison with Nowakowski's one, Eq.\,(\ref{eq:7p}). However, the main difference of our expression (\ref{eq:7o}) from the corresponding expression (\ref{eq:7p}) is that here instead of the original $\beta$-matrices we need to introduce more complicated combinations, namely $\eta$\hspace{0.01cm}-matrices. The physical meaning of this fact so far is not clear for us.

\section{The general structure of a solution of the first order differential equation (\ref{eq:6ww})}
\setcounter{equation}{0}

In this section we analyze the general structure of a solution of the equation of the first order in the derivatives, Eq.\,(\ref{eq:6ww}). With an external electromagnetic field in the system, the equation takes the form
\begin{equation}
\hat{\cal L}(z,D)\hspace{0.02cm}\psi(x;z) = 0.
\label{eq:77q}
\end{equation}
Here, we have introduced a short-hand notation for the first order differential operator
\begin{equation}
\hat{\cal L}(z,D) \equiv
A\hspace{0.02cm}\biggl(\frac{\!i}{\,\varepsilon^{1/3}(z)}\,\eta_{\mu}(z)
\hspace{0.03cm}D^{\mu} - m\hspace{0.02cm}I\biggr).
\label{eq:77w}
\end{equation}
In the notation of this operator we have explicitly separated out the dependence on the deformation parameter $z$. A solution of Eq.\,(\ref{eq:77q}) can be unambiguously presented in the following form:
\begin{equation}
\psi(x;z) = \bigl[\hat{\cal L}(z,D)\bigr]^{2}\varphi(x;z),
\label{eq:77e}
\end{equation}
where in turn the function $\varphi(x;z)$ is a solution of the third order wave equation
\begin{equation}
\bigl[\hat{\cal L}(z,D)\bigr]^{3}\varphi(x;z) = 0.
\label{eq:77r}
\end{equation}
Such a representation of the general solution of the main first order equation (\ref{eq:77q}) is the most convenient in practice by virtue of the fact that the wave function $\varphi (x;z)$ is a regular function of the parameter $z$ in the limit $z \rightarrow q$, whereas $\psi (x;z)$, generally speaking, is not regular (see below). The regularity of the function $\varphi (x;z)$ is a consequence of the existence of a well-defined limit of the cube of the operator $\hat{\cal L}$:
\[
\lim_{z\hspace{0.02cm}\rightarrow\hspace{0.03cm} q}\hspace{0.02cm}
\bigl[\hat{\cal L}(z,D)\bigr]^{3} = \mbox{rhs\! of Eq.\,(\ref{eq:7o})}.
\]
\indent
Let us analyze in more detail the structure of the solution $\psi(x;z)$ in the form (\ref{eq:77e}). For simplicity we restrict our attention to the interaction free case. We introduce the following notation:
\[
\delta \equiv z - q.
\]
Then one can present the matrices\footnote{\,It should be noted that the matrices $\eta_{\mu}(z)$, generally speaking, are defined up to an arbitrary matrix function ${\cal T}_{\mu}(z)$ such that $\lim_{z\hspace{0.02cm}\rightarrow\hspace{0.03cm} q}\hspace{0.02cm} {\cal T}_{\mu}(z)=0$. However, we set for simplicity ${\cal T}_{\mu}(z)\equiv 0$ throughout this paper.} $\eta_{\mu}(z)$ in the form of an expansion in terms of $\delta$:
\begin{equation}
\eta_{\mu}(z) = \biggl(1 + \frac{1}{2}\,z\biggr)\beta_{\mu} + z\hspace{0.03cm}\biggl(\frac{i\sqrt{3}}{2}\biggr)\hspace{0.02cm}\hspace{0.02cm}\xi_{\mu}
= \eta_{\mu} + \delta\hspace{0.03cm} \eta^{\prime}_{\mu}.
\label{eq:77t}
\end{equation}
Here, the matrices $\eta_{\mu}$ are defined by the formula (\ref{eq:5y}), and the matrices $\eta_{\mu}^{\prime}$ have the form
\begin{equation}
\eta^{\prime}_{\mu} \equiv \frac{1}{2}\,\beta_{\mu} + \biggl(\frac{i\sqrt{3}}{2}\biggr)\hspace{0.02cm}\xi_{\mu}.
\label{eq:77y}
\end{equation}
Taking into account the expansion (\ref{eq:77t}) and the definition of the function $\varepsilon(z)$, Eq.\,(\ref{eq:4e}), the first order differential operator $\hat{\cal L}(z, \partial)$ can be rewritten as follows:
\begin{equation}
\hat{\cal L}(z,  \partial) =
\biggl[A\hspace{0.02cm}\biggl(\frac{\!1}{\delta^{1/3}}\,\frac{\!i}{\,\varrho^{1/3}}\,\eta_{\mu}
\hspace{0.03cm}
\partial^{\mu} +\,
\delta^{\hspace{0.02cm}2/3}\frac{\!i}{\,\varrho^{1/3}}\,\eta^{\prime}_{\mu}\hspace{0.03cm}
\partial^{\mu}
-
m\hspace{0.02cm}I\biggr)\biggr], \quad \varrho\equiv q - q^{2}.
\label{eq:77u}
\end{equation}
From an explicit form of this operator it is clear that a solution of the equation (\ref{eq:77q}) can be obtained in the form of a formal series in positive and {\it negative} powers of the parameter $\delta^{1/3}$:
\begin{equation}
\psi(x;z) =\, \ldots\,+ \frac{1}{\delta}\,\psi_{-1}(x) +  \frac{1}{\delta^{\hspace{0.02cm}2/3}}\,\psi_{-2/3}(x) +
\frac{1}{\delta^{\hspace{0.01cm}1/3}}\,\psi_{-1/3}(x) + \psi_{0}(x) +\delta^{\hspace{0.01cm}1/3}\psi_{1/3}(x) +\, \ldots\,.
\label{eq:77i}
\end{equation}
It is naturally to be expected that the wave function $\psi(x;z)$ will be singular in the limit $z\rightarrow q$. It can be seen more precisely from analysis of the representation (\ref{eq:77e}). Really, an expansion of the square of the operator $\hat{\cal L}(z,\partial)$ has the form
\begin{equation}
\bigl[\hat{\cal L}(z,\partial)\bigr]^{2} =
\label{eq:77o}
\end{equation}
\[
\begin{split}
&-\frac{\!1}{\delta^{\hspace{0.02cm}2/3}}\,\frac{\!1}{\,\varrho^{\hspace{0.01cm}2/3}}\,
\bigl(A\hspace{0.02cm}\eta_{\mu}
\hspace{0.02cm}A\hspace{0.02cm}\eta_{\nu}\bigr)
\hspace{0.02cm}\partial^{\mu}\partial^{\nu}
-
\frac{\!1}{\delta^{\hspace{0.01cm}1/3}}\,\frac{\!i}{\,\varrho^{\hspace{0.01cm}1/3}}\,m\hspace{0.03cm}
\bigl(A\hspace{0.02cm}\eta_{\mu}\hspace{0.02cm}A + A^{2}\hspace{0.02cm}\eta_{\mu}\bigr)
\hspace{0.02cm}\partial^{\mu}
+
m^{2}A^{2} \\
&-\,
\delta^{\hspace{0.01cm}1/3}\,\frac{\!1}{\,\varrho^{\hspace{0.01cm}2/3}}\,
\bigl(A\hspace{0.02cm}\eta_{\mu}\hspace{0.02cm}A\hspace{0.02cm}\eta^{\prime}_{\nu}
+
A\hspace{0.02cm}\eta^{\prime}_{\mu}\hspace{0.02cm}A\hspace{0.02cm}\eta_{\nu}
\bigr)\hspace{0.02cm}\partial^{\mu}\partial^{\nu}
-
\delta^{\hspace{0.02cm}2/3}\,\frac{\!i}{\,\varrho^{1/3}}\,m\hspace{0.03cm}
\bigl(A\hspace{0.02cm}\eta^{\prime}_{\mu}\hspace{0.02cm}A + A^{2}\hspace{0.02cm}\eta^{\prime}_{\mu}\bigr)\hspace{0.02cm}\partial^{\mu} \\
&-\,
\delta^{\hspace{0.02cm}4/3}\,\frac{\!1}{\,\varrho^{2/3}}\,
\bigl(A\hspace{0.02cm}\eta^{\prime}_{\mu}\hspace{0.02cm}A\hspace{0.02cm}\eta^{\prime}_{\nu}
\bigr)\hspace{0.02cm}\partial^{\mu}\partial^{\nu}.
\end{split}
\]
Further, by virtue of the fact that the solution $\varphi(x; z)$ is regular at $z = q$,
it can be presented in the form of a formal series in positive powers of $\delta^{1/3}$:
\begin{equation}
\varphi(x;z) =  \varphi_{0}(x) +\delta^{\hspace{0.02cm}1/3}\hspace{0.02cm}\varphi_{1/3}(x) + \delta^{\hspace{0.02cm}2/3}\hspace{0.02cm}\varphi_{2/3}(x) + \delta\hspace{0.02cm}\varphi_{1}(x)\, + \,\ldots\,.
\label{eq:77p}
\end{equation}
Substituting the expansions (\ref{eq:77i})\,--\,(\ref{eq:77p}) into the relation (\ref{eq:77e}) and collecting terms of the same power in $\delta^{\hspace{0.01cm}1/3}$, we obtain that $\psi_{-1}(x)=\psi_{-4/3}(x) =\,\ldots\,= 0$ and
\begin{align}
%
%
&\psi_{-2/3}(x) = -\,\frac{\!1}{\,\varrho^{\hspace{0.02cm}2/3}}\,
\bigl(A\hspace{0.02cm}\eta_{\mu} \hspace{0.02cm}A\hspace{0.02cm}\eta_{\nu}\bigr)
\hspace{0.02cm}\partial^{\mu}\partial^{\nu} \varphi_{0}(x), \label{eq:77pp}\\
%
%
&\psi_{-1/3\;\,}(x) = -
\frac{\!i}{\,\varrho^{\hspace{0.02cm}1/3}}\,m\hspace{0.03cm}
\bigl(A\hspace{0.02cm}\eta_{\mu}\hspace{0.02cm}A + A^{2}\hspace{0.02cm}\eta_{\mu}\bigr)
\hspace{0.02cm}\partial^{\mu\!}\hspace{0.04cm}\varphi_{0}(x)
-
\,\frac{\!1}{\,\varrho^{\hspace{0.02cm}2/3}}\,
\bigl(A\hspace{0.02cm}\eta_{\mu} \hspace{0.02cm}A\hspace{0.02cm}\eta_{\nu}\bigr)
\hspace{0.02cm}\partial^{\mu}\partial^{\nu\!}\hspace{0.04cm}\varphi_{1/3}(x), \notag\\
%
%
&\psi_{0}(x) = m^{2\!}A^{2} \varphi_{0}(x) -\!
\frac{\!i}{\,\varrho^{\hspace{0.02cm}1/3}}\,m\hspace{0.03cm}
\bigl(A\hspace{0.02cm}\eta_{\mu}\hspace{0.02cm}A + A^{2}\hspace{0.02cm}\eta_{\mu}\bigr)
\hspace{0.02cm}\partial^{\mu} \varphi_{1/3}(x)
-\!\frac{\!1}{\,\varrho^{\hspace{0.02cm}2/3}}\,
\bigl(A\hspace{0.02cm}\eta_{\mu} \hspace{0.02cm}A\hspace{0.02cm}\eta_{\nu}\bigr)
\hspace{0.02cm}\partial^{\mu}\partial^{\nu\!}\hspace{0.03cm} \varphi_{2/3}(x), \notag
%
%
\end{align}
and so on. Thus, if $\varphi_{0}(x) \neq 0$ and/or $\varphi_{1/3}(x) \neq 0$, then the solution of the first order differential equation (\ref{eq:77q}) is singular with respect to the parameter $\delta^{\hspace{0.02cm}1/3}$ in the $\delta \rightarrow 0$ limit. The maximal power of the singularity is equal to $2$.\\
\indent
The differential equations to which the functions $\varphi_{0}(x),\,\varphi_{1/3}(x),\,\ldots$ must satisfy are defined by the corresponding expansion of the cube of the operator $\hat{\cal L}(z,\partial)$. With allowance for the expressions (\ref{eq:77u}) and (\ref{eq:77o}), we get the following:
\vspace{-0.3cm}
\begin{flushleft}
1. the singular contributions:
\end{flushleft}
\vspace{-0.7cm}
%
%
\[
\begin{split}
\;\delta^{-1}\!:\hspace{0.45cm}    -\,&\frac{i}{\,\varrho}\,
\bigl(A\hspace{0.02cm}\eta_{\mu}\hspace{0.02cm}A\hspace{0.02cm}\eta_{\nu}A\hspace{0.02cm}
\eta_{\lambda}\bigr)
\hspace{0.02cm}\partial^{\mu}\partial^{\nu}\partial^{\lambda},\\
%
%
\delta^{-2/3\,}:\hspace{0.5cm}  &\frac{\!1}{\,\varrho^{\hspace{0.02cm}2/3}}\,m\hspace{0.03cm}
\bigl(A\hspace{0.02cm}\eta_{\mu}\hspace{0.02cm}A^{2}\hspace{0.02cm}\eta_{\nu} + A^{2}\hspace{0.02cm}\eta_{\mu}A\hspace{0.02cm}\eta_{\nu} + A\hspace{0.02cm}\eta_{\mu}A\hspace{0.02cm}\eta_{\nu}\hspace{0.02cm}A\bigr)
\hspace{0.02cm}\partial^{\mu}\partial^{\nu},\\
%
%
\delta^{-1/3\,}:\hspace{0.5cm}  &\frac{\!i}{\,\varrho^{\hspace{0.02cm}1/3}}\,m^{2}\hspace{0.03cm}
\bigl(A^{3}\hspace{0.01cm}\eta_{\mu} + A\hspace{0.02cm}\eta_{\mu}\hspace{0.02cm}A^{2} +
A^{2}\hspace{0.02cm}\eta_{\mu}\hspace{0.02cm}A\bigr)
\hspace{0.02cm}\partial^{\mu}.
\end{split}
\]
The first expression vanishes by virtue of the nilpotency property: $(\eta \cdot \partial)^3 = 0$. The other two vanish on the strength of the properties (\ref{eq:5u}) and (\ref{eq:1a}). Further, we have
\vspace{-0.3cm}
\begin{flushleft}
2. the regular contributions:
\end{flushleft}
\vspace{-0.7cm}
\[
\begin{split}
%
%
\delta^{\hspace{0.02cm}0}:\hspace{0.45cm}    -\,&\frac{i}{\,\varrho}\,
\bigl[A\hspace{0.02cm}\eta^{\prime}_{\mu}\hspace{0.02cm}A\hspace{0.02cm}\eta_{\nu}
A\hspace{0.02cm}\eta_{\lambda}
+
A\hspace{0.02cm}\eta_{\mu}\hspace{0.02cm}A\hspace{0.02cm}\eta^{\prime}_{\nu}
A\hspace{0.02cm}\eta_{\lambda}
+
A\hspace{0.02cm}\eta_{\mu}\hspace{0.02cm}A\hspace{0.02cm}\eta_{\nu}
A\hspace{0.02cm}\eta^{\prime}_{\lambda}\hspace{0.02cm}\bigr]
\hspace{0.02cm}\partial^{\mu}\partial^{\nu}\partial^{\lambda}
- m^{2}I \equiv \hat{\cal U}_{\hspace{0.03cm}0}(\partial),\!\\
%
%
\delta^{\hspace{0.02cm}1/3\;\;}:\hspace{0.5cm}  &\frac{\!1}{\,\varrho^{\hspace{0.02cm}2/3}}\,m\hspace{0.03cm}
\bigl[\bigl(A\hspace{0.02cm}\eta^{\prime}_{\mu}\hspace{0.02cm}A^{2}\hspace{0.02cm}\eta_{\nu} + A^{2}\hspace{0.02cm}\eta^{\prime}_{\mu}A\hspace{0.02cm}\eta_{\nu}\bigr)
+
\bigl(A\hspace{0.02cm}\eta_{\mu}\hspace{0.02cm}A^{2}\hspace{0.02cm}\eta^{\prime}_{\nu} + A^{2}\hspace{0.02cm}\eta_{\mu}A\hspace{0.02cm}\eta^{\prime}_{\nu}\bigr)\\
&\quad\hspace{5.2cm}+
\bigl(A\hspace{0.02cm}\eta^{\prime}_{\mu}A\hspace{0.02cm}\eta_{\nu}\hspace{0.02cm}A +
A\hspace{0.02cm}\eta_{\mu}A\hspace{0.02cm}\eta^{\prime}_{\nu}\hspace{0.02cm}A\bigr)\bigr]
\hspace{0.02cm}\partial^{\mu}\partial^{\nu}\equiv \hat{\cal U}_{\hspace{0.03cm}1/3}(\partial),\\
%
%
\hspace{-0.02cm}\delta^{\hspace{0.02cm}2/3\;\,}:\hspace{0.5cm}  &\ldots\,.
\end{split}
\]
Substituting this expansion of the cube of the operator $\hat{\cal L}(z; \partial)$ and the expansion (\ref{eq:77p}) in (\ref{eq:77r}), we obtain the desired equations for the functions $\varphi_0(x),\, \varphi_{1/3}(x),\,\ldots\,,$
\begin{align}
%
%
\delta^{\hspace{0.02cm}0\quad}:\hspace{0.5cm}
&\hat{\cal U}_{\hspace{0.03cm}0}(\partial)\hspace{0.02cm}\varphi_{0}(x) = 0, \label{eq:77a}\\
%
%
\delta^{\hspace{0.02cm}1/3\;\,}:\hspace{0.5cm}  &\hat{\cal U}_{\hspace{0.03cm}0}(\partial)\hspace{0.02cm}\varphi_{1/3}(x)
+
\hat{\cal U}_{\hspace{0.03cm}1/3}(\partial)\hspace{0.02cm}\varphi_{0}(x) = 0, \label{eq:77s}\\
%
%
\delta^{\hspace{0.02cm}2/3\;\,}:\hspace{0.5cm}  &\ldots\,. \notag
\end{align}
By using the following relations\footnote{\,The first pair of the relations is a direct consequence of (\ref{eq:4u}) and (\ref{eq:5u}). The third relation is easiest to obtain from Eq.\,(\ref{eq:6t}) by differentiating with respect to $z$ and setting then $z = q$ or by a straightforward calculation with the use of the original definitions of the $\eta$\hspace{0.02cm}- and $\eta^{\prime}$-matrices
(Eqs.\,(\ref{eq:5y}), (\ref{eq:77y})) and of the relations (\ref{ap:C2})\,--\,(\ref{ap:C4}) from Appendix \ref{appendix_C}.} between the $\eta$\hspace{0.02cm}- and $\eta^{\prime}$-matrices,
\[
\left\{
\begin{array}{ll}
A^{2}\hspace{0.02cm}\eta^{\prime}_{\mu}\hspace{0.02cm}A = \displaystyle\frac{1}{m}\,
\bigl(\hspace{0.02cm}\eta_{\mu} + q^{2}\hspace{0.02cm}\eta^{\prime}_{\mu}\bigr),\\[1.5ex]
A\hspace{0.02cm}\eta^{\prime}_{\mu}\hspace{0.02cm}A^{2} = \displaystyle\frac{1}{m}\,
\bigl(\hspace{0.02cm} - \eta_{\mu} + q\hspace{0.03cm}\eta^{\prime}_{\mu}\bigr),
\end{array}
\right.
\]
\[
\bigl(\hspace{0.03cm}\eta^{\prime}_{\mu}\eta_{\lambda}\eta_{\nu} + \eta_{\nu}\eta_{\lambda}\eta^{\prime}_{\mu}\bigr)
+
\bigl(\hspace{0.03cm}\eta_{\mu}\eta^{\prime}_{\lambda}\eta_{\nu} + \eta_{\nu}\eta^{\prime}_{\lambda}\eta_{\mu}\bigr)
+
\bigl(\hspace{0.03cm}\eta_{\mu}\eta_{\lambda}\eta^{\prime}_{\nu} + \eta^{\prime}_{\nu}\eta_{\lambda}\eta_{\mu}\bigr)
=
\varrho\hspace{0.02cm}\bigl(g_{\mu\lambda}\hspace{0.02cm}\eta_{\nu} +
g_{\nu\lambda}\hspace{0.02cm}\eta_{\mu}\bigr)
\]
and the properties (\ref{eq:5u}), it is not difficult to show that the operator $\hat{\cal U}_{\hspace{0.03cm}0}(\partial)$ is reduced to our third order wave operator
\[
\hat{\cal U}_{\hspace{0.03cm}0}(\partial) = \biggl(
-i\,\frac{1}{m}\;\Box\hspace{0.04cm}\eta_{\mu}\hspace{0.02cm}\partial^{\mu} - m^{2\!}\hspace{0.02cm}I
\biggr).
\]
The corresponding first ``correction'' to the operator can result to the following simple form:
\[
\hat{\cal U}_{\hspace{0.03cm}1/3}(\partial) = -\varrho^{1/3}\bigl[\hspace{0.02cm}
\eta_{\mu}\eta_{\nu} - q\hspace{0.03cm}\eta^{\prime}_{\mu}\eta_{\nu} +
q^{2}\hspace{0.02cm}\eta_{\mu}\eta^{\prime}_{\nu}\hspace{0.02cm}\bigr]
\hspace{0.02cm}\partial^{\mu}\partial^{\nu}.
\]
\indent
By this means in this section we have presented a simple scheme of calculating the wave function $\psi(x;z)$ satisfying the basic first order matrix equation (\ref{eq:6ww}). The scheme is based on using the solution $\varphi(x;z)$ of well-defined third order wave equation (\ref{eq:77r}). We have shown that the required solution $\psi(x; z)$ exhibits a singular character in the limit $z \rightarrow q$. This singularity has a finite (the second) order in the small expansion parameter $\delta^{\hspace{0.02cm}1/3}$. The crucial equation in all schemes of calculations is the equation (\ref{eq:77a}). It is the solution $\varphi_{0}(x)$ of the third order wave equation that enables us  to define a complete solution $\varphi(x;z)$ by means of the relations of (\ref{eq:77s}) type and then via the relations of (\ref{eq:77pp}) type to define a complete solution $\psi(x;z)$ with any degree of accuracy in the parameter $\delta^{1/3}$. The generalization of the results of this section for the case of the presence of an external electromagnetic field in the system under consideration is straightforward.

\section{The Fock-Schwinger proper-time representation}
\setcounter{equation}{0}

In this section we discuss in more detail another difficulty (it has already been mentioned to some extent in Introduction) closely related to noncommutativity of the Duffin-Kemmer-Petiau operator in the presence of an external electromagnetic field
\begin{equation}
L_{D\hspace{0.03cm}\!K\hspace{0.02cm}\!P}(D) = i\hspace{0.02cm}\beta_{\mu}D^{\mu} - m\hspace{0.02cm} I,
\label{eq:8q}
\end{equation}
and the proper divisor $d_{D\hspace{0.03cm}\!K\hspace{0.02cm}\!P}(D)$, Eq.\,(\ref{eq:1r}) among themselves. This difficulty is associated with the impossibility of defining the path integral representation for the spin-1 particle propagator interacting with a background gauge field within the standard DKP theory only. To understand why this is so, we turn again to the Dirac theory. For the spin-$\frac{1}{2}$ case there are a number of well-developed techniques of deriving the path integral representation for the Green's function of a spinor particle in background Abelian \cite{fradkin_1991, fradkin_1966} or non-Abelian \cite{borisov_1982, fradkin_1988} gauge fields. Our main interest here is with the very first step in such a construction. It is connected with the Fock-Schwinger proper-time representation of the inverse Dirac operator
$L^{-1}_{Dirac}(D) = (i\hspace{0.02cm}\gamma_{\mu}\hspace{0.02cm}D^{\mu} - m\hspace{0.02cm}I)^{-1}$. This step consists in ``squaring'' the denominator through multiplying the latter by the corresponding Klein-Gordon-Fock divisor
\begin{equation}
\frac{1}{L_{Dirac}(D)} = \frac{d_{Dirac}(D)}{d_{Dirac}(D)L_{Dirac}(D)},
\label{eq:8w}
\end{equation}
where $d_{Dirac}(D) = (i\gamma_{\mu}\hspace{0.02cm}D^{\mu} + m\hspace{0.02cm}I)$ followed by the Fock-Schwinger proper-time representation (see below). It is worthy of special emphasis that the basis for this ``obvious'' passage is a simple, but very important fact: {\it commutativity} of the operators $d_{Dirac}(D)$ and $L_{Dirac}(D)$ among themselves.\\
\indent
Let us return to the Duffin-Kemmer-Petiau theory. Given the explicit expressions for the operators $L_{D\hspace{0.02cm}\!K\hspace{0.02cm}\!P}(D)$ and $d_{D\hspace{0.02cm}\!K\hspace{0.02cm}\!P}(D)$, Eqs.\,({\ref{eq:8q}) and (\ref{eq:1r}) correspondingly, by analogy with the Dirac case, we seemingly could write at once
\[
\frac{1}{L_{D\hspace{0.02cm}\!K\hspace{0.02cm}\!P}(D)} = \frac{d_{D\hspace{0.02cm}\!K\hspace{0.02cm}\!P}(D)}
{d_{D\hspace{0.02cm}\!K\hspace{0.02cm}\!P}(D)L_{D\hspace{0.02cm}\!K\hspace{0.02cm}\!P}(D)},
\]
and further follow the known procedure. However, by virtue of noncommutativity of these two operators among themselves, the expression on the right-hand side is clearly meaningless. For the construction of the needed path integral representation for the $L^{-1}_{D\hspace{0.03cm}\!K\hspace{0.02cm}\!P}(D)$ operator we inevitably come to the necessity of introducing into consideration a divisor that would commute with $L_{D\hspace{0.03cm}\!K\hspace{0.02cm}\!P}(D)$ and eventually result in the third order wave equation. Below we will briefly describe our approach to the problem under consideration.\\
\indent
We return again to the spin-$\frac{1}{2}$ case. Following the paper by Fradkin and Gitman \cite{fradkin_1991} instead of the initial Dirac operator $L_{Dirac}(D)$ we introduce the operator transformed by the factor $i\hspace{0.02cm}\gamma_{5}$,
\[
\hat{\cal L}\equiv\hat{\cal L}_{Dirac}(D) = i\hspace{0.02cm}\gamma_{5}\hspace{0.02cm}(\hspace{0.02cm}
i\hspace{0.02cm}\gamma_{\mu}\hspace{0.02cm}D^{\mu} - m\hspace{0.02cm}I).
\]
We know this operator to be the square root of the Klein-Gordon-Fock operator (more exactly, one of its roots), as it is defined by means of Eq.\,(\ref{eq:1i}) with the replacement $\partial_{\mu} \rightarrow D_{\mu}$. It is precisely for the inverse operator $\hat{\cal L}^{-1}$ that it is natural to determine the Fock-Schwinger proper-time representation. We believe the operator $\hat{\cal L}$ to be an odd (Fermi) one by definition. Instead of (\ref{eq:8w}) we have now
\begin{equation}
\frac{1}{\hat{\cal L}} \equiv \frac{\hat{\cal L}}{\hat{\cal L}^{2}} =
i\!\int\limits_{0}^{\infty}\!d\hspace{0.02cm}\tau\!
\int\!\frac{d\chi}{\tau}\;{\rm e}^{\displaystyle{-\hspace{0.02cm}i\hspace{0.04cm}\tau
(\hat{H} - i\hspace{0.01cm}\epsilon) + \tau\chi\hat{\cal L}}},\quad
\epsilon\rightarrow +\hspace{0.01cm}0,
\label{eq:8e}
\end{equation}
where
\[
\hat{H} \equiv \hat{\cal L}^{2} = -\bigl(D^{2} + m^{2}\bigr)I + \frac{e}{2}\;\sigma_{\mu\nu}F^{\mu\nu}(x),
\quad
\sigma_{\mu\nu} = \frac{1}{2\hspace{0.02cm}i}\,
[\hspace{0.02cm}\gamma_{\mu},\gamma_{\nu}\hspace{0.02cm}],
\]
$\tau$ is an even variable and $\chi$ is an odd (Grassmann) variable, anticommuting by definition with $\hat{\cal L}$. The pair $(\tau,\chi)$ is treated as a proper supertime. By virtue of the fact that the Hamilton operator $\hat{H}$ is represented by a product of two Fermi operators, it is an effective Bose operator as it must be.\\
\indent
For the case of DKP theory as the operator $\hat{\cal L}$  we take the cubic root of the third order wave operator, namely the expression (\ref{eq:77w}). Let us assume that this operator is a para-Fermi operator (parastatistics of order two). In this case it is not difficult to write an analog of the Fock-Schwinger proper-time representation for the inverse operator $\hat{\cal L}^{-1}$  similar to (\ref{eq:8e}),
\begin{equation}
\frac{1}{\hat{\cal L}} \equiv \frac{\hat{\cal L}^{2}}{\hat{\cal L}^{3}} =
i\!\int\limits_{0}^{\infty}\!d\hspace{0.02cm}\tau\!
\int\!\frac{d^{\,2}\chi}{\tau^{2}}\;\hspace{0.02cm}
{\rm e}^{\displaystyle{-\hspace{0.02cm}i\hspace{0.01cm}\tau\bigl (\hat{H}(z) - i\hspace{0.01cm}\epsilon\hspace{0.02cm}\bigr) +
\frac{1}{2}\,\bigl(\hspace{0.02cm}\tau\hspace{0.02cm}[\hspace{0.03cm}\chi,\hat{\cal L}\hspace{0.02cm}] + \frac{1}{4}\,\tau^{2\,}[\hspace{0.03cm}\chi,\hat{\cal L}\hspace{0.02cm}]^{\hspace{0.02cm}2}\hspace{0.03cm}\bigr)}},
\label{eq:8r}
\end{equation}
where now
\[
\hat{H}(z) \equiv \hat{\cal L}^{\hspace{0.01cm}3}(z,D)
\]
and $\chi$ is a para-Grassmann variable of order $p = 2$ (i.e., $\chi^{3} = 0$) with the rules of an integration \cite{omote_1979}
\[
\int\!d^{\,2}\chi  = 0 = \int\!d^{\,2}\chi\,[\hspace{0.03cm}\chi,\hat{\cal L}\hspace{0.03cm}],  \quad
\int\!d^{\,2}\chi\,[\hspace{0.03cm}\chi,\hat{\cal L}\hspace{0.03cm}]^{\,2} = 4\hspace{0.02cm}\hat{\cal L}^{2}.
\]
We consider that the para-Grassmann variable $\chi$ and the operator $\hat{\cal L}$ conform to the
following rules of commutation:
\[
[\hspace{0.03cm}[\hspace{0.03cm}\chi,\hat{\cal L}\hspace{0.04cm}], \hat{\cal L}\hspace{0.03cm}] = 0,
\quad
[\hspace{0.03cm}[\hspace{0.02cm}\chi,\hat{\cal L}\hspace{0.04cm}], \chi\hspace{0.03cm}] = 0.
\]
As a proper para-supertime here it is necessary to take a triple $(\tau, \chi, \chi^2)$. The Hamilton operator $\hat{H}(z)$ represents a product of three para-Fermi operators; therefore, in this case, too, the $\hat{H}(z)$ is an effective Bose operator. Of course, only its limiting value has a physical meaning
\begin{align}
\hat{H} &= \lim_{z\hspace{0.02cm}\rightarrow\hspace{0.03cm} q}\hspace{0.02cm} \hat{H}(z)  \notag\\
&= \lim_{z\hspace{0.02cm}\rightarrow\hspace{0.03cm} q}\hspace{0.02cm} \hat{\cal L}^{3}(z,D) =
\lim_{z\hspace{0.02cm}\rightarrow\hspace{0.03cm} q}\hspace{0.02cm}
\biggl[\hspace{0.03cm}A\hspace{0.02cm}\biggl(\frac{\!i}{\,\varepsilon^{1/3}(z)}\,\eta_{\mu}(z)
\hspace{0.03cm}D^{\mu} - m\hspace{0.02cm}I\biggr)\biggr]^{3}, \notag
\end{align}
where the most right-hand side limit is defined by the expression (\ref{eq:7o}).\\
\indent
However, it should be specially noted that the expression (\ref{eq:8r}) for an arbitrary value of the deformation parameter $z$ is meant here as a purely formal one, since the fact of the presence of supersymmetry corresponding to parastatistics of order two has not been demonstrated by us explicitly. In addition, the indicated parasupersymmetry most likely does not take place for $z$ distinct from $q$ (or from $q^2$). The final conclusion about the existence of this symmetry can be made only after constructing the appropriate path integral representation and passage to the limit $z \rightarrow q$. In this connection one should point out a similar situation taking place in the case of the so-called deformed Heisenberg algebra with reflection $R$ introduced by Wigner \cite{wigner_1950}. The algebra includes a real-valued parameter $\nu$, and reveals very peculiar properties for special, {\it discrete} values of this parameter $(\nu = - (2\hspace{0.02cm}p\,+\,1),\; p=1,\,2,\,\ldots\,)$. This very nontrivial fact was first found by Plyushchay \cite{plyushchay_1997}, who pointed out the relationship of the $(2p\,+\,1)$\,-\,dimensional representations of the $R$\,-\,deformed Heisenberg algebra to parafermions of order $2\hspace{0.02cm}p$. These special values of the parameter $\nu$ also reveal themselves in the context of supersymmetry as well as field theory \cite{klishevich_1999}.\\
\indent
Besides the presence of (local) parasupersymmetry of order two in the system under consideration will ensure the consistency of the minimal coupling prescription for the case of the DKP theory considered in our paper, as it is, for example, for the spin-$1/2$ charged field. In the latter case a local supersymmetric structure associated with the Dirac equation guarantees a consistency of the minimal coupling prescription when a free theory is generalized for the case of interactions with an external electromagnetic field.\\
\indent
The expression (\ref{eq:8r}) can be taken as the starting one for the construction of the desired path integral representation with the use of an appropriate system of coherent states in a close analogy with the approach proposed by Borisov and Kulish \cite{borisov_1982} for the spin-$\frac{1}{2}$ case. Here, it would be possible to make good use of the known for a long time \cite{volkov_1959, chernikov_1962, ryan_1963} connection between the trilinear algebra of $\beta$-matrices and the para-Fermi algebra of order two. In fact all apparatus needed for the construction of the path integral representation (coherent states, formulas of orthonormality and completeness, and so on) can be found in the papers by Kamefuchi and coworkes \cite{omote_1979}. However, in our case instead of the matrices $\beta_{\mu}$ we have the matrices $\eta_{\mu}(z)$. The trilinear relations for these matrices formally coincide only in the limit $z \rightarrow q$. As a hint of what we shall do in this more complicated situation, the unpublished paper by Dunne \cite{dunne_1997_2} can serve. In the latter it was shown how one can define the creation and annihilation operators explicitly depending on the deformation parameter $z$ and the corresponding relations of commutation with the subsequent passage to the limit $z \rightarrow q$ resulting in the finite expressions.\\
\indent
In closing let us mention another fact closely related to the subject matter of this section. In the literature there are very few papers dealing with the problem of construction of an action for a relativistic classical spinning particle using the para-Grassmann variables with subsequent quantization of the classical model \cite{gershun_1985, korchemsky_1991, fleury_1996_2}. Here we note only that in the action suggested in these papers there are the linear and quadratic in para-Grassmann variable $\chi$ (in our notations) terms, which are similar to those in the exponential function in the expression (\ref{eq:8r}). In our case these terms automatically appear in defining the Fock-Schwinger proper-time representation, and in the works \cite{gershun_1985, korchemsky_1991, fleury_1996_2} they insure the invariance of the action under the local world line para-SUSY transformation. However, the kinetic part of the action in \cite{gershun_1985, korchemsky_1991, fleury_1996_2} was chosen in a complete analogy with the kinetic one for the classical models of a Dirac particle, whereas we expect based on the general formula (\ref{eq:8r}) that the situation here may be more complicated.

\section{Conclusion}
\setcounter{equation}{0}
\label{section_12}

In this paper, we have set up the formalism needed to construct a cubic root of the third order wave operator within the framework of Duffin-Kemmer-Petiau theory. One of the key points in our approach is the introduction into consideration of the so-called deformed relation of commutation, Eq.\,(\ref{eq:4q}). On the basis of the latter a new set of the spin matrices $\eta_{\mu}$ was defined instead of the standard DKP matrices $\beta_{\mu}$. It was shown that the third order wave operator is obtained as a formal limit of the cube of a certain first order differential operator. This operator is singular with respect to the deformation parameter $z$ when the latter approaches the primitive cubic root of unity $q$. Finally, we suggested a way to apply the derived expression for the cubic root to the problem of the construction of the path integral representation for the Green's function of a spin-1 particle in an external electromagnetic field.\\
 \indent
A few words may be said here about the para-Grassmann variables, which will be used in the construction of the desired path integral representation. Although the para-Grassmann algebra of order $p=2$ is still quite visible for concrete calculations, however, probably in the situation under consideration the use of its bilinear version \cite{kwasniewski_1985, baulieu_1991, fleury_1996_1, filippov_1992, filippov_1996, isaev_1997} (sometimes it is named the generalized Grassmann algebra) is more suitable. It is connected with the fact that on the one hand the primitive cubic root $q$ explicitly enters into the definition of the $\eta$\hspace{0.02cm}-\hspace{0.02cm}matrices, into the commutation relations and so on, and on the other hand the use of a primitive $n$th root of unity (in particular for $n=3$) is directly laid in the basis of the new para-Grassmann calculus.\\
\indent
It only remains for us to say a few words about the massless limit of the third order wave operator. Throughout this paper we have considered that the parameter $m$, which by convention is responsible for the mass of a particle, was not equal to zero. As it is known \cite{harish-chandra_1946} in the massless variant of DKP theory the scalar ``mass matrix'' $m\hspace{0.02cm}I$ must be replaced by a singular matrix $M \omega^2$, where $M$ is an arbitrary constant with the dimension of mass. The matrix $M \omega^2$ does not commute any more with everything, as this occurs for $m\hspace{0.02cm}I$ (see the text after Eq.\,(\ref{eq:1s})). The last circumstance qualitatively changes the whole picture of calculations in comparison with the massive case. Thus, for example, here there is no analog of the formula (\ref{eq:1p}). Therefore, now we can only speculate that in the massless case instead of the expression (\ref{eq:1d}) there should be something of the type
\[
\bigl[\hspace{0.02cm}{\cal B}\hspace{0.03cm}(\hspace{0.02cm}i\hspace{0.02cm}\beta_{\mu}
\hspace{0.02cm}\partial^{\mu} - M\omega^{2})\bigr]^{3}
=
-i\,\frac{1}{M}\,\hspace{0.02cm}\Box\hspace{0.03cm}\beta_{\mu}\hspace{0.02cm}\partial^{\mu},
\]
where ${\cal B}$ is a certain matrix depending on $\omega$ and additional parameters, with a possible replacement of the matrices $\beta_{\mu}$ by a more complicated combination. Preliminary consideration has shown that, probably, one of the crucial factors here is the use of the cubic roots of {\it minus} unity rather than of unity. This problem requires separate careful consideration.


\section*{\bf Acknowledgments}

The authors are grateful to the referee for valuable remarks that helped us to improve the manuscript appreciably. This work was supported in part by the grant of the President of Russian Federation for the support of the leading scientific schools (Grant No. NSh-5007.2014.9).

\newpage

\begin{appendices}
\numberwithin{equation}{section}

\section{The $\omega$\hspace{0.02cm}-$\beta_{\mu}$ matrix algebra}
\label{appendix_A}

In this Appendix we give some necessary formulas of the $\omega$\hspace{0.02cm}-$\beta_{\mu}$ matrix algebra for the spin-1 case, which are used throughout in the text. Details of the proof of these formulas and also their generalizations to higher dimensions can be found in the papers by Harish-Chandra \cite{harish-chandra_1946} and Fujiwara \cite{fujiwara_1953}. We use the metric $g_{\mu\nu}={\rm diag}(1,-1,-1,-1)$. Let us recall the definition of the $\omega$ matrix:
$$
\omega = \displaystyle\frac{i}{4}\,\epsilon^{\hspace{0.02cm}\mu\nu\lambda\sigma}\beta_{\mu}\beta_{\nu}
\beta_{\lambda}\beta_{\sigma}.
$$
Then in view of the above definition and the trilinear relation for $\beta$-matrices, Eq.\,(\ref{eq:1e}), we have
\begin{align}
&\omega^3 = \omega,  &\label{ap:A1}\\
&\omega^{2\!}\hspace{0.035cm}\beta_{\mu} + \beta_{\mu\,}\omega^2 = \beta_{\mu}, \label{ap:A2}\\
&\omega\beta_{\mu\,}\omega = 0, \label{ap:A3}\\
&\beta_{\mu}\beta_{\nu\,}\omega + \omega\beta_{\nu}\beta_{\mu} = \omega\hspace{0.02cm} g_{\mu\nu},  \label{ap:A4}\\
&\omega^{2\!}\hspace{0.04cm}\beta_{\mu}\beta_{\nu} = \beta_{\mu}\beta_{\nu\,}\omega^2,  \label{ap:A5}\\
&\beta_{\mu\,}\omega\beta_{\nu} + \beta_{\nu\,}\omega\beta_{\mu} = 0.  \label{ap:A6}
\end{align}
The next formulas
\begin{align}
&\{\beta_{\mu},\beta_{\nu}\}\,\omega + \omega\hspace{0.02cm}\{\beta_{\mu},\beta_{\nu}\} =
2\hspace{0.02cm}\omega\hspace{0.02cm} g_{\mu\nu},  &\label{ap:A7}\\
&[\hspace{0.02cm}\beta_{\mu},\beta_{\nu}\hspace{0.02cm}]\,\omega - \omega\,[\hspace{0.02cm}\beta_{\mu},\beta_{\nu}\hspace{0.02cm}] = 0 \label{ap:A8}
\end{align}
are an obvious consequence of (\ref{ap:A4}). If one defines the matrix $B\equiv\beta^{\mu}\beta_{\mu}$, then the following relations are also valid:
\begin{align}
\omega^2 = 3 - B, \quad B\hspace{0.025cm}\omega = \omega B = 2\hspace{0.025cm} \omega.
 \label{ap:A9}
\end{align}
Besides the useful contractions are
\[
\beta^{\mu}\omega^{2\!}\hspace{0.02cm}\beta_{\mu} = 3\hspace{0.025cm}(1 - \omega^2), \quad
\beta^{\mu\!}\hspace{0.01cm}\beta^{\nu\!}\hspace{0.01cm}\beta_{\mu}\beta_{\nu} = 3 - \omega^2,
\quad
\beta^{\mu}\beta_{\nu}\beta_{\mu} = \beta_{\nu}.
 \]

\section{\bf Construction of the matrix ${\cal A}$}
\label{appendix_B}
\numberwithin{equation}{section}

We write down once again an explicit form of the matrices $(A,\hspace{0.02cm}A^{2\!}, A^{3})$ obtained in section 2
\begin{equation}
\begin{split}
&A = \alpha\hspace{0.02cm}\Bigl[\,I+ i\hspace{0.02cm}\frac{\sqrt{3}}{2}\,\omega - \frac{3}{2}\,\hspace{0.02cm}\omega^2\Bigr],\\
&A^{2} = \alpha^{2} \Bigl[\,I- i\hspace{0.02cm}\frac{\sqrt{3}}{2}\,\omega - \frac{3}{2}\,\hspace{0.02cm}\omega^2\Bigr],\\
&A^{3} = \alpha^{3} I \equiv \frac{1}{m}\, I.
\end{split}
\label{ap:B1}
\end{equation}
An immediate consequence of (\ref{ap:B1}) is the following relation:
\begin{equation}
\frac{1}{\alpha}\,A + \frac{1}{\alpha^{2}}\,A^{2} + \frac{1}{\alpha^{3}}\,A^{3} =
3\hspace{0.025cm}(I - \omega^{2}).
\label{ap:B2}
\end{equation}
Let us construct such a matrix ${\cal A}$ that simultaneously satisfies two requirements:
\begin{equation}
\frac{1}{\alpha}\,{\cal A} + \frac{1}{\alpha^{2}}\,{\cal A}^{2} + \frac{1}{\alpha^{3}}\,{\cal A}^{3} = 0
\label{ap:B3}
\end{equation}
and
\begin{equation}
{\cal A}^{3} = \frac{1}{m}\, I.
\label{ap:B4}
\end{equation}
We search for this matrix in the following form
$$
{\cal A} = A + x\hspace{0.02cm}(I - \omega^{2}),
$$
where $x$ is an unknown parameter. It follows that
\begin{equation}
{\cal A}^{2} = A^{2} + x\hspace{0.02cm}(2\alpha + x)\hspace{0.02cm}(I - \omega^{2})
\label{ap:B5}
\end{equation}
and
\begin{equation}
{\cal A}^{3} = \frac{1}{m}\, I  + x\hspace{0.02cm}(x^{2} + 3\hspace{0.02cm}\alpha x + 3\hspace{0.02cm}\alpha^{2} \hspace{0.02cm})(I - \omega^{2}).
\label{ap:B6}
\end{equation}
Substituting the obtained expressions into the left-hand side of the expression (\ref{ap:B3}) and considering (\ref{ap:B2}), we obtain
$$
\frac{1}{\alpha}\,{\cal A} + \frac{1}{\alpha^{2}}\,{\cal A}^{2} + \frac{1}{\alpha^{3}}\,{\cal A}^{3} =
\frac{1}{\alpha^{3}}\,(x^{3} + 4\hspace{0.02cm}\alpha x^{2} + 6\hspace{0.02cm}\alpha^{2}x + 3\hspace{0.02cm}\alpha^{3})\hspace{0.02cm} (I - \omega^{2}).
$$
The requirement of vanishing the expression on the right-hand side results in the following equation for the parameter $x$
$$
x^{3} + 4\hspace{0.02cm}\alpha\hspace{0.02cm}x^{2} + 6\hspace{0.02cm}\alpha^{2}x + 3\hspace{0.02cm}\alpha^{3}
\equiv
(x + \alpha)(x^{2} + 3\hspace{0.02cm}\alpha\hspace{0.02cm} x + 3\hspace{0.02cm}\alpha^{2}) = 0.
$$
It is apparent that the given algebraic equation has three roots but only two of them are compatible to the additional requirement (\ref{ap:B4}). Actually, by virtue of (\ref{ap:B6}) we have another equation for the unknown parameter $x$
$$
x\hspace{0.02cm}(x^{2} + 3\hspace{0.02cm}\alpha x + 3\hspace{0.02cm}\alpha^{2} \hspace{0.02cm}) = 0.
$$
The two roots needed are
\begin{align}
&x = \biggl( - \frac{3}{2} \,+\, i\,\frac{\sqrt{3}}{2}\biggr)\alpha, \notag\\
&x^{\ast} = \biggl( - \frac{3}{2} \,-\, i\,\frac{\sqrt{3}}{2}\biggr)\alpha. \notag
\end{align}
Thus, we find the required expressions for a new set of matrixes $({\cal A},\hspace{0.02cm}{\cal A}^{2},{\cal A}^{3}):$
\begin{align}
&{\cal A} \hspace{0.02cm}=\hspace{0.02cm} A \hspace{0.02cm} +
\hspace{0.02cm}x\hspace{0.03cm}(I - \omega^{2}), \notag\\
&{\cal A}^{2} = A^{2} + \alpha\hspace{0.02cm}x^{\ast}\hspace{0.02cm}(I - \omega^{2}),  \notag\\
&{\cal A}^{3} = \frac{1}{m}\, I. \notag
\end{align}
In deriving the expression for ${\cal A}^{2}$ here, we have considered in (\ref{ap:B5}) the identity
$$
x\hspace{0.03cm}(2\alpha + x)  = \alpha\hspace{0.03cm} x^{\ast}.
$$


\section{\bf The proof of vanishing (\ref{eq:4r})}
\label{appendix_B1}
\numberwithin{equation}{section}

By using the relations (\ref{eq:2e}) and (\ref{eq:2r}) one can rewrite the first, second, fifth and seventh terms in (\ref{eq:4t}) in an identical form:
\[
\begin{split}
&A\hspace{0.02cm}\beta_{\mu}\hspace{0.02cm}A\hspace{0.02cm}\beta_{\nu}
\hspace{0.02cm}A\hspace{0.02cm}\beta_{\lambda}
=
- A\hspace{0.02cm}\beta_{\mu}\hspace{0.02cm}\beta_{\nu}\hspace{0.02cm}A^{2}
\beta_{\lambda}
-
A\hspace{0.02cm}\beta_{\mu}\hspace{0.02cm}A^{2}\beta_{\nu}\beta_{\lambda}\hspace{0.02cm}A, \\
&\beta_{\mu}A\hspace{0.02cm}\beta_{\nu}\hspace{0.02cm}A\hspace{0.02cm}\beta_{\lambda}
\hspace{0.02cm}A
=
- \beta_{\mu}A^{2}\beta_{\nu}\hspace{0.02cm}\beta_{\lambda}\hspace{0.02cm}A
-
\beta_{\mu}\hspace{0.02cm}\beta_{\nu}A^{2}\beta_{\lambda}\hspace{0.02cm}A, \\
&A\hspace{0.02cm}\beta_{\mu}\beta_{\nu}A\hspace{0.02cm}\beta_{\lambda}\hspace{0.02cm}A
=
- A\hspace{0.03cm}\beta_{\mu}\hspace{0.02cm}\beta_{\nu}\hspace{0.02cm}\beta_{\lambda}
\hspace{0.02cm}A^{2}
-
A\hspace{0.02cm}\beta_{\mu}\hspace{0.02cm}\beta_{\nu}\hspace{0.02cm}A^{2}
\beta_{\lambda}, \\
&\beta_{\mu}A\hspace{0.02cm}\beta_{\nu}\hspace{0.02cm}A^{2}\beta_{\lambda}
=
- \hspace{0.02cm}\beta_{\mu}\hspace{0.03cm}A^{2}\beta_{\nu}\hspace{0.02cm}A\hspace{0.02cm}
\beta_{\lambda}
-
\frac{1}{m}\;\beta_{\mu}\hspace{0.02cm}\beta_{\nu}\hspace{0.02cm}\beta_{\lambda}. \notag \\
\end{split}
\]
Substituting these relations into (\ref{eq:4t}) and collecting similar terms, we obtain then
\[
\begin{split}
&-\frac{1}{m}\,z^{2}\beta_{\mu}\hspace{0.02cm}\beta_{\nu}\hspace{0.02cm}\beta_{\lambda}
+
\bigl(\hspace{0.02cm}\beta_{\mu}\hspace{0.02cm}\beta_{\nu}\hspace{0.02cm}A^{2}
\beta_{\lambda}\hspace{0.02cm}A - A\hspace{0.02cm}\beta_{\mu}\hspace{0.02cm}A^{2\!}\hspace{0.02cm}\beta_{\nu}\beta_{\lambda}
\hspace{0.02cm}\bigr)
-
(z + z^{2}\hspace{0.02cm})\beta_{\mu}A^{2}\beta_{\nu}\hspace{0.02cm}A\beta_{\lambda}\\
&+\bigl(\hspace{0.02cm}z\hspace{0.02cm}A^{2}\beta_{\mu}\hspace{0.02cm}
\beta_{\nu}\hspace{0.02cm}\beta_{\lambda}\hspace{0.02cm}A
-
z^{2\!}A\hspace{0.02cm}\beta_{\mu}\hspace{0.02cm}\beta_{\nu}\hspace{0.02cm}\beta_{\lambda}
\hspace{0.02cm}A^{2}\hspace{0.02cm}\bigr)
+
\varepsilon(z)\bigl(\hspace{0.02cm}\beta_{\mu}\hspace{0.02cm}A^{2}\beta_{\nu}
\hspace{0.02cm}\beta_{\lambda}\hspace{0.02cm}A
-
A\hspace{0.02cm}\beta_{\mu}\hspace{0.02cm}\beta_{\nu}\hspace{0.02cm}A^{2\!}
\beta_{\lambda}\hspace{0.02cm}\bigr).
\end{split}
\]
The last term here can be turned into zero if one sets $z=q$. Taking into account another identity
\[
-\,A\hspace{0.02cm}\beta_{\mu}\hspace{0.02cm}A^{2}\beta_{\nu}\beta_{\lambda}
=
A^{2}\beta_{\mu}\hspace{0.02cm}A\hspace{0.02cm}\beta_{\nu}\beta_{\lambda}
+
\frac{1}{m}\,\beta_{\mu}\hspace{0.02cm}\beta_{\nu}\hspace{0.02cm}\beta_{\lambda}
\]
and the equality $q+q^2=-1$, we derive further
\[
\frac{1}{m}\;(1 - q^{2})\hspace{0.02cm}
\beta_{\mu}\hspace{0.02cm}\beta_{\nu}\hspace{0.02cm}\beta_{\lambda}
+
\bigl(\hspace{0.02cm}q\hspace{0.02cm}A^{2}\beta_{\mu}\hspace{0.02cm}\beta_{\nu}\hspace{0.02cm}
\beta_{\lambda}\hspace{0.02cm}A
-
q^{2\!}A\hspace{0.02cm}\beta_{\mu}\hspace{0.02cm}\beta_{\nu}\hspace{0.02cm}\beta_{\lambda}
\hspace{0.02cm}A^{2}\hspace{0.02cm}\bigr)
\]
\[
+\,
\bigl(\hspace{0.02cm}
A^{2}\beta_{\mu}\hspace{0.02cm}A\hspace{0.02cm}\beta_{\nu}\beta_{\lambda}
+
\beta_{\mu}A^{2}\beta_{\nu}\hspace{0.02cm}A\beta_{\lambda}
+
\beta_{\mu}\hspace{0.02cm}\beta_{\nu}A^{2}
\beta_{\lambda}\hspace{0.02cm}A\hspace{0.02cm}\bigr).
\]
The final step is a contraction of this expression with $ \partial^{\mu} \partial^{\nu} \partial^{\lambda}$. Making use again of the relation $q = -1-q^2$, the identity (\ref{eq:1s}) and the property (\ref{eq:2r}), we obtain
\begin{equation}
\frac{1}{m}\;\Box\hspace{0.02cm}\beta_{\mu}\partial^{\mu}
-
\Box\hspace{0.03cm}(A^{2\!}\hspace{0.02cm}\beta_{\mu}A\hspace{0.02cm})\hspace{0.02cm}
\partial^{\mu}
\label{eq:4y}
\end{equation}
\[
+\,
\bigl(\hspace{0.02cm}
A^{2}\beta_{\mu}\hspace{0.02cm}A\hspace{0.02cm}\beta_{\nu}\beta_{\lambda}
+
\beta_{\mu}\hspace{0.02cm}A^{2}\beta_{\nu}\hspace{0.02cm}A\hspace{0.02cm}\beta_{\lambda}
+
\beta_{\mu}\hspace{0.02cm}\beta_{\nu}\hspace{0.02cm}A^{2}
\beta_{\lambda}\hspace{0.02cm}A\hspace{0.02cm}\bigr)
\partial^{\mu}\partial^{\nu}\partial^{\lambda}.
\]
\indent
Now we are concerned with an analysis of terms containing the matrices $A$ and $A^2$. With allowance made for (\ref{eq:4u}), the expression (\ref{eq:4y}) turns to
\begin{equation}
\frac{1}{m}\;\Box\hspace{0.03cm}\beta_{\mu}\hspace{0.02cm}\partial^{\mu}
-
\frac{3}{2\hspace{0.02cm}m}\,\hspace{0.02cm}
\beta_{\mu}\hspace{0.02cm}\beta_{\nu}\hspace{0.02cm}\beta_{\lambda}\hspace{0.02cm}
\partial^{\mu}\partial^{\nu}\partial^{\lambda}
+
\frac{1}{2m}\;\Box\hspace{0.03cm}\beta_{\mu}\hspace{0.02cm}\partial^{\mu}
\label{eq:4i}
\end{equation}
\[
-\,
\frac{1}{2m}\;i\hspace{0.02cm}\sqrt{3}\,\bigl(\hspace{0.02cm}
\xi_{\mu}\hspace{0.02cm}\beta_{\nu}\hspace{0.02cm}\beta_{\lambda}
+
\beta_{\mu}\hspace{0.02cm}\xi_{\nu}\hspace{0.02cm}\beta_{\lambda}
+
\beta_{\mu}\hspace{0.02cm}\beta_{\nu}\hspace{0.02cm}\xi_{\lambda}\hspace{0.02cm}
\bigr)\hspace{0.03cm}
\partial^{\mu}\partial^{\nu}\partial^{\lambda}
+
\frac{1}{2m}\;i\hspace{0.02cm}\sqrt{3}\,\hspace{0.02cm}
\Box\hspace{0.04cm}\xi_{\mu}\hspace{0.03cm}\partial^{\mu}.
\]
By using once more the identity (\ref{eq:1s}) we see that the first three terms here mutually cancel. Furthermore, it is not difficult to verify that the following equality holds
\begin{equation}
\xi_{\mu}\hspace{0.02cm}\beta_{\nu}\hspace{0.02cm}\beta_{\lambda}
+
\beta_{\mu}\hspace{0.02cm}\xi_{\nu}\hspace{0.02cm}\beta_{\lambda}
+
\beta_{\mu}\hspace{0.02cm}\beta_{\nu}\hspace{0.02cm}\xi_{\lambda}
=
\omega\hspace{0.02cm}\beta_{\mu}\hspace{0.02cm}\beta_{\nu}\hspace{0.02cm}\beta_{\lambda}
-
\beta_{\mu}\hspace{0.02cm}\beta_{\nu}\hspace{0.02cm}\beta_{\lambda}\hspace{0.03cm}\omega.
\label{eq:4o}
\end{equation}
In view of (\ref{eq:4o}) the last but one term in (\ref{eq:4i}) takes the form
\[
-\,
\frac{1}{2m}\;i\hspace{0.02cm}\sqrt{3}\,\bigl(\hspace{0.02cm}
\omega\hspace{0.02cm}\beta_{\mu}\hspace{0.02cm}\beta_{\nu}\hspace{0.02cm}\beta_{\lambda}
-
\beta_{\mu}\hspace{0.02cm}\beta_{\nu}\hspace{0.02cm}\beta_{\lambda}\hspace{0.03cm}
\omega\hspace{0.02cm}\bigr)\hspace{0.02cm}\partial^{\mu}\partial^{\nu}\partial^{\lambda}
\equiv
-\,\frac{1}{2m}\;i\hspace{0.02cm}\sqrt{3}\,\hspace{0.02cm}
\Box\hspace{0.04cm}\xi_{\mu}\hspace{0.02cm}\partial^{\mu}.
\]
This term is canceled precisely by the last term in (\ref{eq:4i}).

\section{\bf Trilinear relation for the $\eta$-matrices}
\label{appendix_C}
\numberwithin{equation}{section}

By a direct multiplication of the matrices $\eta_{\mu}$, Eq.\,(\ref{eq:5y}), we derive the starting expression
\begin{align}
&\eta_{\mu}\hspace{0.01cm}\eta_{\nu}\hspace{0.01cm}\eta_{\lambda} + \eta_{\lambda}\hspace{0.01cm}\eta_{\nu}\hspace{0.01cm}\eta_{\mu}
=
\biggl(1 + \frac{1}{2}\;q\biggr)^{\!\!3}
(\beta_{\mu}\hspace{0.02cm}\beta_{\nu}\hspace{0.02cm}\beta_{\lambda} + \beta_{\lambda}\hspace{0.02cm}\beta_{\nu}\hspace{0.02cm}\beta_{\mu}) \label{ap:C1}\\
&+ \biggl(1 + \frac{1}{2}\;q\biggr)^{\!\!2}\biggl(\frac{i\sqrt{3}}{2}\biggr)q
\bigl[\hspace{0.02cm}(\beta_{\mu}\hspace{0.02cm}\beta_{\nu}\hspace{0.02cm}\xi_{\lambda} + \beta_{\mu}\hspace{0.02cm}\xi_{\nu}\hspace{0.02cm}\beta_{\lambda}
+ \xi_{\mu}\hspace{0.02cm}\beta_{\nu}\hspace{0.02cm}\beta_{\lambda})
+ (\mu\leftrightarrow\lambda)\hspace{0.02cm}\bigr] \notag\\
&+ \biggl(1 + \frac{1}{2}\;q\biggr)\biggl(\frac{i\sqrt{3}}{2}\biggr)^{\!\!2}q^{2}
\bigl[\hspace{0.02cm}(\xi_{\mu}\hspace{0.02cm}\xi_{\nu}\hspace{0.02cm}\beta_{\lambda} + \xi_{\mu}\hspace{0.02cm}\beta_{\nu}\hspace{0.02cm}\xi_{\lambda}
+ \beta_{\mu}\hspace{0.02cm}\xi_{\nu}\hspace{0.02cm}\xi_{\lambda})
+ (\mu\leftrightarrow\lambda)\hspace{0.02cm}\bigr] \notag\\
&+ \biggl(\frac{i\sqrt{3}}{2}\biggr)^{\!\!3}q^{3}
(\hspace{0.02cm}\xi_{\mu}\hspace{0.02cm}\xi_{\nu}\hspace{0.02cm}\xi_{\lambda} + \xi_{\lambda}\hspace{0.02cm}\xi_{\nu}\hspace{0.02cm}\xi_{\mu}). \notag
\end{align}
For the first term on the right-hand side we use the basic relation for the $\beta$-matrices,
Eq.\,(\ref{eq:1e}). For the second term it is necessary to use the relation (\ref{eq:4o}), which in view of Eq.\,(\ref{eq:1e}) leads to
\begin{equation}
(\beta_{\mu}\hspace{0.02cm}\beta_{\nu}\hspace{0.02cm}\xi_{\lambda} + \beta_{\mu}\hspace{0.02cm}\xi_{\nu}\hspace{0.02cm}\beta_{\lambda}
+ \xi_{\mu}\hspace{0.02cm}\beta_{\nu}\hspace{0.02cm}\beta_{\lambda})+ (\mu\leftrightarrow\lambda)
=
g_{\mu\nu\,}\xi_{\lambda} + g_{\lambda\nu\,}\xi_{\mu}.
\label{ap:C2}
\end{equation}
In analysis of the third term in (\ref{ap:C1}) we first note that
$$
\xi_{\mu}\xi_{\nu}
=
\{\hspace{0.02cm}\beta_{\mu},\beta_{\nu}\}\hspace{0.03cm}\omega^{2} - g_{\mu\nu}\hspace{0.025cm}\omega^{2} -\beta_{\mu}\beta_{\nu}
$$
and
$$
 \xi_{\mu}\beta_{\nu\,}\xi_{\lambda}
 =
g_{\mu\nu}\hspace{0.03cm}\omega^{2\!}\hspace{0.02cm}\beta_{\lambda} + g_{\lambda\nu}\hspace{0.025cm}\beta_{\mu}\hspace{0.025cm}\omega^{2}
-
(\hspace{0.02cm}\omega^{2}\beta_{\nu}\beta_{\mu}\beta_{\lambda} + \beta_{\mu}\beta_{\lambda}\hspace{0.02cm}\beta_{\nu} \hspace{0.025cm}\omega^{2}\hspace{0.02cm})
-
\omega\beta_{\mu}\beta_{\nu}\beta_{\lambda}\hspace{0.025cm}\omega,
$$
then
\begin{align}
&(\xi_{\mu}\hspace{0.02cm}\xi_{\nu}\hspace{0.02cm}\beta_{\lambda} + \xi_{\mu}\hspace{0.02cm}\beta_{\nu}\hspace{0.02cm}\xi_{\lambda}
+ \beta_{\mu}\hspace{0.02cm}\xi_{\nu}\hspace{0.02cm}\xi_{\lambda}) +
(\mu\leftrightarrow\lambda)
=
-\hspace{0.01cm}2\hspace{0.03cm}(\beta_{\mu}\hspace{0.02cm}\beta_{\nu}\hspace{0.02cm}
\beta_{\lambda} + \beta_{\lambda}\hspace{0.02cm}\beta_{\nu}\hspace{0.02cm}\beta_{\mu}) \label{ap:C3}\\
&+
\omega^{2}
\Bigl[\hspace{0.01cm}\{\beta_{\mu},\beta_{\nu}\}\hspace{0.02cm}\beta_{\lambda} + \{\beta_{\lambda},\beta_{\nu}\}\hspace{0.02cm}\beta_{\mu}
 -
\beta_{\nu}\hspace{0.02cm}\{\beta_{\mu},\beta_{\lambda}\}\hspace{0.01cm}\Bigr] \notag\\
&+\Bigl[\hspace{0.01cm}\beta_{\lambda}\hspace{0.02cm}\{\beta_{\mu},\beta_{\nu}\} + \beta_{\mu}\hspace{0.02cm}\{\beta_{\lambda},\beta_{\nu}\}
 -
\{\beta_{\mu},\beta_{\lambda}\}\hspace{0.02cm}\beta_{\nu}\hspace{0.01cm}\Bigr]\omega^{2} \notag \\
&=
-\hspace{0.01cm}2\hspace{0.03cm}(g_{\mu\nu}\hspace{0.02cm}
\beta_{\lambda} + g_{\lambda\nu}\hspace{0.02cm}\beta_{\mu})
+
\omega^{2}\hspace{0.01cm}(g_{\mu\nu}\hspace{0.02cm}
\beta_{\lambda} + g_{\lambda\nu}\hspace{0.02cm}\beta_{\mu})
+
(g_{\mu\nu}\hspace{0.02cm}
\beta_{\lambda} + g_{\lambda\nu}\hspace{0.02cm}\beta_{\mu})\hspace{0.02cm}\omega^{2} \notag\\
&=
-\hspace{0.01cm}(\hspace{0.02cm}g_{\mu\nu}\hspace{0.02cm}
\beta_{\lambda} + g_{\lambda\nu}\hspace{0.02cm}\beta_{\mu}).\notag
\end{align}
In the last step we have used the property (\ref{ap:A2}).\\
\indent
Finally, for the last term in (\ref{ap:C1}) we have
\begin{align}
\xi_{\mu}\hspace{0.02cm}&\xi_{\nu}\hspace{0.02cm}\xi_{\lambda} + \xi_{\lambda}\hspace{0.02cm}\xi_{\nu}\hspace{0.02cm}\xi_{\mu}
=
\beta_{\nu}\hspace{0.01cm}(\hspace{0.01cm}\beta_{\mu\,}\omega\beta_{\lambda} + \beta_{\lambda}\hspace{0.02cm}\omega\beta_{\nu})
-\omega\hspace{0.02cm}(g_{\mu\nu}\hspace{0.02cm}\beta_{\lambda}
+ g_{\lambda\nu}\hspace{0.02cm}\beta_{\mu}) \label{ap:C4}\\
&+ (g_{\mu\nu}\hspace{0.02cm}\beta_{\lambda} + g_{\lambda\nu}\hspace{0.02cm}\beta_{\mu})\hspace{0.03cm}\omega
\equiv
-\hspace{0.01cm}(\hspace{0.02cm}g_{\mu\nu}\hspace{0.02cm}
\xi_{\lambda} + g_{\lambda\nu}\hspace{0.02cm}\xi_{\mu}).  \notag
\end{align}
Here, for the first term on the right-hand side we have used the property (\ref{ap:A6}). Gathering the expressions calculated above and collecting similar terms, we obtain instead of (\ref{ap:C1})
$$
\eta_{\mu}\eta_{\nu}\eta_{\lambda} + \eta_{\lambda}\eta_{\nu}\eta_{\mu}
$$
$$
=
\biggl[\biggl(1 + \frac{1}{2}\,q\biggr)^{\!\!2} -\, \biggl(\frac{i\sqrt{3}}{2}\biggr)^{\!\!2\!}q^{2}
\biggr]\biggl\{\biggl(1 + \frac{1}{2}\,q\biggr)\bigl(\hspace{0.02cm}g_{\mu\nu}\hspace{0.02cm}
\beta_{\lambda} + g_{\lambda\nu}\hspace{0.02cm}\beta_{\mu}\bigr)
+
\biggl(\frac{i\sqrt{3}}{2}\biggr)\hspace{0.02cm}q\hspace{0.02cm}
\bigl(\hspace{0.02cm}g_{\mu\nu}\hspace{0.025cm}
\xi_{\lambda} + g_{\lambda\nu}\hspace{0.025cm}\xi_{\mu}\bigr)\!\biggr\}
$$
$$
\equiv
\varepsilon(q)\hspace{0.02cm}(\hspace{0.02cm}g_{\mu\nu}\hspace{0.02cm}\eta_{\lambda} + g_{\lambda\nu}\hspace{0.02cm}\eta_{\mu}),
$$
whence it follows the trilinear relation (\ref{eq:6t}).

\newpage


\section{Proof of the identity (\ref{eq:7e})}
\label{appendix_D}
\setcounter{equation}{0}

Let us present a product of three covariant derivations $D^{\mu}$ in an identical form
\begin{equation}
D^{\mu}D^{\nu}D^{\lambda} = \{D^{\mu\!},D^{\nu\!},D^{\lambda}\}
\label{ap:D1}
\end{equation}
\[
-\,
\bigl(D^{\lambda}D^{\nu}D^{\mu} + D^{\nu}D^{\mu}D^{\lambda} + D^{\lambda}D^{\mu}D^{\nu}
+
D^{\mu}D^{\lambda}D^{\nu} + D^{\nu}D^{\lambda}D^{\mu}\bigr).
\]
By the symbol $\{D^{\mu\!}, D^{\nu\!}, D^{\lambda}\}$ one means a completely symmetrized expression defined by the formula (\ref{eq:7r}). We transform each term in parentheses in such a manner so that the expression obtained contains the term $D^{\mu}D^{\nu}D^{\lambda}$. Here, we have
\[
\begin{split}
&D^{\lambda}D^{\nu}D^{\mu} = D^{\mu}D^{\nu}D^{\lambda} -
D^{\mu}\hspace{0.02cm}[\hspace{0.02cm}D^{\nu\!},D^{\lambda}\hspace{0.02cm}] - [\hspace{0.02cm}D^{\mu\!},D^{\lambda}\hspace{0.02cm}]\hspace{0.02cm}D^{\nu} - D^{\lambda}\hspace{0.02cm}[\hspace{0.02cm}D^{\mu\!},D^{\nu}\hspace{0.02cm}],\\
&D^{\nu}D^{\mu}D^{\lambda} = D^{\mu}D^{\nu}D^{\lambda} -
[\hspace{0.02cm}D^{\mu\!},D^{\nu}\hspace{0.02cm}]\hspace{0.02cm}D^{\lambda},\\
&D^{\lambda}D^{\mu}D^{\nu} = D^{\mu}D^{\nu}D^{\lambda} -
[\hspace{0.02cm}D^{\mu\!},D^{\nu}\hspace{0.02cm}]\hspace{0.02cm}D^{\lambda} - D^{\nu}\hspace{0.02cm}[\hspace{0.02cm}D^{\mu\!},D^{\lambda}\hspace{0.02cm}],\\
&D^{\mu}D^{\lambda}D^{\nu} = D^{\mu}D^{\nu}D^{\lambda} -
D^{\mu}\hspace{0.02cm}[\hspace{0.02cm}D^{\nu\!},D^{\lambda}\hspace{0.02cm}],\\
&D^{\nu}D^{\lambda}D^{\mu} = D^{\mu}D^{\nu}D^{\lambda} -
[\hspace{0.02cm}D^{\mu\!},D^{\nu}\hspace{0.02cm}]\hspace{0.02cm}D^{\lambda} - D^{\nu}\hspace{0.02cm}[\hspace{0.02cm}D^{\mu\!},D^{\lambda}\hspace{0.02cm}].
\end{split}
\]
Substituting these expressions into (\ref{ap:D1}), taking into account the equality
\[
[\hspace{0.02cm}D^{\mu\!},D^{\nu}\hspace{0.02cm}] = i\hspace{0.015cm}e\hspace{0.005cm}F^{\mu\nu}(x),
\]
and collecting similar terms, leads to (\ref{eq:7e}).

\end{appendices}


\newpage
\begin{thebibliography}{}
%
%
\bibitem{nowakowski_1998}
M. Nowakowski, {\it Phys. Lett. A} {\bf 244} (1998) 329.
\bibitem{kemmer_1939}
N. Kemmer, {\it Proc. R. Soc. A} {\bf 173} (1939) 91.
%
%
\bibitem{umezawa_1956}
H. Umezawa and A. Visconti, {\it Nucl. Phys. B} {\bf 1} (1956) 348.
\bibitem{takahashi_book}
Y. Takahashi, {\it An Introduction to Field Quantization}, Pergamon Press, 1969.
%
%
\bibitem{nagpal_1974}
A.K. Nagpal, {\it Nucl. Phys. B} {\bf 80} (1974) 206.
\bibitem{cox_1976}
W. Cox, {\it J. Phys. A: Math. Gen.} {\bf 9} (1976) 1025.
\bibitem{krajcik_1976}
R.A. Krajcik and M.M. Nieto, {\it Phys. Rev. D} {\bf 13} (1976) 924.
%
\bibitem{borisov_1982}
N.V. Borisov and P.P. Kulish, {\it Theor. Math. Phys.} {\bf 51} (1982) 535.
\bibitem{fradkin_1991}
E.S. Fradkin and D.M. Gitman, {\it Phys. Rev. D} {\bf 44} (1991) 3230.
%
%
\bibitem{finkelstein_1986}
D. Finkelstein, {\it Phys. Rev. Lett.} {\bf 56} (1986) 1532;
D. Finkelstein, S.R. Finkelstein, and C. Holm, {\it Int. J. Theor. Phys.} {\bf 25} (1986) 441.
%
%
\bibitem{solovyov_2001}
A.V. Solov'yov and Yu.S. Vladimirov, {\it Int. J. Theor. Phys.} {\bf 40} (2001) 1511;
A.V. Solov'yov, {\it J. Math. Sci.} {\bf 172} (2011) 894.
\bibitem{yamaleev_1987}
R.M. Yamaleev, {\it Analogs of Clifford Algebra for Forms Above Quadratical and Their Matrix Realization}, preprint P5-87-766, Dubna, 1987.
\bibitem{yamaleev_1988}
R.M. Yamaleev, {\it Elements of Cubic Quantum Mechanics}, preprint P2-88-147, Dubna, 1988.
\bibitem{yamaleev_1989}
R.M. Yamaleev, {\it On Constraction of Quantum Mechanics on Cubic Forms}, preprint E2-89-326, Dubna, 1989.
%
%
\bibitem{pais_1950}
A. Pais and G.E. Uhlenbeck, {\it Phys. Rev.} {\bf 79} (1950) 145.
\bibitem{barut_1970}
A.O. Barut, P. Cordero, G.C. Ghirardi, {\it Nuovo Cimento} {\bf 66} (1970) 36.
%
\bibitem{kruglov_2007}
S.I. Kruglov, {\it Can. J. Phys.} {\bf 85} (2007)  887.
%
%
\bibitem{joos_1962}
H. Joos, {\it Fortschr. Phys.} {\bf 10} (1962) 65.
\bibitem{weinberg_1964}
S. Weinberg, {\it Phys. Rev.} {\bf 133} (1964) B1318.
\bibitem{shay_1965}
D. Shay, H.S. Song, and R.H. Good jr. {\it Nuovo Cimento Suppl.} {\bf 3} (1965) 455.
%
%
\bibitem{bhabha_1945}
H.J. Bhabha, {\it Rev. Mod. Phys.} {\bf 17} (1945) 200; {\it Proc. Indian Acad. Sci.} {\bf 21}
(1945) 241.
\bibitem{krajcik_1974}
R.A. Krajcik and M.M. Nieto, {\it Phys. Rev. D} {\bf 10} (1974) 4049; {\it Am. J. Phys.} {\bf 45} (1977) 818.
%
%
\bibitem{kerner_1992}
R. Kerner, {\it J. Math. Phys.} {\bf 33} (1992) 403; {\it Classical Quantum Gravity} {\bf 9} (1992) 137.
%
%
\bibitem{plyushchay_2000}
M.S. Plyushchay and M. Rausch de Traubenberg, {\it Phys. Lett. B} {\bf 477} (2000) 276.
\bibitem{schrodinger_1943}
E. Schr\"odinger, {\it Proc. Roy Irish. Acad. A} {\bf 48} (1943) 135; {\it ibid.} {\bf 49} (1943) 29.
\bibitem{harish-chandra_1946}
Harish-Chandra, {\it Proc. R. Soc. A} {\bf 186} (1946) 502;
{\it Proc. Cambridge Philos. Soc.} {\bf 43} (1947) 414; {\it Phys. Rev.} {\bf 71} (1947) 793.

%
%

%
%
\bibitem{azimov_1995}
Ya.I. Azimov and R.M. Ryndin, {\it JETP Lett.} {\bf 61} (1995) 453.
\bibitem{nikitin_1976}
A.G. Nikitin, Yu.N. Segeda, V.I. Fushchich, {\it Theor. Math. Phys.} {\bf 29} (1976) 943;
V.I. Fushchich, A.G. Nikitin, {\it Phys. Elemt. Part. Atom. Nucl.} {\bf 14} (1983) 5.

%
%
\bibitem{fujiwara_1953}
I. Fujiwara, {\it Soryushiron Kenkyu} {\bf 4} (1952) 1; {\it Prog. Theor. Phys.} {\bf 10} (1953) 589.
\bibitem{fischbach_1973}
E. Fischbach, M.M. Nieto, and C.K. Scott, {\it J. Math. Phys.} {\bf 14} (1973) 1760;
E. Fischbach, J.D. Louck, M.M. Nieto, and C.K. Scott, {\it J. Math. Phys.} {\bf 15} (1974) 60.
%
%
%
%
%
%
%
%
%
\bibitem{fradkin_1966}
E.S. Fradkin, {\it Nucl. Phys. B} {\bf 76} (1966) 588; A.O. Barut and I.H. Duru, {\it Phys. Rev. Lett.}
{\bf 53} (1984) 2355; F. Cooper, A. Khare, R. Musto and A. Wipf, {\it Ann. Phys.} {\bf 187} (1988) 1;
V.Ya. Fainberg and A.V. Marshakov, {\it Nucl. Phys. B} {\bf 306} (1988) 659; T.M. Aliev, V.Ya. Fainberg, N.K. Pak, {\it Nucl. Phys. B} {\bf 429} (1994) 321; {\it Phys. Rev. D} {\bf 50} (1994) 6594; J.W. van Holten, {\it Nucl. Phys. B} {\bf 457} (1995) 375.
%
%
\bibitem{fradkin_1988}
E.S. Fradkin, Sh.M. Shvartsman, {\it Fortschr. Phys.} {\bf 36} (1988) 831; Sh.M. Shvartsman,
{\it Mod. Phys. Lett. A} {\bf 5} (1990) 943.
%
\bibitem{omote_1979}
M. Omote, S. Kamefuchi, {\it Lett. Nuovo Cimento} {\bf 24} (1979) 345; Y. Ohnuki and S. Kamefuchi, {\it J. Math. Phys.} {\bf 21} (1980) 609.
%
%
\bibitem{wigner_1950}
E.P. Wigner, {\it Phys. Rev.} {\bf 77} (1950) 711.
\bibitem{plyushchay_1997}
M.S. Plyushchay, {\it Nucl. Phys. B} {\bf 491} (1997) 619.
\bibitem{klishevich_1999}
S. Klishevich and M.S. Plyushchay, {\it Mod. Phys. Lett. A} {\bf 14} (1999) 2739; M.S. Plyushchay, {\it Int. J. Mod. Phys. A} {\bf 15} (2000) 3679; {\it Ann. Phys.} {\bf 245} (1996) 339;
P.A. Horv\'athy, M.S. Plyushchay, M. Valenzuela, {\it Ann. Phys.} {\bf 325} (2010) 1931.
%
%
\bibitem{volkov_1959}
D.V. Volkov, {\it Sov. Phys. -- JET\!\hspace{0.02cm}P\,} {\bf 9} (1959) 1107; {\bf 11} (1960) 375.
\bibitem{chernikov_1962}
N.A. Chernikov, {\it Acta Phys. Pol.} {\bf 21} (1962) 51.
\bibitem{ryan_1963}
C. Ryan, E.C.G. Sudarshan, {\it Nucl. Phys.} {\bf 47} (1963) 207.
%
\bibitem{dunne_1997_2}
R.S. Dunne, {\it Intrinsic anyonic spin through deformed geometry}, preprint hep-th/9703137 (1997).
%
%
\bibitem{gershun_1985}
V.D. Gershun, V.I. Tkach, {\it Problems of Nuclear Physics and Cosmic Rays} (Kharkov University Press) {\bf 23} (1985) 42 (in Russian).
\bibitem{korchemsky_1991}
G.P. Korchemsky, {\it Phys. Lett. B} {\bf 267} (1991) 497; {\it Int. J. Mod. Phys. A} {\bf 07} (1992) 3493. \bibitem{fleury_1996_2}
N. Fleury and M. Rausch de Traubenberg, {\it Mod. Phys. Lett. A} {\bf 11} (1996) 899.
%
%
\bibitem{kwasniewski_1985}
A.K. Kwasniewski,  {\it J. Math. Phys.} {\bf 26} (1985) 2234.
%
\bibitem{baulieu_1991}
L. Baulieu and E.G. Floratos, {\it Phys. Lett. B} {\bf 258} (1991) 171.
%
\bibitem{fleury_1996_1}
N. Fleury and M. Rausch de Traubenberg, {\it J. Math. Phys.} {\bf 33} (1992) 3356; M. Rausch de Traubenberg, {\it Adv. Appl. Clifford Alg.} {\bf 4} (1994) 131.
%
\bibitem{filippov_1992}
A.T. Filippov, A.P. Isaev and A.B. Kurdikov, {\it Mod. Phys. Lett. A} {\bf 07} (1992) 2129; {\it Int. J. Mod. Phys. A} {\bf 08} (1993) 4973; {\it Theor. Math. Phys.} {\bf 94} (1993) 150.
\bibitem{filippov_1996}
A.T. Filippov, A.P. Isaev and A.B. Kurdikov, in ``Problems in Modern Theoretical Physics'', Dubna 96-212 (1996) p. 83 (in Russian).
\bibitem{isaev_1997}
A.P. Isaev, {\it Int. J. Mod. Phys. A} {\bf 12} (1997) 201.
%
%
\end{thebibliography}
\end{document}